%% file: ms_redline.tex
\documentclass[preprint]{aastex}

\shorttitle{Survey of IR SNRs in the LMC}
\shortauthors{Seok, Koo, and Onaka}

\begin{document}

\title{A SURVEY OF INFRARED SUPERNOVA REMNANTS IN THE LARGE MAGELLANIC CLOUD}

\author{Ji Yeon Seok}
\email{jyseok@asiaa.sinica.edu.tw}
\affil{Academia Sinica Institute for Astronomy and Astrophysics, P. O. Box 23-141, Taipei 10617, Taiwan}
\affil{Department of Physics and Astronomy, Seoul National University, Seoul 151-742, Korea}
\author{Bon-Chul Koo}
\affil{Department of Physics and Astronomy, Seoul National University, Seoul 151-742, Korea}
\and
\author{Takashi Onaka}
\affil{Department of Astronomy, Graduate School of Science,
University of Tokyo, Bunkyo-ku, Tokyo 113-0033, Japan}

\begin{abstract}

We present a comprehensive infrared study of supernova remnants (SNRs) in the Large Magellanic Cloud (LMC) using near- to mid-infrared images taken by Infrared Array Camera (IRAC; 3.6, 4.5, 5.8, and 8 \micron) and Multiband Imaging Photometer (MIPS; 24 and 70 \micron) onboard the {\it Spitzer Space Telescope}. Among the 47 bona fide LMC SNRs, 29 were detected in infrared, giving a high detection rate of 62\%. All 29 SNRs show emission at 24 \micron, and 20 out of 29 show emission in one or several IRAC bands. We present their 4.5, 8, 24, and 70 \micron~images and a table summarizing their $Spitzer$ fluxes. We find that the LMC SNRs are considerably fainter than the Galactic SNRs, and that, among the LMC SNRs, Type Ia SNRs are significantly fainter than core-collapse SNRs. We conclude that the MIPS emission of essentially all SNRs originates from dust emission, whereas their IRAC emissions originate from ionic/molecular lines, polycyclic aromatic hydrocarbons emission, or synchrotron emission. The infrared fluxes show correlation with radio and X-ray fluxes. For SNRs that have similar morphology in infrared and X-rays, the ratios of 24 to 70 \micron~fluxes have good correlation with the electron density of hot plasma. The overall correlation is explained well by the emission from collisionally-heated silicate grains of 0.1 \micron~size, but for mature SNRs with relatively low gas temperatures, the smaller-sized grain population is favored more. For those that appear different between infrared and X-rays, the emission in the MIPS bands is probably from dust heated by shock radiation.

\end{abstract}


\section{INTRODUCTION}

Before the {\it Infrared Astronomical Satellite} ($IRAS$) was launched, studies of supernova remnants (SNRs), in particular, surveys of a large number of SNRs had been carried out mainly in the radio, optical, and X-ray regimes. The All-Sky survey of the $IRAS$ enabled us to complete an infrared (IR) survey of Galactic SNRs in 10 to 100 \micron~bands for the first time. \citet{arendt89} examined 157 objects and found 51 SNRs with probable IR emission. Later on, \citet{saken92} carried out an independent survey of the IR emission for 161 Galactic SNRs and could find clear IR emission from 35 SNRs with nine additional possible detections. Both studies showed that about 30\% of the known SNRs exhibit some evidence of IR emission related to the SNRs. The IR spectra of SNRs were regarded in general to be dominated by thermal emission from dust. Based on the measured IR fluxes of the SNRs, it is found that young SNRs show stronger emission at 12 and 25 \micron~while older SNRs do at 60 and 100 \micron~\citep{saken92}. This suggested the relation between the evolution of SNRs and dust properties in the SNRs. The IR emission was compared to radio and X-ray emission, both of which generally showed good agreement in their morphologies. Also, some Galactic SNRs showed the morphological similarity between the IR and optical emission implying significant contributions from line emission although any direct evidence for the contributing line emission could not be clearly confirmed \citep{arendt89}. Since the moderate-to-severe source confusion in or near the Galactic plane was inevitable toward the Galactic SNRs, the detection of IR emission in them and the interpretation of it would be limited.

Subsequently, the {\it Infrared Space Observatory (ISO)} observed several SNRs by ISOCAM (2.5--17 \micron), ISO-SWS (2.4--45.4 \micron) and/or ISOPHOT (between 2.5 and 240 \micron), and its high spatial resolutions and the spectroscopic capabilities provided detailed IR structures and the physical origin of the IR emission in the SNRs \citep[e.g.,][]{gall99,tuff99,dou01a, dou01b}. \citet{laga96} revealed a very good spatial correlation between gaseous ionic emission and dust continuum in Cassiopeia A (Cas A) using the ISOCAM, which implies dust formation in the ejecta. Later on, \citet{dou01a} derived composition of the SN-origin dust using ISOCAM and ISO-SWS spectra and discussed the dust heating in Cas A. For three other young Galactic SNRs, the ISOCAM observations showed that the mid-IR (MIR) emissions in the Kepler and Tycho SNRs are dominated by circumstellar and/or interstellar dust rather than the SN-origin, and that the MIR emission in the Crab is dominated by synchrotron radiation \citep{dou01b}. However, the $ISO$ observations were made only for selected historical SNRs (e.g., Tycho, Cas A, and SN 1987A) or plerionic SNRs such as the Crab and SNR 0540--69.3 in the Large Magellanic Cloud (LMC).

Almost two decades later, the {\it Spitzer Space Telescope} offers another opportunity to attempt IR surveys of SNRs. Toward the inner Galactic plane (10$\degr<l<65\degr$ and 285$\degr<l<350\degr$, $|b|<1\degr$), two $Spitzer$ surveys, GLIMPSE \citep{ben03} and MIPSGAL \citep{car09} provide IR data with higher angular resolution and sensitivity in the four bands (3.6, 4.5, 5.8, and 8.0 \micron) of the Infrared Array Camera (IRAC) and the first two bands (24 and 70 \micron) of the Multiband Imaging Photometer (MIPS), respectively. In particular, the IRAC bands cover near-IR (NIR) wavelengths that the $IRAS$ did not, so we can investigate diverse origins of NIR emission, including ionic/molecular line emission and polycyclic aromatic hydrocarbon (PAH) emission. \citet{hglee05} and \citet{reach06} presented new NIR surveys of 100 and 95 SNRs in GLIMPSE, respectively, and as a complement, \citet{pin11} searched for MIR counterparts of 121 SNRs in MIPSGAL. In the area covered by these $Spitzer$ surveys, the previous $IRAS$ surveys by \citet{arendt89} and \citet{saken92} found possible emission from 12 and 14 remnants, respectively, with only seven in common between each other. On the other hand, \citet{hglee05} and \citet{reach06} identified 16 and 18 Galactic SNRs in the IRAC bands respectively, and \citet{pin11} did 39 Galactic SNRs in the MIPS bands. Using GLIMPSE, it is found that the remnants interacting with dense ambient medium lose most of their energy in the shock through molecular or ionic lines. Using MIPSGAL, dust temperatures and masses in the Galactic SNRs are estimated to range from 45 to 70 K and from 0.06 and 2.60 $M_{\sun}$. A correlation is found between the total MIR fluxes (24 and 70 \micron) and the 1.4 GHz non-thermal radio fluxes as seen in external galaxies \citep{helou85}. However, the detection rates of the IR emission still remain about 19\% and 32\% for the IRAC and the MIPS bands albeit the improvement of the data quality, and it is probably because of the inevitable confusion with back/foreground sources and diffuse emission in the line of sight.

In this context, the LMC is the best place to study IR emission from SNRs as it is devoid of bright IR sources from the Galactic plane. Currently, there are more than 50 SNRs with several SNR candidates reported in the literature \citep[e.g.,][]{bade10,desai10}, and new identifications of SNRs are still ongoing \citep[e.g.,][]{gro12}. Using $IRAS$ observations, \citet{graham87} discovered IR emission from three out of four LMC SNRs, of which a considerable fraction could originate from dust grains heated by collisions with hot plasma. As a survey of the LMC SNRs, \citet{schwe89} identify five out of 25 SNRs showing good quality IR emission as well as eight with possible IR emission. While these early studies are constrained by the limited resolution and sensitivity of the $IRAS$, more detailed studies of LMC SNRs have recently been carried out by using $Spitzer$ imaging and spectroscopic data \citep[e.g.,][]{will06, borko06, bwill06}. Besides the $Spitzer$ data, there is an IR survey of the LMC by the $AKARI$ satellite \citep{ita08,kato12}. The $AKARI$ LMC survey covers $\sim10$ deg$^2$ of the LMC including 21 SNRs in five bands (3, 7, 11, 15, and 24 \micron). Using the $AKARI$ LMC survey data, \citet{seok08}\footnote{This paper missed one SNR, N158A (SNR 0540--69.3), of which pulsar wind nebula actually shows IR emission in the $AKARI$ bands. Hence, nine SNRs, in total, show IR emission in the $AKARI$ LMC survey.} identified eight SNRs that have associated IR emission in the NIR and/or MIR bands including three Type Ia SNRs and six core-collapse SNRs (CCSNRs). However, all these works are restricted to limited numbers of SNRs, instead of the entire SNRs in the LMC. By surveying the entire sample of LMC SNRs and examining their properties statistically, we may characterize the IR emission from the LMC SNRs and investigate their dependence on environments. Furthermore, the LMC SNRs can be compared with the Galactic SNRs to assess systematic differences caused by differing dust composition and metallicity.

In this paper, we present the investigation of 47 confirmed LMC SNRs in the 3.6 to 70 \micron~bands using all available $Spitzer$ data. In addition, we compile the fundamental properties of the SNRs such as size, SN type, and age to interpret these IR properties. We find that 29 out of 47 SNRs show associated IR emission and that IR emission is generally correlated with X-ray/optical/radio emission. We discuss differences between the LMC and Galactic SNRs, the origin of their IR emission, and the global perspective of shock processing in LMC SNRs based on IR colors and spatial coincidence with emission at other wavelengths.

\section{DATA AND SNR IDENTIFICATION}

To search for IR emission from SNRs in the LMC, we examine all available archival data of $Spitzer$ at individual positions of the 47 SNRs. There is an imaging survey of the LMC ($\sim7\degr\times7\degr$), the Surveying the Agents of a Galaxy's Evolution (SAGE) survey \citep{meix06}. The SAGE survey provides uniform and unbiased data covering all known LMC SNRs\footnote{\citet{bade10} combined previous SNR catalogs and listed 54 LMC SNRs in the paper. We checked that all of them are covered by the SAGE survey, but we consider 47 SNRs in this paper because some objects still need further confirmation.} in all IRAC and all MIPS (24, 70, and 160 \micron) bands. In addition to the SAGE data, there are a number of $Spitzer$ observations toward LMC SNRs. Particularly, there is a separate survey of Magellanic Clouds SNRs (PI: K. Borkowski) to investigate both the formation and destruction process of interstellar dust related to SNRs \citep[e.g.,][]{borko06, bwill06}. This survey conducts deep MIPS 24 \micron~imaging observations of 33 LMC SNRs, as well as several IRAC and MIPS 70 \micron~imaging toward some SNRs. All $Spitzer$ data, including the SAGE survey are retrieved from the $Spitzer$ Heritage Archive, and the data set we used are post-Basic Calibrated Data created by the standard $Spitzer$ software (ver. S18). In this paper, we used the deepest $Spitzer$ imaging data of LMC SNRs among the all available data, and the data set we used is summarized in Table \ref{tab:lmcsnr}.

Although SNRs in the LMC have less confusion by IR emission from the Galactic sources compared to Galactic SNRs, it is often required to disentangle IR emission of an SNR from IR emissions of other objects in the LMC. Since a CCSNR can be embedded in an \ion{H}{2} region or an \ion{H}{2} complex, it is sometimes difficult to confirm IR emission associated with the SNR based only on IR images. As IR morphology can show similarities to those seen at other wavelengths, we can be assured by comparing the IR data with other multi-wavelength data such as Australia Telescope Compact Array (ATCA) 4.8 and 8.6 GHz radio survey data\footnote{http://www.atnf.csiro.au/research/lmc\_ctm} \citep{dic05}, optical images from Magellanic Cloud Emission Line Survey.\footnote{http://www.ctio.noao.edu/mcels/} \citep[MCELS;][]{smith00}, and X-ray images from $Chandra$ Supernova Remnant Catalog\footnote{http://hea-www.harvard.edu/ChandraSNR/} Figure \ref{lmc} shows the coverage of the SAGE survey with the positions of 47 LMC SNRs on a true color image made of the MCELS data. Almost all SNRs are included in the SAGE, the MCELS, and the ATCA survey, and a half of them are observed by $Chandra$.

When diffuse IR emission from an SNR is faint compared to nearby point sources, it is also difficult to identify IR emission from the SNR, especially at shorter wavelengths. To avoid confusion by point sources, we apply point source subtraction to all IRAC images using Point Spread Function photometry with an IDL (Interactive Data Language) code, $StarFinder$ \citep{dio00}. If necessary, point-source subtracted MIPS 24 \micron~images are used for further analysis, too. Our strategy to identify IR emission associated with SNRs is very simple. Firstly, we search for a well-defined structure such as a shell, a filament, or a sharp boundary, and so on at the location of each SNR in any band (mainly 24 \micron~band but one of IRAC bands for some cases). Then, we check the morphological similarity between the IR structure and the structures seen in X-ray, optical, or radio to confirm its association to the SNR. When the association is clear, then we search the rest of the IR band images including point-source subtracted images for the same structure. After carrying out all processes described above, we find 29 out of 47 SNRs showing IR emission in several IR bands. IR morphologies of 13 SNRs are reported for the first time in this paper: 0450--70.9, SNR in N4, N86, DEM L72, SNR in N206, DEM L238, DEM L241, DEM L249, DEM L256, SNR in N159, DEM L299, DEM L316A/B. In addition, although the detections of N11L at 4.5 \micron~and N103B at 24 \micron~are previously reported \citep{will06,bwill09}, we can also identify their IR emissions in other $Spitzer$ bands; N11L is additionally detected in IRAC 3.6 and 5.8 \micron~bands and MIPS 24 and 70 \micron~bands, and knotty emission of N103B is also detected in four IRAC bands. Non-detected sources are mostly confused by surrounding objects, and a few are too faint to be compared with the background. 

The results of the IR detection by $Spitzer$ are summarized in Table \ref{tab:lmcsnr}. All detected 29 SNRs show IR emission in the MIPS 24 \micron~band, and 20 out of 29 are also seen in the IRAC bands. Table \ref{tab:lmcsnr} also contains the general properties of all SNRs in the LMC including SN type, SNR age, and the results of the $AKARI$ detection from the literature. Among 47 SNRs, 11 and 21 are known to be Type Ia and CCSNRs, respectively, and the SN types of the rest are still uncertain. The youngest SNR is SN 1987A, and ages of the oldest SNRs can reach hundreds of thousands years (e.g., SNR 0450--70.9, N86, and DEM L72). We measure 4.8 GHz radio fluxes of 44 SNRs within the ATCA survey and list them in Table \ref{tab:lmcsnr} for further analysis. The size of the source aperture to measure the radio fluxes is defined as two times larger than the size of each SNR in Table \ref{tab:lmcsnr}, which is large enough to collect most emission from an SNR taking into account the beam size of the ATCA (HPBW$\sim33\arcsec$). If a remnant was smaller than 1\arcmin, we applied a circular aperture with a diameter of 2\arcmin. The background intensity is estimated from an annulus of each source aperture. When a source is severely confused with its surrounding region, the contaminated parts are excluded. The statistical errors (1 $\sigma$) of the 4.8 GHz fluxes are less than 5\%.

IR morphologies at 4.5, 8.0, 24, and 70 \micron~of all 29 SNRs are shown in Figure \ref{irimg1}--\ref{irimg6}. The 4.5 \micron~and 8.0 \micron~IRAC band images are point source subtracted. MCELS H$\alpha$ and $Chandra$ X-ray images are shown for comparison. Most of the detected SNRs show shell-like structures in the MIPS 24 \micron~band. In particular, nine SNRs identified only in the MIPS bands clearly manifest shell-like morphologies which correspond well to their X-ray morphologies (e.g., DEM L71, N23, SNR 0519--69.0, N132D). In the IRAC bands, we investigate both the original and the point source subtracted images to examine diffuse emission against severe confusion by point sources. Many of the SNRs detected in the IRAC bands have good spatial correlations with their optical morphologies. Some SNRs show complete shell structures that are sometimes different from their morphologies seen in the MIPS bands. For other SNRs, filamentary or patchy emissions are only distinguished, and several knots are also detected. More details on each SNR are described in the Appendix.

\section{IR PROPERTIES OF SNRs} 

\subsection{IR Fluxes and Colors \label{sec:irflx}}

For the newly identified SNRs, their fluxes in each band are estimated using the point-source removed IRAC and original MIPS images. If there are bright point sources in MIPS images, we also removed them. We determine areas with IR emission clearly associated with SNRs for flux measurement, and the extracted SNR regions are marked in Figure \ref{scimg}. Background subtraction is coherently applied by using an annulus of the extracted region with 15$\arcsec$ in width. As SNRs tend to be embedded in bright \ion{H}{2} regions or \ion{H}{2} complexes (e.g., SNR in N159), however, it is sometimes difficult to define their own IR emissions. Also, in some SNRs (e.g., DEM L241), the associated IR emission features vary with wavebands. In those cases, fluxes are measured from restricted regions instead of from the entire remnant. For the previously-known IR SNRs, we adopt the fluxes from the literature. If the adopted 24 \micron~fluxes were measured from limited areas in spite of the presence of IR emission in the whole area, we also derive 24 \micron~fluxes from the entire area for further comparisons. Finally, we accumulate the newly estimated fluxes and the previously estimated fluxes in Table \ref{tab:sstflux}.

Figure \ref{fig:mmd} shows the distribution of SNRs in the 24 \micron~and 70 \micron~flux plane. There is a correlation between the fluxes, which is expected to some extent because the emission at 24 and 70 \micron~are usually dominated by dust continuum, although the dust populations contributing to these wavelengths can be different \citep[e.g., Cas A;][]{hines}. It is worth noting that the median 24 \micron~flux is 31 mJy for Type Ia while it is 46 mJy for CCSNR (71 mJy for the unknown type). The median absolute deviations are 28 mJy, 50 mJy, and 33 mJy for Type Ia, CCSNR, and the unknown type, respectively. Considering that the fluxes of many CCSNRs are derived from limited areas, CCSNRs are significantly brighter than Type Ia SNRs in general. CCSNRs such as SNR in N159 could have contaminations from emission of associated \ion{H}{2} regions in their flux measurements even though we attempted to exclude those regions, but emission from most CCSNRs are likely to originate from themselves. Also, the SNRs interacting with molecular clouds are brighter (Section $\ref{sec:irxr}$). Assuming that both 24 and 70 \micron~fluxes originate from dust continuum, these fluxes can be used to estimate the dust mass and temperature by adopting a modified black-body of single-component dust. Flux density, $F_{\nu}$, can be given by
\begin{equation}
F_{\nu}=\frac{\kappa_{\nu}B_{\nu}(T_d)}{d^2}M_d,
\label{eq1}
\end{equation} 
where $T_d$ is the dust temperature, $M_d$ is the dust mass, $\kappa_{\nu}$ is the dust mass absorption coefficient, $B_{\nu}$ is the Planck function, and $d$ is the distance to the LMC (50 kpc). The absorption coefficient is adopted from the ``average'' LMC model of \citet{wein01}\footnote{http://www.astro.princeton.edu/$\sim$draine/dust/dustmix.html}. According to the ranges of dust temperature and dust mass in Figure \ref{fig:mmd}, the 24 and 70 \micron~fluxes of the SNRs can be explained by $T_d\sim45$ to 80 K and $M_d\lesssim0.001$ to 0.8 $M_{\sun}$. The range of dust temperature in the LMC SNRs is consistent with that in Galactic SNRs (dashed line in Figure \ref{fig:mmd}), which is between $\sim45$ and 85 K, assuming $\kappa_{\nu}\propto\nu^\beta$ when $\beta=2$ \citep{pin11}. Furthermore, the range of dust mass agrees with that of the Galactic SNRs varying from 0.008 $M_\sun$ for Cas A to 2.5 $M_\sun$ for W44 \citep{pin11}.

Similarly, Figure \ref{fig:mmd3} shows the distribution of SNRs in the 8 \micron~and 24 \micron~flux plane. As half of the SNRs do not show 8 \micron~emission, for clarity we do not show their upper limits in the diagram. If we compare this with the black-body model as in Figure \ref{fig:mmd}, the 8 and 24 \micron~fluxes can be reproduced by dust emission with a wide range of mass ($M_d\sim2\times10^{-7}$ to $\ga8\times10^{-4}~M_\sun$) and relatively high temperatures ($T_d\sim$180-440 K). These dust properties are not consistent with those derived from the MIPS emission, which indicates that the 8 \micron~fluxes cannot be explained with the same dust properties from the MIPS emission. Moreover, the dust temperature derived from 8 to 24 \micron~ratio seems to be much higher than the typical ranges ($\sim$40-100 K) for SNRs in the literature \citep[e.g.,][]{borko06,bwill06,seok08,pin11} and in this work. Thus, we suppose that the IRAC band emission is of a different origin from the MIPS emission (see Section \ref{sec:origin}).

Nevertheless, it is interesting that those with both 8 and 24 \micron~fluxes show a rough correlation. Assuming that the IR emission is from radiatively-heated dust, this correlation can be reproduced by models of \cite{DL07}, which are calculated for dust mixtures of amorphous silicate and graphitic grains heated by starlight. We adopt three cases for the LMC with three different mass fractions of PAHs and derive $Spitzer$ fluxes varying starlight intensities ($u_\nu=Uu_\nu^{\rm MMP83}$, $U$: a scaling factor, $u_\nu^{\rm MMP83}$: the interstellar radiation field estimate by \citet{mathi83} for the solar neighborhood) and dust masses ($M_{d}=1$ and 100 $M_{\sun}$) with the dust emissivity ($j_\nu$).\footnote{http://www.astro.princeton.edu/$~$draine/dust/irem.html} As the dust mass or the starlight intensity increases, both 8 \micron~and 24 \micron~fluxes increases. Those very bright SNRs such as N49 or SNR in N159 require either a large amount of dust masses ($M_d\ga100~M_{\sun}$) or very high starlight intensities ($U\ga10^3$), and such large dust masses could be unreasonable for SNRs. \citet{ander11} derive the strengths of the radiation field for Galactic SNRs adopting the ``on-the-spot'' approximation (so called in Case B), which range from about 10 to 4800 relative to the general interstellar radiation field. This indicates that very small grains heated by a very strong radiation field, in addition to possible contamination from line emission, might contribute to the IR emission in the 8 \micron~band of some SNRs. However, it should be further confirmed which one is the dominant origin of the IR emissions in the SNRs between collisionally-heated dust and radiatively heated dust. The SNRs having 8 \micron~fluxes higher than the model calculations such as DEM L72 or DEM L249 could be explained if the PAH mass fraction increases ($q_{\rm{PAH}}>2.37\%$). This is indeed consistent with the PAH-dominated SNRs categorized according by the IRAC colors (see next and Figure \ref{fig:ccd1}). The high PAH abundance has been noticed in several Galactic SNRs showing signs of interactions with a surrounding molecular cloud \citep{ander11}. Their PAH abundances are higher than those observed in the diffuse ISM in the Milky Way, which is interpreted as a result from shattering large grains \citep[e.g.,][]{jones96,seok12}. Hence, the relatively bright 8 \micron~emission of the LMC SNR might also suggest fragmentation of large grains in shock.

For the SNRs detected in the IRAC bands, we examine their IRAC colors ($[F_{3.6}/F_{5.8}]$ and $[F_{4.5}/F_{8}$]) in Figure \ref{fig:ccd1}. It is known that the origin of NIR emission can be inferred by the IRAC colors \citep{reach06}. Presumably, IRAC colors associated with ionic/molecular shocks or PAH bands are adopted from \cite{reach06}, and colors for synchrotron emission are drawn from varying spectral indices. Representative Galactic SNRs (Crab, Cas A, RCW 103, and IC 443) are also shown \citep[and references therein]{tem09,pin11}. IR emission from Crab nebula is dominated by a synchrotron emission, while RCW 103 and IC 443 are well known to be dominated by ionic line emission and molecular line emission, respectively. Most LMC SNRs fall on the area for SNRs associated with molecular shocks, while several fall on the area for SNRs with PAH emission, and only few have IRAC colors associated with ionic shocks. However, note that IRAC colors of N49 and N63A, known to be ionic-line dominated \citep[e.g.,][]{will06}, fall in color ranges for molecular shocks. This indicates that one should be cautious to interpret the origin of the IRAC colors using this diagram, and those having both ionic (e.g., [\ion{Fe}{2}] 5.34 \micron) and molecular line emission (e.g., H$_2$ 1-0 O(5) 3.235 \micron, 0-0 S(7) 5.512 \micron) might have $[F_{3.6}/F_{5.8}]$ similar to the lowest value for molecular shocks.

\subsection{Comparison to Multi-wavelength Data}
\subsubsection{IR versus X-Ray\label{sec:irxr}}

We compare the MIR morphologies of the SNRs to their X-ray morphologies by using the archival $Chandra$ data (Figures \ref{irimg1}--\ref{irimg6} and Figure \ref{fig:corr_irxr}) and {\it XMM} images from the literature \citep{kli10,bam06,mag12}. Among the 29 IR SNRs, nine SNRs, mostly shell-like SNRs, show strong spatial correlations between MIR and X-ray emissions. Four SNRs (N49, DEM L238, N158A, and SNR 0548--70.4) show similarities with some minor differences (Figure \ref{fig:corr_irxr}). 12 SNRs show considerable discrepancies between the IR and X-ray morphologies (Figure \ref{fig:corr_irxr}), and there are no available X-ray data for the other four SNRs. The 12 SNRs with spatial discrepancies and four with both similarities and differences are listed in Table \ref{tab:corr_irxr}, and among the 16 SNRs, the 12 remnants with the archival {\it Chandra} data are shown in Figure \ref{fig:corr_irxr}. There are various mechanisms that deform the IR morphology of the SNR, and the different morphologies in the IR and X-ray seen in these 16 SNRs can be the result of interactions with nearby molecular clouds. When an SNR interacts with an ambient molecular cloud, MIR emission can originate from dust heated by UV radiation from shocks (not by electron collisions) and can have considerable contributions from ionic/molecular emission lines. This suggests that those SNRs are likely to show line emissions in the IRAC bands. In addition, the dense regions having interactions with an SNR can protect PAHs against complete destruction by shocks, so these SNRs might have PAH emission in the IRAC bands. In fact, NIR emissions in one or several IRAC bands are detected for the 12 out of 16 SNRs; The IRAC colors of three SNRs (DEM L72, SNR in N206, and DEM L249) are similar to the colors for PAH emission, and the colors of the others are similar to those for ionic or molecular shocks (Figure \ref{fig:ccd1}).

We compare $Spitzer$ 24 \micron~fluxes ([$\nu F_{\nu}$]$_{24 \micron}$) to $Chandra$ wide band (0.3-10 keV) X-ray fluxes in Figure \ref{fig:irxr}. Available X-ray fluxes of 17 SNRs are taken from the $Chandra$ SNR catalog, but the X-ray fluxes of DEM L205 and DEM L241 are from the {\it XMM-Newton} observations\footnote{Although the energy coverages of {\it XMM-Newton} (0.2-5 keV and 0.5-10 keV) are different from that of $Chandra$, the missing fluxes caused by it could be less than 30\%-40\% based on calculations using PIMMS (http://cxc.harvard.edu/toolkit/pimms.jsp). Thus, we suppose that the {\it XMM-Newton} fluxes are comparable with the $Chandra$ fluxes.}. While all X-ray fluxes are measured from whole SNR regions, 24 \micron~fluxes of some SNRs are estimated from specific parts of the remnants, so only 24 \micron~fluxes from whole SNRs can be used for the direct comparison. The 24 \micron~fluxes show relatively good correlation with the X-ray fluxes, which can be expected by the morphological similarities between the 24 \micron~and X-ray images (e.g., Figures \ref{irimg2} and \ref{irimg3}). However, six SNRs in the diagram have spatial discrepancies in the IR and X-ray emissions, so it is worth to separately examine the SNRs with or without the spatial correlation. We perform a linear fit to the fluxes in a logarithmic scale. While the best fit using the fluxes measured in the whole SNRs is given as [$\nu F_\nu]_{24\micron}=(0.99^{+0.40}_{-0.14})\times F_{\rm{Xray}}$, the fluxes from the SNRs with the morphological similarities can be fitted with [$\nu F_\nu]_{24\micron}=(0.43^{+0.14}_{-0.11})\times F_{\rm{Xray}}$. Also, we estimate correlation coefficients of all SNRs in the diagram and of SNRs with good spatial correlations, which are 0.87 and 0.97, respectively. A strong correlation between IR and X-ray fluxes is expected for the latter SNRs because dust grains in those SNRs are collisionally heated by gas species in the hot X-ray emitting plasma (see Section \ref{dis_dust}).

The $Spitzer$ $[F_{70}/F_{24}]$ ratios are compared with X-ray surface brightness (Figure \ref{fig:m2m1_xr}). There is a trend that the $[F_{70}/F_{24}]$ decreases as the X-ray surface brightness increases. Even SNRs with their IR fluxes extracted from limited areas also follow this trend, while their X-ray brightnesses are measured from the whole. The relation for those with the good spatial correlations can be approximated by a function of $[F_{70}/F_{24}]=12.1^{+3.9}_{-3.0}\times \Sigma_X^{-0.84\pm0.14}$ whereas all data with total fluxes are fitted by $[F_{70}/F_{24}]=7.2^{+1.7}_{-1.4}\times \Sigma_X^{-0.43\pm0.09}$. This trend is expected because the equilibrium temperatures of dust grains in SNRs is physically related to the electron density of X-ray emitting plasma (see Section \ref{dis_dust} for details). In the hot plasma with a higher density producing relatively brighter X-ray emissions, dust grains can be heated up to higher temperatures due to frequent collisions with electrons, which results in a low $[F_{70}/F_{24}]$ ratio. The Galactic SNR, G292.0+1.8 also shows a similar trend \citep[see Figure 5 in][]{ghava12}. For the IR as well as X-ray fluxes extracted from different regions in G292.0+1.8, the $[F_{70}/F_{24}]$ ratio shows a declining trend with increasing X-ray surface brightness. They also interpret this trend due to dust heating by electrons from hot post-shock plasma. Moreover, using this trend, the elevated IR to X-ray flux ratios of several regions in G292.0+1.8 are attributed to higher dust-to-gas ratios than different parts of the SNR. However, one should be cautious when interpreting the correlation, because the MIR emission in some SNRs can have a contribution from ionic line (or synchrotron) emission. Also, as \citet{ghava12} mentioned, the X-ray surface brightness can be overestimated due to ejecta emission which is not related to the shocked circumstellar (or interstellar) dust.

\subsubsection{IR versus CO Emission}

For those 16 SNRs with the IR-X-ray morphological discrepancies, we examine the existence of CO emission around each SNR to find more direct evidence for the interaction with molecular clouds (Figure \ref{fig:corr_irxr}). For the inspection, we use the masked CO intensity map\footnote{The masked intensity map is made by masking by the 3-$\sigma$ contour of a cube that has been smoothed to a 90$\arcsec$ resolution, then integrating from 200 to 305 km s$^{-1}$. The data can be downloaded from http://mmwave.astro.illinois.edu/magma/DR1/.} from the MAGMA Data Release 1 \citep[Magellanic Mopra Assessment;][]{wong11}. The MAGMA survey consists of CO ($J=0\rightarrow1$) observations done by the Mopra 22 m telescope, which covers $\sim100$ giant molecular clouds (GMCs) selected based on a previous survey for GMCs in the LMC, the NANTEN survey \citep{fuk01}. The detection limit of the MAGMA is approximately $I_{\mathrm{CO}}=2$ K km s$^{-1}$ (or $N$(H$_2$)$\sim1.4\times10^{22}$ cm$^{-2}$). Since the MAGMA survey does not cover all LMC SNRs, we also refer to images of the NANTEN survey from Figures 1 and 2 in \citet{desai10}. Among those 16 SNRs, seven remnants actually show the association with molecular clouds in the MAGMA and/or NANTEN surveys; N49, DEM L241, DEM L249, DEM L256, N157B, SNR in N159, and N158A (Table \ref{tab:corr_irxr}). In addition, we detect a CO emission adjacent to the southern boundaries of three SNRs, N11L, SNR in N206, and DEM L316B. Since the NANTEN survey has a large beam ($2\arcmin.6$ half power beam), small clouds that possibly exist around SNRs would not be detected. Among the six remaining SNRs, five (DEM L205, DEM L238A, N63A, DEM L316A, and SNR 0548--70.4) show centrally brightened X-ray emission. This is generally caused by the dense ambient medium for CCSNRs while in the case of Type Ia SNRs, the bright X-ray emission originates from the reverse-shocked SN ejecta. The last remaining DEM L72 does not have any evidence for the interaction. However, considering that the IR emission in DEM L72 is categorized as PAH emission, the existence of PAHs indicates the presence of dense clumps because PAHs are otherwise rapidly destroyed by shocks \citep{seok12,micel10}. In summary, ten SNRs show evidence of association with molecular clouds, and six show indirect indications of interactions with a dense medium in part.

\subsubsection{IR versus Radio}

We compare the $Spitzer$ 24 \micron~fluxes to the 4.8 GHz radio fluxes in the left panel of Figure \ref{fig:irra}. The 24 \micron~fluxes are moderately correlated to the 4.8 GHz fluxes (correlation coefficient $=0.61$), however the correlation is not as good as that seen in the $AKARI$ 24 \micron~fluxes from six SNRs \citep[correlation coefficient $=0.98$;][]{seok08}. The fluxes from whole SNRs (21 SNRs including six in the $AKARI$ study) can be approximated by a linear fit in a logarithmic scale, $F_{24\micron}=(0.89^{+0.27}_{-0.21})\times F_{4.8\mathrm{GHz}}$. The measured slope is lower than that of the $AKARI$ samples \citep[slope $=1.89$;][]{seok08} probably due to more diverse remnants in the present sample. When the IR fluxes are compared with the radio fluxes, there is a well-known property for the ratio of the IR-to-radio flux, $q_{\lambda,\mathrm{IR}}= \mathrm{log}(F_{\lambda,\mathrm{IR}}/F_{1.4\mathrm{GHz}}$) where $F_{\lambda,\mathrm{IR}}$ and $F_{1.4\mathrm{GHz}}$ are continuum flux densities at a given IR wavelength and 1.4 GHz (i.e., 21 cm), respectively. Then, $q_{24}$ can be derived using 4.8 GHz flux as $q_{24}=$ log($F_{24\micron}/F_{1.4\mathrm{GHz}}$) = log($F_{24\micron}/F_{4.8\mathrm{GHz}})- \alpha\mathrm{log}$(4.8/1.4), where $\alpha$ is the spectral index of an SNR ($F_\nu\propto\nu^{-\alpha}$). Since the spectral indices of the LMC SNRs are not well studied, we assume them to be 0.5 en bloc. The $q_{24}$ of the individual SNRs ranges from $-1.17$ to 0.72, and the average $q_{24}$ for the all LMC SNRs defined by, $q_{24,{\rm avg}}\equiv \mathrm{log}[\Sigma (F_{24\micron})/\Sigma (F_{1.4\mathrm{GHz}})]$ is $0.14$.

We investigate the ratio of the IR to the 4.8 GHz radio fluxes ($[F_{24\micron}/F_{4.8\rm{GHz}}]$ and $[F_{70\micron}/F_{4.8\rm{GHz}}]$) for each SNR (Figure \ref{fig:irra}, right). Note that some IR fluxes are extracted from limited regions while all radio fluxes represent the total fluxes. As expected from the wide range of $q_{24}$, the $[F_{24\micron}/F_{4.8\rm{GHz}}]$ ratio varies considerably. However, it is noticeable that while the 24 \micron~flux varies over four orders of magnitude (i.e., a few mJy to $\sim10^4$ mJy), the $[F_{24\micron}/F_{4.8\rm{GHz}}]$ ratios vary over $\sim2$ order of magnitude. There is a correlation between the ratios of 24 \micron~and 70 \micron~fluxes to the 4.8 GHz flux as seen at the correlation between 24 and 70 \micron~fluxes (Figure \ref{fig:mmd}). This correlation is well fitted by a linear function in a logarithmic scale as $[F_{70\micron}/F_{4.8\rm{GHz}}]=6.1^{+2.2}_{-1.6}\times [F_{24\micron}/F_{4.8\rm{GHz}}]$. For this fitting, the IR fluxes estimated from the whole SNRs are only used. We compare this result with the case of Galactic SNRs. The ratios of five representative Galactic SNRs and the slope derived for Galactic SNRs in the MIPSGAL survey are overlaid in Figure \ref{fig:irra} (right). To derive the ratios of the Galactic SNRs, we take the IR properties of four SNRs from \cite{pin11} and of Cas A from \cite{hines} and convert 1 GHz fluxes of the Galactic SNRs to 4.8 GHz fluxes using their spectral indices \citep{green}. A correlation for Galactic SNRs is also measured by \cite{pin11} using 1.4 GHz fluxes instead of 4.8 GHz, of which slope is $1.2\pm0.1$. Assuming the spectral index is 0.5 ($\alpha=0.5$) en bloc, we compare their correlation to ours. It seems that the LMC SNRs reasonably follow the correlation of the Galactic SNRs, too.

Despite of the relatively good agreement between the ratios of Galactic and LMC SNRs, it is found that there are some differences between them. The $q_{24,{\rm avg}}$ ($=0.14$) is smaller than that of the Galactic SNRs in the MIPSGAL survey \citep[$q_{24}=0.39\pm0.11$ excluding objects closer to \ion{H}{2} region,][]{pin11}, so the correlation between 24 \micron~and 4.8 GHz fluxes for the Galactic SNRs ($F_{24 \micron}=4.55^{+1.31}_{-1.02}\times F_{4.8\rm{GHz}}$) is also different from that for the LMC SNRs. This discrepancy is directly depicted in both panels of Figure \ref{fig:irra}. By scaling the 24 \micron~and radio fluxes of Galactic SNRs from \citet{pin11} to the distance of the LMC (50 kpc), the fluxes of the LMC SNRs can be compared to the fluxes of the Galactic SNRs. Figure \ref{fig:irra} (left) clearly shows that there are more IR faint SNRs in the LMC relative to the Galactic SNRs while the radio fluxes of both SNRs are comparable. Except Cas A in Figure \ref{fig:irra} (right), many of the ratios of the LMC SNRs are lower than those of the Galactic SNRs. 3C396 is one of the Galactic SNRs having the lowest ratio \citep{pin11}, and the majority of the LMC SNRs have ratios lower than that of 3C396. Section\ref{sec:irsnr} further discusses a possible explanation for the fact that LMC SNRs are fainter in the MIR, but their radio fluxes are similar to Galactic SNRs. In fact, $q_{24}$ is often measured for galaxies because the MIR/radio correlation is well-known among galaxies \citep[e.g., $q_{24}=0.84\pm0.28$;][]{apple}. The $q_{24}$ of the LMC SNRs is much smaller than that of the extra galaxies. If the 1.4 GHz radio emissions of galaxies mainly originate from synchrotron emission of SNRs, our result implies that only a small portion of IR emission can be directly contributed from the SNRs (6\%-20\%).

\section{DISCUSSION}
\subsection{Characteristics of IR SNRs in the LMC}\label{sec:irsnr}

One of the remarkable results in this study is the high detection rate of IR SNRs. 29 IR counterparts out of 47 LMC SNRs are identified in the IRAC and/or MIPS bands, which yields about a 62\% detection rate. In the case of Galactic SNRs, 18 out of 95 SNRs were clearly detected in the GLIMPSE survey \citep{reach06}, and 39 out of 121 SNRs were detected in the MIPSGAL survey \citep{pin11}. Their detection rates are about 20\% and 32\%, respectively. Even for the previous studies using the $IRAS$ all-sky survey, about a 30\% detection rate is obtained for the Galactic SNRs \citep{arendt89,saken92}. In comparison with these previous results, the IR detection rate of the LMC SNRs seems to be interestingly high.

This high detection rate could be caused by either extrinsic or intrinsic aspects. One of the apparent extrinsic effects is less IR confusion by the Galactic disk. Both the GLIMPSE and the MIPSGAL surveys are restricted to the inner Galactic plane ($|b|<1\degr$), and contamination by other Galactic sources is severe. Even if the $IRAS$ all-sky survey includes whole Galactic SNRs, back/foreground confusion is still a serious problem when disentangling IR emission associated with SNRs from other emissions. Besides, due to the limited spatial resolution of $IRAS$, it was difficult to identify small and/or faint structures. We compare the intensity distribution of the LMC and the MIPSGAL SNRs detected in the MIPS 24 \micron~band (Figure \ref{fig:irhis}). After scaling the fluxes of the Galactic SNRs at the distance of the LMC (50 kpc), either using their known distances or assuming the distances as 3 kpc, it is found that the median flux of the Galactic SNRs is higher than that of the LMC SNRs (284 mJy and 88 mJy, respectively). This indicates that we detect more IR-faint SNRs in the LMC, and in other words, there could be Galactic SNRs with faint IR emission that elude detection.

Since there are physical and chemical differences between the Milky Way and the LMC, those differences could intrinsically affect the IR emission from an SNR. The LMC is known to have a lower dust-to-gas ratio and (lower) metallicity than the Milky Way \citep{pei}. Moreover, the dust composition (graphite to silicate ratio, $r_c$/$r_s$) of the LMC is also known to differ from the Galaxy. While $r_c$/$r_s$ of the LMC is about 0.22, $r_c$/$r_s$ of the Galaxy is about $0.9-0.95$ \citep{pei}. This different dust composition can affect the lifetime of dust in the postshock region. According to calculations by \citet{dwek96}, when a shock of $v=400$ km s$^{-1}$ is propagating into a dusty medium with a preshocked density of $n_0=0.1$ cm$^{-3}$, 21\% of graphite is returned to the gas phase while 29\% of silicates are returned at $N_{\rm H}=1.25\times10^{18}$ cm$^{-2}$. This indicates that silicate is more rapidly destroyed by shocks, and dust in the LMC might be more easily destroyed than the Galactic dust. However, more recent studies \citep[e.g.,][]{jones11, serra08} propose that carbon dust is more readily destroyed than silicate dust. Thus, since there is still an uncertainty regarding the destruction efficiency of dust components (and also the dust compositions in the LMC), it is difficult to directly concatenate the destruction efficiency to the intrinsic characteristics of the LMC.

As we have not found any intrinsic aspects to make a difference for the IR emission between the LMC SNRs and the Galactic SNRs, more detection of IR LMC SNRs, precisely IR-faint LMC SNRs, is most likely due to less confusion. Then, the lower slope of the IR-to-radio ratio (or lower $q_{24}$) for the LMC SNRs than those of the $AKARI$ sample \citep{seok08} or Galactic SNRs \citep{pin11} could be explained by the additional detection of faint IR SNRs. While the intensity of the radio continuum is primarily determined by the ambient density and the magnetic field strength \citep[and references therein]{bandi10}, the intensity of the IR emission could be affected by more diverse conditions such as dust properties, shock velocities, post-shock gas temperature, as well as the ambient density. This distinction can generate the evolution of the IR brightness, different from that of the radio continuum. Hence, the detection of a number of IR faint SNRs relative to their radio brightnesses can result in a lower $q_{24}$ value, although we cannot rule out any influences from the intrinsic differences between the LMC and the Galaxy.

\subsection{Origin of IR Emission in SNRs\label{sec:origin}}

As briefly mentioned in previous sections, there are four primary sources of the IR emission in SNRs; ionic and/or molecular lines, thermal dust continuum emission, PAH bands, and non-thermal synchrotron emission \citep[e.g.,][]{reach06,koo07,seok08}. Synchrotron emission can dominate IR emission over a wide wavelength range, but only for a peculiar case with a strong pulsar wind nebula (PWN) such as Crab, and even then, it is insignificant. Although PAH emission has not been detected toward various SNRs so far, a few SNRs have been reported to show several PAH features \citep[e.g.,][]{tappe06,seok12}. Major PAH features at 3.3 \micron, 6 to 8 \micron, and 11.3 \micron~can contribute to the IR emission, but the PAH bands are often accompanied with strong line emission in SNRs. Eventually, forbidden lines from the elements such as Ne, O, Fe ions and/or pure rotational H$_2$ lines usually dominate in the NIR bands and thermal emission from collisionally-heated dust grains dominates in the MIR bands. In the IRAC bands ($\la 10~\micron$), ionic/molecular line emission is usually a primary contributor. Representative ionic emission lines are Br$\alpha$ 4.05 \micron, [\ion{F}{2}] 5.34 \micron, and [\ion{Ar}{2}] 6.99 \micron, and there are various transitions of molecular hydrogen emission lines (e.g., pure rotational lines such as $v=$0--0 S(19) 3.40 \micron~to S(3) 9.66 \micron). Dust continuum usually dominates IR emission at longer than $\sim15$ \micron~considering that typical temperatures of dust in thermal equilibrium range from $\sim40$ to 100 K.

There are several direct or indirect methods to classify the dominant origin of the IR emission. When IR spectra and/or narrow-band filter images are available, the origin can be identified directly. Otherwise, it is necessary to perform indirect classifications based on their IR colors and/or morphologies in multi-wavebands. We compare the IRAC colors to the theoretical prediction of the main emission mechanisms in the color-color diagram. In addition, we compare IR morphology to X-ray and/or optical. When IR morphology is not well-correlated with its optical morphology but rather with that of an X-ray, this can indicate that thermal dust emission is dominant in the SNR. On the other hand, IR morphology would be similar to the optical when the line radiation from a radiative shock is dominant. However, in case of the Balmer-dominated SNR, resemblance between IR and optical often looks as good as that between IR and X-ray \citep[e.g.,][]{borko06}, so SNR characteristics such as SN type should be considered together. For a given SNR, the origin of the IR emission can be different depending on the band, and thus we discuss IR emission in the IRAC bands and MIPS bands separately. The estimated origins of the emission in the IRAC-bands for individual SNRs are summarized in Table \ref{tab:origin} and discussed in the next subsection.

\subsubsection{Origin of the Emission in the IRAC Bands}

\noindent 1. SNRs dominated by ionic/molecular line emission
\\
\\
\indent To constrain the dominant origins of the IRAC emission, previous observations such as spectroscopy or narrow-band filter imaging are very critical. For SNRs whose IR spectra are obtained, it is straightforward to examine the emission source. IR spectra of several LMC SNRs have been observed by $AKARI$ and/or $Spitzer$, and the archival data of four SNRs (N11L, N49, N63A, and SN 1987A) are available. N11L, N49, and N63A are previously suggested to be line dominated based on their IRAC colors, and the $Spitzer$ IRS spectra of N49 show strong ionic lines \citep{will06}. We also confirm that the archival $Spitzer$ IRS spectra of N11L and N63A show dominant ionic lines. SN 1987A is only detected in the 5.8 \micron~and 8.0 \micron~bands, and its IRS spectra show several ionic emission lines \citep{bouch06}. In addition, previous studies of some SNRs could give hints on the origin of IR emission. NIR spectra (1-5~\micron) of N103B show the signature of [\ion{Fe}{2}] emission \citep{oliva89}, and N157B was previously suggested to be line-dominated SNRs according to the $AKARI$ data \citep[see more details in Appendix \ref{app_n157b};][]{seok08}. Hence, it is most likely that the IRAC band emissions of the above SNRs are ionic-line dominated, but it is interesting that most of them have the IRAC colors similar to those associated with molecular shocks. This suggests that the criteria for the origin of IRAC emission in the color-color diagram is not sufficiently stringent, or that even ionic-line dominated SNRs might have IRAC colors associated with molecular shocks. Even though SNRs are dominated by ionic line emission, it does not necessarily mean that they are not associated with molecular shocks. In fact, the IRS spectra of N49 and N63A clearly show several transitional H$_2$ lines \citep[e.g.,][]{will06}.

For the SNRs without IR spectra nor previous IR studies, we categorize the origins of their IR emission based on the IRAC color-color diagram (Figure \ref{fig:ccd1}). As mentioned above, SNRs with a mixture of ionic and molecular shocks such as N49 and N63A are located at the lower boundary of the IR color range for molecular shocks (i.e., $F_{3.6}/F_{5.8}\simeq0.3$), so only SNRs with the IR color of $F_{3.6}/F_{5.8}\ga 0.3$ are regarded to be molecular line dominated (N86, DEM L241 western region, DEM L256, north lobe of SNR in N159, and DEM L299). These SNRs usually show weak correlation with optical emission. Among SNRs with lower $F_{3.6}/F_{5.8}$ ratios, the remnants of which IR morphologies fairly correspond to the optical morphologies are classified as ionic line dominated SNRs (south shell of SNR in N159 and DEM L316A/B). Although we could not measure the IRAC colors from the whole of DEM L241 due to the absence of the 5.8 \micron~and 8.0 \micron~band emissions, the entire shell-like structure seen in the 4.5 \micron~band corresponds well to that seen in optical (Figure \ref{irimg5}, top row). This suggests that the overall emission is dominated by ionic emission, and the bright western region with the IRAC colors for molecular shocks is locally enhanced by molecular emission. In summary, nine SNRs are dominated by ionic line emission, and four are dominated by molecular line emission. Also, note that there could be several SNRs possibly having both ionic and molecular line emission.  
\\
\\
\noindent 2. SNRs dominated by PAH emission
\\
\\
\indent Five SNRs (SNR 0450--70.9, SNR in N4, DEM L72, SNR in N206, and DEM L249) are considered to be dominated by PAH emission based on the IRAC color-color diagram (Figure \ref{fig:ccd1}). These SNRs show lack of correlation to the optical images and are well defined in the 5.8 and 8.0 \micron~bands. DEM L249 is somewhat contentious because it has the IRAC colors expected for either ionic shocks or PAH emission. In terms of morphology, both IR and optical emissions show shell structures with the enhanced emission in the east (Figure \ref{irimg5}, second row). However, the locations of the IR peaks along the eastern rim differ from those of the optical peaks, which is unlikely to be ionic line-dominated. Further observations will enable us to clarify the origin of IR emission.
\\
\\
\noindent 3. SNRs dominated by synchrotron emission
\\
\\
\indent As the intensity of synchrotron emission follows a spectral energy distribution (SED) with a power-law, $S_\nu\propto\nu^{-\alpha}$, its IRAC colors are tightly constrained by its spectral index (Figure \ref{fig:ccd1}). Also, IR morphology shows very good agreement with radio morphology. There are two confirmed Crab-like SNRs in the LMC, N157B and N158A. Because the PWN of N157B is too faint to be detected in the IR, the PWN of N158A is the only one seen at IR wavelengths. Since the flux of N158A is extracted only from its PWN, its IR emission is naturally considered to be dominated by synchrotron emission. \citet{bwill08} show that its IRS spectrum is dominated by synchrotron emission as well as continuum at wavelengths longer than 20 \micron.

\subsubsection{Origin of the Emission in the MIPS Bands}

The IR emission in the MIPS bands (typically longer than $\sim15~\micron$) are generally dominated by dust continuum but can also be contributed by ionic/molecular emission lines (e.g., [\ion{O}{4}] 25.88 \micron, [\ion{Fe}{2}] 25.98 \micron, or H$_2$ 0-0 S(0) 28.2 \micron~for the MIPS 24 \micron~band, [\ion{O}{1}] 63.0 \micron~for the MIPS 70 \micron~band). The diffuse emission of all IR SNRs seen at 24 \micron~is more likely to originate from dust emission. Two Type Ia SNRs (0509--67.5 and 0519--69.0) and two CCSNRs (N132D and SN 1987A) have $Spitzer$ IRS spectroscopy. SNR 0509--67.5 and 0519--69.0 are young, Balmer-dominated Type Ia SNRs. Their shocks are very fast ($\ge 3,000$ km s$^{-1}$) and non-radiative \citep{ghava07}, thus we do not expect strong IR ionic or molecular line emission. Their IRS spectra show no line emission in both SNRs \citep{bwill11}. The spectra of N132D and SN 1987A (10-30~\micron) also show dominant thermal dust continuum with small contributions from several ionic lines and PAH emission \citep{tappe06, bouch06, dwek08}.

Good spatial correlation between IR and X-ray indicates a dust emission origin. For example, Type II SNRs, SNR 0453--68.5, N23, and N49B are middle-aged (0.5-2$\times10^4$ yr) shell-type SNRs, and their MIR morphologies are almost identical to the X-rays whereas their H$\alpha$ images hardly show any shell structures \citep{will99, bwill06}. The eastern shell-like structure of DEM L205 and the southern shell of DEM L238 also show a similar spatial distribution between the MIR and X-ray emission, which leads to the origin of the IR emission as dust continuum. In addition, three Type Ia SNRs, DEM L71, N103B, and SNR 0548--70.4 show morphological correlations between IR and X-ray. DEM 71 and SNR 0548--70.4 are Balmer-dominated, and the shell-like emission from the three remnants are more likely to be dominated by dust emission. However, the bright central IR emissions of N103B and SNR 0548--70.4 seem to correspond better to optical knots seen in their H$\alpha$ images rather than the X-ray emission. This implies some contribution from ionic-line emission. Similarly, in the case of N49B, the bright portion of the southern shell and the clump in the eastern part show relatively bright H$\alpha$ and [\ion{O}{3}] emission \citep{math83}, thus there might be some contribution from ionic line emission in the 24 \micron~band.

In some cases, the 24 \micron~MIPS-band emission can be dominated by ionic emission lines rather than dust continuum. The IRS spectra of N49 show abundant ionic lines from shocked gas without considerable dust continuum, and it is found that the ionic lines such as the [\ion{O}{4}] line at 25.88 \micron~and the [\ion{Fe}{2}] line at 25.98 \micron~can contribute up to 80\% of the MIPS 24 \micron~emission of N49 \citep{will06}. However, the 24 \micron~image shows a faint shell-like structure as seen in X-ray, and the mid-to-far IR (M/FIR) data can be attributed to dust continuum with two temperature components \citep{otsu10}. Therefore, M/FIR emission at longer wavelengths is most likely to be dominated by dust continuum even though MIR emission of some SNRs ($\lesssim30$ \micron) can be dominated by line emission.

\subsubsection{IR Origin and SN Type}

The origin of the IR emission might be related to the SN type. While many CCSNRs have IRAC band emission that can be dominated by line emission or PAH emission, Type Ia SNRs such as DEM L71, SNR 0509--67.5, and SNR 0519--69.0 are only seen in the MIPS bands dominated by dust emission. In general, Balmer-dominated Type Ia SNRs with fast and non-radiative shocks cannot have associated ionic and/or molecular line emission. Also, complete destruction of PAHs in these remnants is expected, due to fast shocks. However, a Type Ia SNR that had a more massive progenitor, so called a ``prompt'' Type Ia, might have some contribution from other IR mechanisms. While general Type Ia SNRs are isolated from dense environments, prompt Type Ia SNRs can be located in dense circumstellar medium (CSM) like the well-known example for a prompt Type Ia SNR, Kepler \citep{blair}. The bright northwest filaments in Kepler are considered to be dominated by radiative emission from the shocked CSM, and the strong [\ion{Ar}{2}] line is detected at 7.0 \micron~in the IRAC 8~\micron~band \citep[and references therein]{blair}. Because DEM L238 and DEM L249 are suggested to be prompt Type Ia based on the X-ray spectral analysis \citep{borko06}, the environmental conditions could be similar to the case for Kepler. In the case of DEM L249, the association between H$_\alpha$ and 8 \micron (or 4.5 \micron) emission is evident (Figure \ref{irimg5}), which implies the contribution of emission lines from its swept-up CSM. For DEM L238, however, evidence of the contribution of emission lines is not clear because its NIR emissions in the IRAC bands are not detected, and the emission seen in the 24 \micron~band mission shows marginal similarities to its optical emission (Figure \ref{irimg4}). In conclusion, CCSNRs and prompt Type Ia SNRs are more likely to have the IR emission of in the IRAC bands that originates from ionic/molecular lines and/or PAHs while it is difficult to expect that normal Type Ia SNRs have such contribution. Further observations are required to clarify the origin of the IR emission and the existence of the dense CSM around SNRs.

\subsection{Dust Heating and Shock Processing \label{dis_dust}}

Dust in SNRs can be heated by two physical processes, collisions and radiation. When a fast, non-radiative shock is propagating into a dusty plasma, dust grains embedded in the X-ray emitting plasma are heated mainly by electronic collisions \citep[e.g.,][]{dwek87}. In this case, SNRs show IR morphology very similar to X-ray morphology. On the other hand, when shocks become radiative and can produce sufficient radiation in the shock front, the UV photons from the cooling postshock gas are the dominant heating mechanisms for grains \citep{HM79}. The radiative heating could become important particularly for SNRs interacting with a molecular cloud. Further, the relative importance of collisional to radiative heating is dependent on the grain size \citep{ander11}; big grains are dominantly heated radiatively, but very small grains or PAHs could be heated by collisions as well as by radiation.

When dust is collisionally heated by electrons in hot plasma and cools down radiatively, then dust continuum is directly related to the physical properties of X-ray emitting gas. We follow the dust cooling and heating model of \citet{dwek08}. At a given size of a dust grain, $a$, the collisional heating rate, $\mathcal{H}$ (erg s$^{-1}$), is determined by the electron density, $n_e$, and the gas temperature, $T_e$. When most electrons are stopped in a grain, the grain heating rate is given by $\mathcal{H}\propto a^2n_eT^{3/2}_e$. Above a certain gas temperature ($T_c$, critical temperature), electrons go through a grain, and the grain heating rate becomes only dependent on $n_e$, $\mathcal{H}\propto a^3n_e$. The radiative cooling rate, $\mathcal{L}$ (erg s$^{-1}$), simply follows the Stefan-Boltzmann law, which gives $\mathcal{L}=\pi a^2\sigma T^4_d\langle\mathcal{Q}\rangle$, where $\sigma$ is the Stefan-Boltzmann constant, $T_d$ is the dust temperature, and $\langle\mathcal{Q}\rangle$ is the averaged dust emissivity ($\langle\mathcal{Q}\rangle\propto aT_d^\beta$). Consequently, when dust is in thermal equilibrium, $\mathcal{H}=\mathcal{L}$, dust temperature can be expressed by
\begin{eqnarray}
T_d \propto \left\{ \begin{array}{ll}
 (n_e/a)^{\gamma}T_e^{3\gamma/2}, & \textrm{electrons stopped in grain ($T_e\la T_c$),}  \\
 n_e^{\gamma}, & \textrm{electrons go through grain ($T_e\ga T_c$),} 
  \end{array} \right.
\label{eq:tequil}
\end{eqnarray}
where $\gamma\equiv1/(4+\beta)$ with the emissivity index, $\beta \approx$1-2.

Figure \ref{fig:nete} depicts the equilibrium dust temperature ($T_d$) as a function of the electron density ($n_e$) and the gas temperature ($T_e$) for single-sized silicate dust grains (top panel for $a=0.01$ and bottom panel for $a= 0.1$ \micron).\footnote{Contours in Figure \ref{fig:nete} are actually simplified, and the original calculation can be found in Figure 1 of \citet{dwek08}.} The critical temperatures ($T_c$) are $5\times10^6$ K and $ 3\times10^7$ K for $a=0.01$ and 0.1 \micron, respectively. Taking their X-ray properties from the literature, $T_e$ and $n_e$ of LMC SNRs are overlaid. As a comparison, four Galactic SNRs are also shown. Their X-ray properties are taken from \citet{bam05} for the Cas A and Kepler ($Chandra$ data), \citet{hwang} for Tycho ($ASCA$ data), and \citet{miceli} for W49B south ({\it XMM-Newton} data). Uncertainties in $T_e$ and $n_e$ may be due to several factors. Spectral modeling is generally required to derive the electron density and the temperature from observed X-ray spectra, which needs various input parameters. Since the plasma properties of the LMC SNRs have not been studied coherently, methods used for the model calculation can differ among previous studies. Moreover, an explicit value of the electron density is not always given, and instead, the ionization timescale, $\tau=n_e\times t$ (cm$^{-3}$ s), is often provided, where $n_e$ is the electron density and $t$ is the elapsed time after the hot gas was heated up. In those cases, we derive the electron density using the ionization age (or SNR age). Also, the density is sometimes given in terms of $\int n_e^2dV$ so that $n_e$ is proportional to the volume filling factor ($\sim f^{-0.5}$). We take a $f$-factor if a preferential value is given in the literature, otherwise, we just assume $f=1$. Detailed information of the electron density and the temperature for LMC SNRs that we refer to is summarized in Table \ref{tab:nete}.

We examine any tendency between IR properties and X-ray properties in Figure \ref{fig:nete}. We use colors of symbols to represent $[F_{24}/F_{70}]$ of LMC SNRs and Galactic SNRs. The $[F_{24}/F_{70}]$ ratios are derived from the IR fluxes in Table \ref{tab:sstflux}, and the ratios of the Galactic SNRs are derived from \citet{hines} for Cas A, \citet{blair} for Kepler, \citet{ishi} for Tycho ($AKARI$ 24 and 65 \micron~fluxes), and \citet{pin11} for W49B. In addition, SNRs without detected MIR emission are marked with $triangles$. SNRs associated with a high equilibrium dust temperature, $T_d$, tend to have high $[F_{24}/F_{70}]$ ratios while those with a low $T_d$ tend to have low $[F_{24}/F_{70}]$. This is not surprising because the dust continuum has a peak at a shorter wavelength as dust temperature increases. Note that some SNRs have high $[F_{24}/F_{70}]$ but have low equilibrium temperatures or vice versa, and most of those SNRs have inconsistent IR morphologies with the X-ray morphologies. This can occur if the MIR emission has contributions from line emission and/or dust continuum by radiative heating rather than collisional heating, which is most likely related to interactions between SNRs and nearby dense materials (see Section\ref{sec:irxr}). Moreover, most SNRs with undetected IR emission are located in areas below $T_d$ of 50 K in the diagram based on their $n_e$ and $T_e$. Even if the SNR has a dust continuum at such a low temperature, it could be too faint to be currently detected at 24 (and 70) \micron. Although $n_e$ and $T_e$ that we adopt have uncertainties and dependencies on the shock models and the assumptions in the references (see Table \ref{tab:nete}), Figure \ref{fig:nete} shows that the equilibrium dust temperature derived from the plasma properties and the one inferred from the observed MIR colors ($[F_{24}/F_{70}]$) are compatible (see also Figure \ref{fig:tdust}).

Figure \ref{fig:irne} shows the dependence of dust emission on the gas density directly, where we compare the measured $[F_{24}/F_{70}]$ with the electron density ($n_e$). The general trend is in agreement with model predictions in terms of good correlation between the ratios and the electron density, particularly for those showing good spatial association between IR and X-ray emission. To compare with theoretical models, we derive three cases according to the relations between the dust temperature and the X-ray properties: (I) when $T_e\ga T_c$ (e.g., $T_e\ga 5\times10^6$ K for $a=0.01$ \micron, or $T_e\ga3\times10^7$ K for $a=0.1$ \micron), the dust temperature only depends on the electron density given as $T_{\rm dust}\simeq 57\times (n_e/\rm{cm^{-3}})^{\gamma}$ K (solid line in Figure \ref{fig:irne}). If $T_e< T_c$, dust temperature depends on the electron density, temperature, and grain size. When grain size is (II) $a=0.01~\micron$ at $10^6\leq T_e\leq 5\times10^6$ K (line-filled region between the dotted lines) or (III) $a=0.1~\micron$ at $10^6\leq T_e \leq 3\times10^7$ K (solid-filled region between the dashed lines), the dust temperature is given as $T_{\rm dust}\simeq 0.6\times [(n_e/$cm$^{-3})/(a/\micron)]^{\gamma}T_e^{3\gamma/2}$ K. Assuming $\beta=2$, we adopt $\gamma\simeq0.17$ for the all calculations. The temperature range is different for two cases because the critical temperature is different depending on the grain size. Finally, we calculate a flux ratio according to $F_{24}/F_{70}=\kappa_{24}B_{24}(T_d)/\kappa_{70}B_{70}(T_d)$ from Equation (\ref{eq1}) with the equilibrium dust temperature derived above for comparison to the observed ratios.

For small grains ($\sim0.01$ \micron), the critical temperature ($T_c$) is $\sim5\times10^6$ K, which is lower than the derived gas temperature for most SNRs. Thus, Case I is supposed to hold in this situation. However, Figure \ref{fig:irne} shows data scattering around the Case I prediction (solid line), which suggests that this might not be the case. Then, the dust size must be large, and assuming the grain size of 0.1 \micron~(Case III), the IR ratios of most SNRs can be well explained by $T_e\simeq10^7$ K. In fact, many SNRs show gas temperatures in agreement with $\sim10^7$ K (Figure \ref{fig:nete} and Table \ref{tab:nete}). Previously, the dust destruction in four Type Ia SNRs (DEM L71, 0509--67.5, 0519--69.0, and 0548--70.4) and four CCSNRs (N132D, N49B, N23, 0453--68.5) were examined \citep[respectively]{borko06,bwill06}. They described the $Spitzer$ 70 \micron~to 24 \micron~flux ratios by applying the same one-dimensional shock models taking into account the grain size distributions appropriate for the LMC. Both showed that the models can reproduce the flux ratios only if they include the effects of sputtering, destroying small grains \citep[destroying most grains smaller than 0.03-0.04 \micron;][]{borko06}. In fact, the SNRs in their samples are mostly well-aligned with the Case III at $T_e=10^7$ K in Figure \ref{fig:irne}.

Furthermore, we directly compare the dust temperatures derived from the plasma properties to those from the observed flux ratios (Figure \ref{fig:tdust}). For the equilibrium dust temperature, the grain sizes of both 0.01 \micron~and 0.1 \micron~are considered (shown as $circle$ and $triangle$ in the diagram, respectively). Figure \ref{fig:tdust} shows that the dust temperatures derived from two different methods are compatible with each other. In particular, the observed color gives values of the dust temperatures closer to those derived from the dust model with a grain size of 0.1 \micron~(when $T_{\rm dust}\ga60$ K), which supports that the dust model with a grain size of 0.1 \micron~(Case III) better explains the shocked dust properties, rather than the model with a grain size of 0.01 \micron~(Case II).

However, it is worth noting that our conclusion is obtained based on the simplified dust properties. In particular, we only consider a single-sized dust population for simplicity, but the typical interstellar dust grains have a power-law size distribution \citep{mathi77}. When dust grains undergo SN shocks, the grain size distributions are also modified. If the shock is fast ($\ge400$ km s$^{-1}$), small grains are preferentially destroyed by sputtering \citep[e.g.,][]{dwek96}. On the contrary, if the shock is slower ($\le200$ km s$^{-1}$), grain shattering becomes important and large grains are preferentially fragmented into small grains \citep[e.g.,][]{jones96}. In any case, dust population with the grain size distribution results in a range of the dust temperature rather than a single dust temperature. Another caveat in interpreting our data is that not all SNRs have fast shocks, quite unlike the previous samples \citep{borko06,bwill06}. Some SNRs with low $[F_{24}/F_{70}]$ such as DEM L241 and DEM L316B in Figure \ref{fig:irne} can predominantly experience slower shocks (or radiative shocks). As they have relatively low gas temperatures ($T_e<10^7$ K from Table \ref{tab:nete}), smaller grains are more likely to be preferred, which implies that dust shattering would affect the grain size distribution. This is in agreement with the results that Galactic SNRs interacting with molecular clouds show an overabundance of small grains due to dust shattering \citep[e.g.,][]{ander11}.

We also investigate the correlation between the grain size and SNR age. The SNR age is designated by the size of the symbol in Figure \ref{fig:irne}. As SNRs get evolved and/or have interaction with the ambient medium, the size distribution of the dust swept-up by SNR shocks could differ from that of the dust in young SNRs. As mentioned above, the grain size distribution can be affected by sputtering as well as shattering. One of the key physical properties to determine which of the processes becomes dominant is the shock velocity. Thus, it is expected that sputtering is more efficient for young SNRs while shattering is more efficient for mature SNRs. Since dust sputtering is more effective in smaller size grains because of their larger surface-to-volume ratio \citep[e.g.,][]{sank10}, destruction of small grains is expected for young SNRs. Even very young CCSNRs with the age of $10-100$ yr, or so called transitional-phase SNe, show IR emission, which can originate from circumstellar dust grains that have already experienced dust destruction by sputtering \citep{tana12}. On the other hand, as SNRs evolve or interact with a dense ambient material, shocks become slower, and shattering can produce smaller grains from larger grains \citep[e.g.,][]{ander11}. In Figure \ref{fig:irne}, we find that mature SNRs ($\ga10^4$ yr) mostly have low $[F_{24}/F_{70}]$. However, it is not straightforward to correlate the grain sizes in those SNRs and the SNR ages. In fact, there is a wide range for $[F_{24}/F_{70}]$ for mid-age SNRs ($\ga10^3$ yr). We found that both the $n_e$ and $T_e$ tend to decrease with the SNR ages, which indicates that there is degeneracy in the attribution of the declining trend of the flux ratios with the SNR age to the evolution of the grain size. Further studies of the grain size distributions in various SNRs are required to understand the evolution of the grain size in an SNR.

\section{SUMMARY}

We present a statistical study of IR SNRs in the LMC. In this multi-wavelength analysis, we use a vast amount of archival {\it Spitzer} data (from IRAC and MIPS instruments), to coherently study the IR emission in LMC SNRs and to correlate it with other wavelength data for the first time. 

1.  29 out of 47 SNRs in the LMC show detectable IR emission in the IRAC (3.6, 4.5, 5.8, and 8.0 \micron) and/or MIPS (24 and 70 \micron) bands. IR morphologies of 13 SNRs are firstly shown in the paper. All 29 SNRs show emission features in the MIPS 24 \micron~band, and most of them also have similar morphologies at 70 \micron. 19 SNRs show NIR emission in one or more IRAC bands. The detection rate of IR SNRs in the LMC is remarkably high ($\sim62\%$) compared to that of Galactic SNRs ($\sim30\%$). The major reason is likely to be less IR confusion by the Galactic disk, and we cannot find any intrinsic difference between the LMC and the Galaxy that augments the detection rate.

2.  We found a linear correlation between the $Spitzer$ 24 \micron~and 70 \micron~fluxes with a large scatter. Assuming the MIR emission is mainly due to dust grains in thermal equilibrium, we compare the observed 24 and 70 \micron~fluxes of the LMC SNRs with a modified blackbody (Equation (\ref{eq1})) using the dust mass absorption coefficient from the ``average'' LMC model \citep{wein01}. Dust temperatures from 45 to 80 K and corresponding dust masses ranging from 0.001 to 0.8 $M_\sun$ can reproduce the observed MIR fluxes of the LMC SNRs. The wide range of the dust properties (i.e., the large scatter) may be related to the different dust heating mechanisms (collision or radiation) among SNRs. The MIR emissions of most SNRs in both MIPS bands (24 and 70 \micron) are more likely to be emission from dust. For some remnants such as N49 or N63A, however, previous spectroscopic studies have revealed a significant contribution from ionic lines to the 24 \micron~emission, and these SNRs have morphologies in the IRAC bands similar to the morphology in the 24 \micron~band associated with the line emission.

3.  Among 19 SNRs detected in the IRAC bands, we classify the origin of the IRAC-band emission into four groups; nine by ionic line emission, four by molecular line emission, five by PAH emission, and one by synchrotron emission. The classification is based on the IRAC colors and their morphological comparison to X-ray, optical, and radio.  Taking into account the previous spectroscopic data, the origins of the IRAC-band emission of five SNRs (N11L, N103B knot, N49, N63A, and N157B) and SNR N158A are classified as ionic lines and synchrotron, respectively, even though all of them show IRAC colors for molecular shocks in the IRAC color-color diagram. This indicates that the criteria to discern the origin of IRAC emission in the color-color diagram might not be sufficiently stringent.

4.  The MIR fluxes show a rough correlation with radio fluxes, and the ratios of 24 \micron~to 21 cm radio continuum fluxes ($q_{24}$) are measured. The average $q_{24}$ ($q_{24,{\rm avg}}=0.14$) for LMC SNRs is smaller than that of Galactic SNRs (0.39), which reflects that many LMC SNRs have relatively fainter IR emission with respect to their radio brightness. In addition, $q_{24,avg}$ of the LMC SNRs (0.14) is much lower than those measured from extragalactic galaxies (typically, $\ga0.8$). This indicates that only small portion of the total 24 \micron~emission in the LMC is contributed from the SNRs.

5.  Using the $Chandra$ data and the {\it XMM} images in the literature, IR and X-ray morphologies of 25 SNRs (out of 29) are compared. Nine SNRs show strong spatial correlations, four show similarities with some minor differences, and 12 have considerable discrepancies. The discrepancies are likely to result from the interaction with a dense ambient material. The 24 \micron~fluxes have a good correlation with X-ray fluxes, suggesting that the dust heating mechanism is physically related to X-ray emitting hot plasma. The $Spitzer$ $[F_{70}/F_{24}]$ ratios tend to decrease as the X-ray brightness increases. This can be explained by the variation of the dust temperature depending on the electron density of hot plasma. In the hot plasma with a higher density emitting brighter X-ray emission, dust grains are heated up to higher equilibrium temperatures due to more collisions.

6.  The $[F_{24}/F_{70}]$ ratios show strong correlation with the electron density, and the dust model with a grain size of 0.1 \micron~can reproduce the observed ratios nicely. This could support destruction of small grains by sputtering in these SNRs. Meanwhile, the small-sized grain population may be favored for the mature SNRs with relatively low gas temperatures as dust shattering becomes efficient. Further observational studies of the grain size distribution in SNRs of various evolutionary stages will shed light on our understanding of the evolution of dust grains in an SNR.

\acknowledgments
This work is based in part on observations made with the {\it Spitzer Space Telescope}, obtained from the NASA/IPAC Infrared Science Archive, both of which are operated by the Jet Propulsion Laboratory, California Institute of Technology, under contract with the National Aeronautics and Space Administration. We thank Sean Points for providing the flux-calibrated MCELS data and Jae-Joon Lee for processing the archival $Chandra$ data excluded in the $Chandra$ LMC SNR catalog. J.Y.S. thanks Hiroyuki Hirashita for his helpful discussions on the 8 and 24 \micron~flux correlation and You-Hua Chu for useful comments to improve the manuscript. This work was partially supported by Basic Science Research
program (NRF-2011-0007223) through the National Research Foundation of
Korea (NRF)
funded by the Ministry of Education, Science and Technology.

\input{figures_low}

\input{table1}
\input{table2}
\input{table3}
\input{table4}

\input{table5}

\appendix

\section{BRIEF DESCRIPTION ON THE INDIVIDUAL SNRs}

\subsection{SNR 0450--70.9 (Figure \ref{irimg1})\label{app_0450}}
SNR 0450--70.9 is one of the largest SNRs in the LMC. Its optical extent is $6\arcmin.5\times4\arcmin.7$ \citep[$98\times70$ pc at 50 kpc;][]{will04}, and its IR size is seen even larger than those seen in optical or radio. The IR emission corresponds well to the rim of the radio emission and can be detected in all IRAC and MIPS bands. The overall morphology shows an elliptical shell. While the IR emission in the west is continuous, only patchy emission is identified in the east. The brightest region in the northwest contains a point source, which was identified as a source of IR cirrus \citep[IRAS 04505--7052;][]{stra92}. To avoid any contamination by this IR source, the area near the source is excluded for the flux estimation of the SNR (see Figure \ref{scimg}).  

In optical, outer shell and interior filaments are shown, and a high [\ion{S}{2}]/H$\alpha$ ratio is detected \citep[][and references therein]{will04}. All observations are consistent with 0450--70.9 as a mature SNR. Using theoretical models for thermal X-ray emission of the SNR, a lower limit of the SNR age is derived as $\sim45,000$ yr. No evidence of a pulsar/PWN has been detected at X-ray nor radio, and there is no \ion{H}{2} region or OB stars near this SNR \citep{chu88}. These facts might prefer the origin of Type Ia SN, but more direct evidence is required to confirm the origin of the remnant. 

\subsection{SNR in N4 (Figure \ref{irimg1})\label{app_n4}}
The IR images of SNR in N4 show a bright northeast region and patchy emission at south. The fragmentary emission seen at 24 \micron~is circularly distributed. An inward boundary of the bright northeast region is noticeably sharp, which adjoins the boundary of the optical emission. The IR emission of this region is distributed along the rim of the radio emission. In optical, an incomplete shell with filaments is seen, but no emission related to the SNR is identified at south. Also, no radio continuum is detected at south.

\subsection{0453--68.5 (Figure \ref{irimg1})\label{app_0453}}

This middle-aged SNR ($\sim13,000$ yr) shows a clear shell structure at 24 \micron~that well corresponds to the X-ray morphology. There are several bright knots along the rim, and the shell is filled with diffuse emission. Inside the shell, some bright emission is also seen at the center. Recently, \citet{gaen03} found the presence of a PWN at the center based on the radio and X-ray observations. The PWN properties most closely resemble those of the Galactic SNR Vela, and the elongated morphology of the PWN can indicate compression by the SNR reverse shock. The enhanced IR emission at the center might be associated with the PWN, but it is necessary to confirm it further. The bright emission seen at the southwest of the SNR is not related to the SNR.   

\subsection{N11L (Figure \ref{irimg1})\label{app_n11l}} 
\citet{will06} report the IR emission of N11L at 4.5 \micron~of which morphology is very similar to that seen in optical, while no emission associated with the SNR is detected at 8.0 \micron. Due to the limited spatial coverage of the observation, emission in the 3.6 and 5.8 \micron~IRAC bands and all MIPS bands could not be distinguished. Using the SAGE data, we could examine all IRAC and MIPS images and detect IR emissions in all bands except 8.0 \micron. The IR shell of N11L appears most clearly in the IRAC 4.5 \micron~band with a good spatial correlation with the optical emission, which suggests that Br$\alpha$ $\lambda4.05~\micron$ can be dominant. We cannot confirm the existence of Br$\alpha$ line emission because the IRS spectra do not cover wavelengths shorter than $\sim5$ \micron, but the IRS spectra at longer wavelengths do only show ionic emission lines with weak continuum, which is consistent with the ionic-line origin for the IRAC band emission.

This SNR shows a shell morphology with signatures of a breakout toward the northeast. Filamentary emissions protruding from the shell are seen in optical and IR bands, while diffuse emission is distributed on the north of the shell in X-ray and radio. Besides the morphological structures, optical echelle spectra of N11L suggest kinematic evidences that velocity profiles through the shell and filamentary loops seen in an optical show rapidly moving material along the filaments \citep{will99}. The breakout in N11L can implicate a low-density cavity outside the SNR surrounded by a structure on the western periphery of the N11 \ion{H}{2} complex.

\subsection{N86 (Figure \ref{irimg2})}
N86 shows faint IR emission in all $Spitzer$ bands, but the emission in the west of the SNR is difficult to be discriminated from the emission of other sources. However, hook-shaped IR emission in the east is discernible and shows a good correlation with optical and radio emissions. Similarly to N11L, N86 has a large breakout to the north as well as smaller outflow structures around the SNR, which can be seen in optical images \citep{will99}. However, these features are not detected in the IR bands. 

\citet{will99} found that a well-defined spherical expansion pattern and faint diffuse X-ray emission in the breakout. The expansion velocity of the material breaking out to the north is measured up to 100 km s$^{-1}$ from the optical spectroscopy. The existence of hot gas and the faster expending velocity indicate that material from the SNR is moving to a less dense medium through the breakout. Although N86 is relatively away from \ion{H}{2} complexes or superbubbles, the presence of the breakout could imply a complicated structure of its local surrounding medium.

\subsection{DEM L71 (Figure \ref{irimg2}) \label{app_l71}}

DEM L71 is one of the Balmer-dominated SNRs in the LMC. It has a well-defined shell morphology with the bright emission at the east and west limbs. The 24 \micron~morphology is in good agreement with that of the X-ray. This SNR shows very faint emission in radio relative to the IR emission unlike the other Type Ia Balmer-dominated SNRs (Table \ref{tab:lmcsnr}). $Chandra$ X-ray observations reveal a double-shock morphology consisting of a blast-wave interacting with the surrounding ISM and a reverse shock heating ejecta at the central region \citep{hugh03}. By optical spectroscopy, broad and narrow components of H$\alpha$ emission have been confirmed, which is a characteristic of nonradiative shocks in partially neutral gas \citep{ghava03}.

\subsection{N23 (Figure \ref{irimg2}) \label{app_n23}}
The IR emission seen at 24 \micron~is highly asymmetric, showing bright emission in the southeast but faint (or no) emission in the northwest. Such morphology is also seen in X-ray and optical bands. This can be explained by a considerable gradient of the ambient medium density probably due to the proximity to the open cluster HS114 which is located toward the southeastern side of the SNR \citep[see ][and references therein]{hugh06}. Recent $Chandra$ observations reveal the existence of a compact source detected in the hard X-ray band \citep[$>2$ keV;][]{hugh06,haya06}. The position of the source is nearly at the center of the remnant, which makes its association with the SNR more plausible. This object can be a pulsar (or PWN) or a neutron star related to the remnant although any observational evidence at different wavelengths, including IR, cannot clarify the origin of the source.

\subsection{DEM L72 (Figure \ref{irimg2}) \label{app_l72}}

IR images of DEM L72 show the emission of ``L'' shape in 8, 24, and 70 \micron~bands. The IR emission is well enclosed with an optical shell. Since this SNR is outside of the $ATCA$ survey, and the morphology of the IR emission cannot be compared to that of radio continuum. Besides, although \cite{kli10} found that optical shell is filled with diffuse X-ray emission by using {\it XMM-Newton} data, however, the distribution of the X-ray emission is more enhanced in the northwest where there is no IR emission detected. DEM L72 is one of the largest SNR in the LMC, and faint filaments extending to the northeast from the bright optical shell are visible, especially in [\ion{O}{3}]. Taking them into account, the angular size of this SNR can reach up to 9$\arcmin$ (or 135 pc at the distance of 50 kpc). Its age is estimated to $\sim100$ kyr, well into the relatively late stage of SNR evolution. This SNR seems to have similarities to another evolved object, SNR 0450--70.9 (see Appendix \ref{app_0450}), and further studies on these mature SNRs could improve our knowledge about the last stage of SNR evolution.

\subsection{N103B (Figure \ref{irimg2})\label{app_n103b}}
In IRAC bands, only IR knots are detected above the background level (denoted by black crosses in Figure \ref{irimg2}, fifth row). The locations of the knots are well coincident with those of the knots seen in H$\alpha$. Moreover, its IRAC colors are smaller than those for molecular shocks (Figure \ref{fig:ccd1}), and its NIR spectra show the signature of [\ion{Fe}{2}] emission \citep{oliva89}. These facts indicate that the NIR emission of N103B is dominated by ionic-line emission. IR emission of N103B at 24 \micron~has a well-defined shell morphology, which shows good correlation to that seen in X-ray. The IR emission is enhanced in the west, which can be seen in X-ray and H$\alpha$, too. The 70 \micron~image of the SNR also shows a shell-like structure, but the distribution of the emission is somewhat different from that of the emission in 24 \micron. The shell of N103B seen at 24 \micron~and longer wavelengths is considered to be dominated by dust emission although ionic line emission originating from the IR knots could contribute to some extent.

N103B is a young SNR \citep[$\sim860$ yr based on the light echo][]{rest05} and is one of the interesting objects among the LMC SNRs. Its proximity to the young star cluster NGC 1850 ($\sim40$ pc) led that the progenitor of the SNR was a massive star from the cluster \citep{chu88}. Moreover, overabundance of O, Ne, and Mg seen in the {\it XMM-Newton} emission-line spectra also supports the Type II SN origin \citep{vander}. However, a $Chandra$ ACIS observation reveals the distribution of the ejecta and the large mass of Fe (0.34 $M_\sun$) that are in favor of a Type Ia SN, rather than of a Type II SN \citep{lewis}. More recently, \cite{bade09} suggested that N103B is associated with recent star formation and could be a ``prompt'' Type Ia SN of which progenitor is relatively younger and more massive before the explosion. It is expected that more detailed investigation on the IR emission could provide convincing evidence for its progenitor.

\subsection{SNR 0509--67.5 (Figure \ref{irimg3}) \label{app_0509}}
This remnant is one of the youngest SNRs in the LMC \citep[$\sim410$ yr,][]{rest05}, and its X-ray spectra and Balmer-dominated optical emission indicate that it originated from Type Ia SN explosion \citep{hugh95}. The MIPS 24 \micron~image shows a bright peak in southwest with a diffuse shell, which coincides with the shell structure seen in both the X-ray and optical images \citep{borko06}. No emission is detected in the IRAC bands.

\subsection{SNR 0519--69.0 (Figure \ref{irimg3}) \label{app_0519}}
Another young Balmer-dominated SNR 0519--69.0 originates from Type Ia SN explosion \citep{hugh95, ghava07}. A complete shell with three bright knots is clearly seen at 24 \micron~(and 70 \micron), which has moderately good agreement with optical and X-ray morphologies. 

\subsection{N132D (Figure \ref{irimg3}) \label{app_n132d}}
N132D is a young, oxygen-rich SNR which is a subset of core-collapse SNe \citep{morse}. It is one of the brightest LMC SNRs in IR and shows a shell-like morphology as seen in the X-rays. Enhanced emission is seen in southeast which is likely to result from the interaction with a molecular cloud \citep{banas97, tappe06}. It is difficult to distinguish IR emission from the SNR in the IRAC bands, but the northwestern knot named ``West Complex'' by the previous optical observation can be noticed \citep{tappe06}. Besides, $Spitzer$ IRS observations reveal the dominant dust continuum with broad PAH emission at $15-20$ \micron~\citet{tappe06}.

\subsection{N49B (Figure \ref{irimg3}) \label{app_n49b}}
This remnant is a CCSNR with an age of $\sim10,000$ yr \citep{hugh98, park03}. Its shell is clearly seen in the MIPS bands, which is very similar to the X-ray morphology. IR emission along the limb is quite patchy, and the brightest peak is located at the southern rim. In addition, some diffuse emission is also seen in the central region. No distinct emission is detected in the IRAC bands.

\subsection{N49 (Figure \ref{irimg3})\label{app_n49}}
N49 is one of the IR bright SNRs in the LMC and shows bright emission over wide wavelengths. Its $Spitzer$ IRS spectra show abundant emission lines from shocked gas including strong ionic lines (e.g., [\ion{Fe}{2}] 5.34 \micron, [\ion{Ar}{2}] 7.03 \micron) as well as modest molecular hydrogen lines \citep{will06}. Although its IRAC colors are close to what is expected to molecular shocks, the IR spectra clearly show that it is dominated by ionic line emission. Even IR emission in the 24 \micron~band, the contribution from ionic lines can be up to $\sim80\%$, but dust continuum is a dominant emission source at longer wavelengths \citep{otsu10}. IR spectra taken by $AKARI$ and $Spitzer$ also show several PAH features \citep{seok12}, which is likely to relate to the interaction with a molecular cloud \citep{banas97}.

\subsection{DEM L205 (Figure \ref{irimg4}) \label{app_l205}}
DEM L205 is recently confirmed by new X-ray and radio data as an SNR \citep{mag12}. Although the remnant is confused by the bright emission of the \ion{H}{2} complex (LHA 120-N 51) on the western side, the optical image clearly shows an arc of a shell-like structure which can also be discernible in the 24 \micron~band. The shorter wavelength images seem to show some associated emission, but it is not clear because of the confusion with other emission. \citet{mag12} measured 24 and 70 \micron~fluxes, but the result depends strongly on the background fluxes. We measured the background from both sides of the arc region, and our 24 and 70 \micron~fluxes (115 and 734 mJy) are smaller compared to the fluxes of \citet{mag12}.  

\subsection{SNR in N206 (Figure \ref{irimg4})\label{app_n206}}
  
IR emission of SNR in N206 is not isolated from that of its surrounding medium, which makes it difficult to disentangle IR emission associated with the SNR from other emission. While no IR emission associated with the SNR is detected in the north, patchy shell-like emission is identified, which spatially corresponds to the well-defined optical shell. The southern areas are confused by the bright ambient medium (possibly periphery structures of N206 \ion{H}{2} region), so the eastern rim is only used for its flux estimation. While bright X-ray emission is observed at the center of the SNR, IR and optical (and radio) emissions mostly show shell-like morphology. This fact makes the SNR categorized as ``mixed morphology'' SNR. Recently, \cite{will05a} performed new radio observations by using the $ATCA$ and found an elongated, radially oriented feature across the SNR. In the $Chandra$ images, X-ray emission associated with this radio feature is also detected, which suggests that this feature originates from a PWN. 

\subsection{DEM L238 (Figure \ref{irimg4})\label{app_l238}}

We could detect distinct IR emission of this SNR only at 24 \micron. Although the emission at 24 \micron~is somewhat confused by surrounding emission, the southeastern shell and the northwestern shell are distinctly identified showing good correlation to a shell seen in optical. As the northern shell in optical is brighter than the southern, the IR shell also shows the same tendency. This enhancement might be explained by the fact that SNR shocks encounter denser mediums in the north. Besides, the IR shell is in moderate agreement with the faint X-ray shell, but the X-ray emission has a bright central region unlike the IR. 
 
This is a moderately old ($\sim10^4$ yr) SNR with high Fe abundances in X-ray spectra \citep{bork06}. This Fe overabundance suggests that DEM L238 is a result of a Type Ia explosion. In view of a Type Ia SN, the presence of Fe-rich ejecta at the center of this SNR is unexpected, because usual Type Ia SNRs have faint ejecta emission in the middle stage of their evolution. This fact might indicate that a dense CSM is present around the progenitor, which is one of the characteristics for more massive Type Ia progenitors (``prompt'' Type Ia).

\subsection{SN 1987A (Figure \ref{irimg4})\label{app_87a}}
SN 1987A is only detected in the 5.8 \micron, 8.0 \micron, and 24 \micron~bands, and the detail structure seen in optical cannot be resolved by $Spitzer$. Its IRS spectra show a dust continuum with several ionic emission lines such as [\ion{Ne}{2}] 12.81 \micron~and [\ion{Ne}{3}] 15.56 \micron~\citep{bouch06}. The dust emission is most likely to arise from astronomical silicate particles with the dust temperature of $\sim160$ K. Recent observations by {\it Herschel Space Observatory} reveal FIR emission with a peak at $150-200$ \micron. The detection of the FIR emission suggests a population of cold dust grains with a temperature of $\sim20$ K and a dust mass of $0.4-0.7$ $M_\sun$. This amount of dust mass is sufficient to support that SNe can produce the large dust masses observed in dusty galaxies at high redshifts.

\subsection{N63A (Figure \ref{irimg4})\label{app_n63a}}
N63A is located in the periphery of a lager \ion{H}{2} region, N63, and it is a relatively young SNR \citep[2000--5000 yr;][]{hugh98}. In all $Spitzer$ bands, a three-lobed structure is seen, which is also observed in optical \citep[e.g.,][]{will06}. We can confirm that the archival IRS spectra of the three-lobed region shows dominant ionic lines with several H$_2$ lines and PAH features. Characteristics of the three lobes are interesting, because the two eastern lobes show evidence of shock ionization (e.g., high [\ion{S}{2}]/H$\alpha$ ratio) while the western lobe shows that of photoionization \citep{leven95}. However, the IR SED of the western lobe is unexpectedly similar to the SED of the eastern lobe, and the northeastern lobe shows the most different SED from the others \citep{will06}. N63A also shows a diffuse but clear shell-like structure that well corresponds to the morphologies seen in radio or X-ray \citep[e.g.,][]{dick93,warr03}. In particular, crescent features seen in X-ray faintly, but apparently, appear at 24 \micron. The faint shell emission probably arises from dust grains collisionally heated by shocks. 

\subsection{DEM L241 (Figure \ref{irimg5})\label{app_l241}}
IR images of DEM L241 show a shell-like structure with looped filaments inside, which is similar to that seen in optical. In particular, the 4.5 \micron~image depicts the IR shell most clearly. The enhanced emission in the western rim is visible in all $Spitzer$ bands, while other structures are distinct in only few bands. The western rim is largely brighter than the eastern rim, which might result from the denser medium in the west. The bright emission region in the north is an \ion{H}{2} region, so that area is excluded for the flux estimation.   

Recent {\it XMM-Newton} data reveal the head-tail structure of X-ray emission with a point source in it \citep{bam06}. The luminosity and spectrum of the source indicates that the source could be a PWN of DEM L241. Moreover, overabundance in O and Ne with the existence of a nearby OB association (LH 88) implies a very massive progenitor of the SNR with more than 20 $M_\sun$.

\subsection{DEM L249 (Figure \ref{irimg5})\label{app_l249}}
DEM L249 shows a circular shell morphology with some extended emission to the north from the shell in all $Spitzer$ bands. In the 3.6 and 4.5 \micron~images, faint diffuse emission can be identified after subtracting emission from point sources. The eastern rim is brighter than the western rim, and several bright knots are present along the eastern rim. Asymmetric brightness between the east and the west is also seen in the optical morphology, but the locations of IR peaks on the rim are different from those of optical peaks. Although overall distribution of X-ray emission shows similar extent to the IR shell, bright central emission surrounded by diffuse emission is observed.  

This SNR has similar features in X-ray emission to DEM L238 such as bright central emission with high Fe abundances \citep[; see also \S\ref{app_l238}]{bork06}. These facts suggest that DEM L249 is a result of Type Ia explosion with a relatively massive progenitor. Further investigation is required to convince the origin of this SNR, and studies of DEM L249 together with N103B and DEM L238 could disclose the evolution of ``prompt'' Type Ia SNR and their influences to ambient media.

\subsection{DEM L256 (Figure \ref{irimg5})\label{app_l256}}
Since there are several \ion{H}{2} regions (DEM L251, 253, and 264) in the vicinity of DEM L256, the IR emission associated with the SNR is somehow confusing. This SNR has well-defined double shell morphology in optical, and the IR emission corresponding to the northern part of the outer shell can be designated in all $Spitzer$ bands. In the IR images, a bright clump lies on the extended emission in the west, which has a counterpart in optical. This clump is located just outside of the inner shell and is along an extension of the outer shell. In fact, the clump is known to have lower [\ion{S}{2}]/H$\alpha$ ratio ($<0.4$) unlike the other filamentary shells have much higher values \citep[$>0.7$;][]{kli10}. This indicates that the clumpy emission is not associated with the SNR, so the bright emission seen in IR bands also originates from other sources. Although the northern tip of the extended emission in the west has a good coincidence with the inner optical shell, we only used the northern shell for the flux estimation to avoid any contamination from the bright source in the west.       

The two nested shells of DEM L256 invoke the bilobed structure for this SNR. Moreover, two expansion velocities observed in an echelle spectrum support this scenario \citep{kli10}. X-ray images taken by {\it XMM-Newton} show much enhanced emission toward the outer half-shell (the eastern shell). The X-ray emission generally trace the H$\alpha$ filaments. The edge where strong H$\alpha$ emissions at the systemic velocity are seen together with corresponding X-ray emission, which might indicate that the SNR is encountering dense materials.

\subsection{N157B (Figure \ref{irimg5})\label{app_n157b}}
The Crab-like SNR N157B was previously suggested to be a line-dominated SNR, according to the $AKARI$ data \citep{seok08}. IR emission associated with the PWN is not detected because the emission expected from the synchrotron emission is lower than the detection limit. The brightest IR emission in the SNR region does not originate from the SNR itself, but from the nearby molecular cloud, though some filamentary structures are similar to its H$\alpha$-emitting nebula. This can infer that the IR emission has the origin of ionic-line emission.

\subsection{SNR in N159 (Figure \ref{irimg5})\label{app_n159}}
SNR in N159 is located in the \ion{H}{2} complex N159 that contains several bright emission regions, which makes it difficult to define the SNR emission. The optical morphology of the \ion{H}{2} complex is similar to the shape of the number ``8'' consisting of northern (N) and southern (S) lobes, and only the SNR emission in the N lobe was previously known by detecting the fast expending shell \citep{chu97}. Recent $Chandra$ observations with the MCELS data, \cite{sew10} can define larger SNR emission including the emission in the S lobe. After removing scattered emission from the nearby bright source LMC X-1, X-ray emissions overlapping with the N and the S lobes are identified. 

IR emission in N159 shows very complicated morphology; it largely has a circular shape with several sharp filaments and some bright clumps in the middle. The two bright regions in the middle are not associated with this SNR, which are well defined by radio continuum and no X-ray emission \citep{sew10}. The clump in the east is a bright \ion{H}{2} region called Papillon Nebula. A thin filament correlated with the S lobe is shown in all $Spitzer$ bands, and a small shell-like structure with several filaments is found in the N lobe. For the flux estimation, we masked out the two bright regions in the middle and measured fluxes from the N lobe and the filament in the S lobe separately (see Figure \ref{scimg}).

\subsection{N158A (Figure \ref{irimg6})\label{app_n158a}}
N158A is one of the well-known Crab-like SNRs in the LMC. There is a PWN at the center of the remnant, which is detected as a point source in the all bands of $AKARI$ and $Spitzer$. No distinct emission from the shell is found unlike that seen in X-ray or radio, but some diffuse emission might exist along the southern rim of the X-ray shell. The $Spitzer$ IRS spectroscopy to the PWN reveals an excess of IR emission at longer than 20 \micron~from synchrotron emission expected from the radio and optical power laws \citep{bwill08}. The continuum is attributed to a warm dust component with a temperature of $\sim50-65$ K, and the origin of the excess is more likely to SN synthesized dust heated by shocks generated by the PWN. 

Its IRAC ratios slightly differ from the theoretical calculation (Figure \ref{fig:ccd1}). This might result from the contribution of emission lines, for instance, [\ion{Fe}{2}] $\lambda 5.34$ \micron~in the 5.8 \micron~band and [\ion{Ar}{2}] $\lambda 6.99~\micron$ in the 8.0 \micron~band while there is no strong line contribution to the 3.6 \micron~and 4.5 \micron~bands \citep[see IRS spectrum in][]{bwill08}. Such emission lines are also seen in the Crab nebular, and the origin of line emission is most likely to be dense clumps of ejecta swept up by highly radiative shocks.


\subsection{DEM L299 (Figure \ref{irimg6})\label{app_l299}}
SNR DEM L299 is located on the northwestern periphery of the ring-shaped nebula, DEM L299. IR emissions of DEM L299 in the IRAC bands are sporadic, and extended emissions from other sources in the west possibly obscure the emission from the SNR. The cavity seen in the 24 \micron~image is well confined by the shell seen in H$\alpha$. The patchy emission in the east has correspondence with the eastern rim in optical, which leads us to associate the emission with the SNR. The bright extended emission in the northwest originates from an \ion{H}{2} region photoionized by a star, Sk--68 155 and a young stellar object \citep[and references therein]{desai10}. Since this emission is not associated with the SNR and might be extended to the south, we only measure fluxes from the eastern rim.

\subsection{DEM L316A/B (Figure \ref{irimg6})\label{app_l316}}
The DEM L316 system shows an interesting double shell structure. Both SNRs show clear shell-type morphologies in optical, while their X-ray images show centrally brightened emission with diffuse emission and several clumps. Their IR emissions are quite confusing; L316A (the northeastern shell) shows a wedge-shape structure in the east with some confusion in the north. In the case of L316B (the southwestern shell), the southern rim has good correspondence with the optical shell while the northern region of L316B is hardly discriminated from contamination by other emissions.  

DEM L316A and DEM L316B were firstly thought to be colliding with each other because of their spatial adjacency \citep{will97}. Later, their progenitor types were, however, suggested to be different based on the spectral X-ray features \citep[L316A: Type Ia and L316B: Type II,][]{will05}, which makes the collision between two SNRs doubtful. \cite{tole09} performed hydrodynamic simulations to delineate the observed X-ray features. It is found that a non-colliding condition including thermal conduction can reproduce X-ray emission in good agreement with observational results. Their results are also in favor of the non-colliding case, so one of the remaining questions is how two different types of SNRs can be formed in vicinity.

\subsection{SNR 0548--70.4 (Figure \ref{irimg6})\label{app_0548}}
The IR emission of this remnant is only detected in the MIPS bands, and no emission can be distinguished at shorter wavelengths. A shell-like structure seen in X-ray can be discerned in the 24 \micron~image, and a bright clump is also shown inside. The clump might be an IR counterpart of the bright X-ray emission on the western side, but its location has a better agreement with the optical emission rather than with the X-ray emission. SNR 0548--70.4 is a middle-aged Type Ia SNR based on the Balmer-dominated optical emission and Fe-overabundant X-ray spectra \citep{hend03}.

\input{reference}
\clearpage

\end{document}

%% file: figures_low.tex
\begin{figure}
\plotone{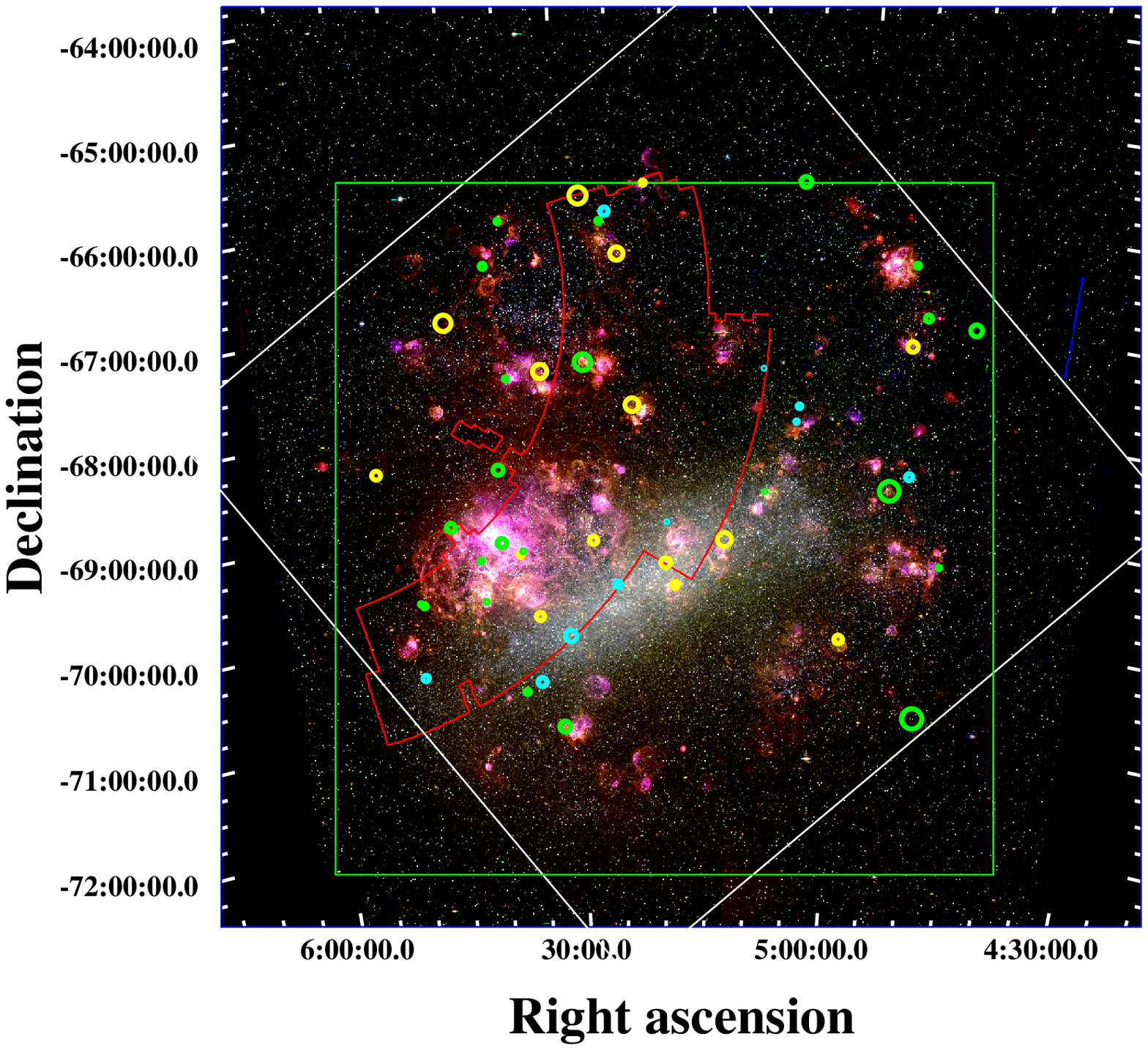}
\caption{Distribution of 47 bona fide SNRs marked on the MCELS composite image of the LMC (H$\alpha$: red, [\ion{S}{2}]: green, [\ion{O}{3}]: blue). Circles represent positions of the SNRs, and their sizes reflect the sizes of the SNRs. Colors of symbols indicate detection of IR emission from each SNR (yellow: no IR detection, cyan: MIR only, green: N/MIR emission). The spatial coverages of the $AKARI$ LMC survey (red), the ATCA radio survey (green), and the $Spitzer$ SAGE survey (white) are also overlaid.   \label{lmc}}
\end{figure}

\begin{figure}
\epsscale{1}
\plotone{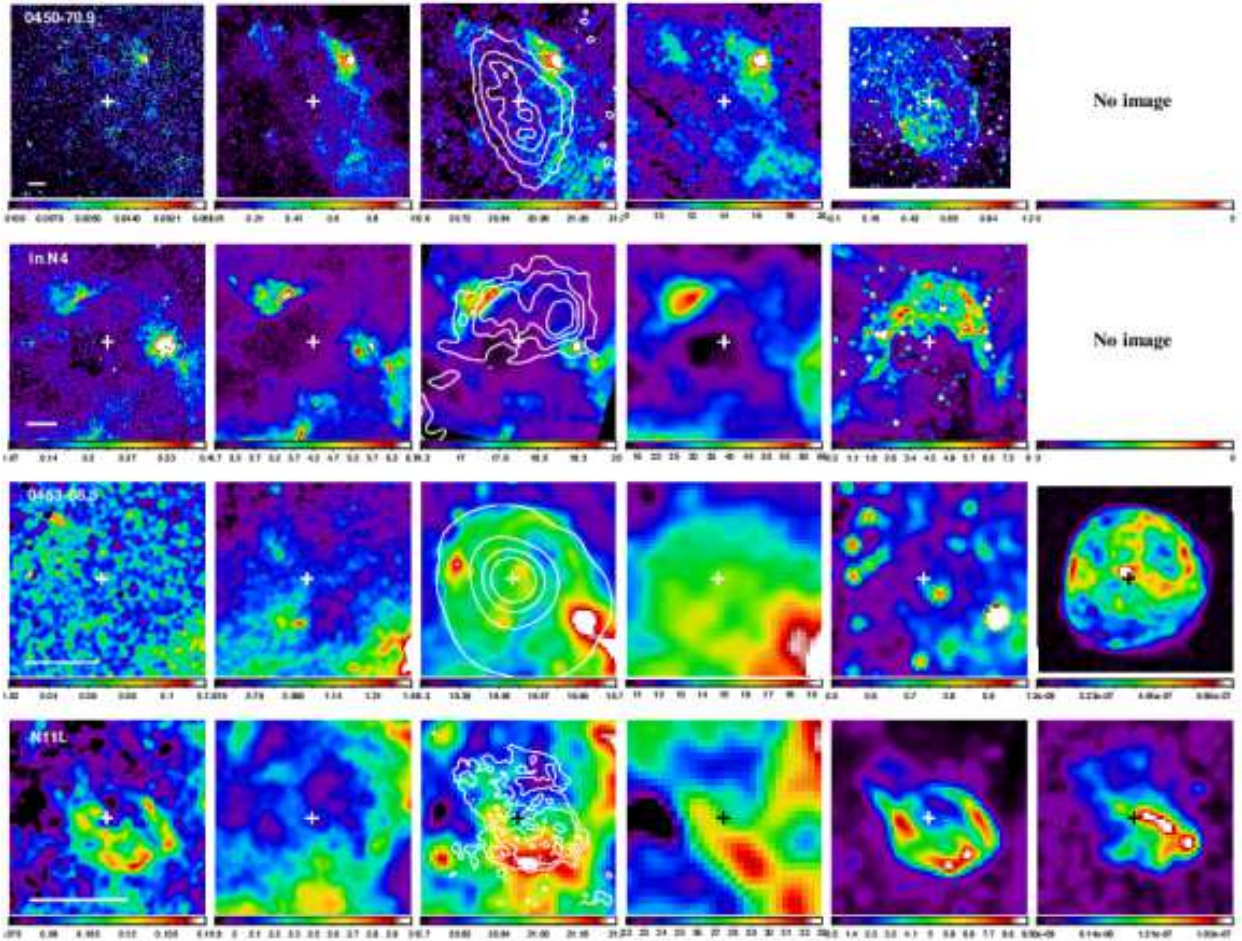}
\caption{From top to bottom: $Spitzer$ images of SNR 0450--70.9, SNR in N4, SNR 0453--68.5, and N11L. For each SNR, point source subtracted IRAC 4.5 and 8.0 \micron, and MIPS 24 and 70 \micron~images are shown together with (flux-calibrated) MCELS H$\alpha$ and $Chandra$ X-ray images for comparison. When there is no available X-ray data, the last column is left blank. Cross marks in all frames denote the center position of each SNR listed in Table \ref{tab:lmcsnr}, and the scale bar in the 4.5 \micron~images (first column) corresponds 1\arcmin~or 15 pc at 50 kpc). Contours in the 24 \micron~images (third column) are mostly from ATCA 4.8 GHz data which have half-power beam width (HPBW) of 33\arcsec \citep{dic05}, levels are stated individually in Appendix. All images are smoothed with a three-pixel Gaussian. The units on the colorbar are MJy sr$^{-1}$ for the $Spitzer$ images, ADU s$^{-1}$ for the MCELS, and counts cm$^{-2}$ s$^{-1}$ for the $Chandra$. The MCELS images for three SNRs (SNR 0450, SNR in N4, and N11L) are flux-calibrated, so that they are in units of 10$^{-5}$ erg cm$^{-2}$ s$^{-1}$. Contours of N11L are from 5 GHz ATCA data with HPBW of 3\arcsec.0 \citep{will99}. North is up, and east is to the left. \label{irimg1}
}
\end{figure}

\begin{figure}
\epsscale{1}
\plotone{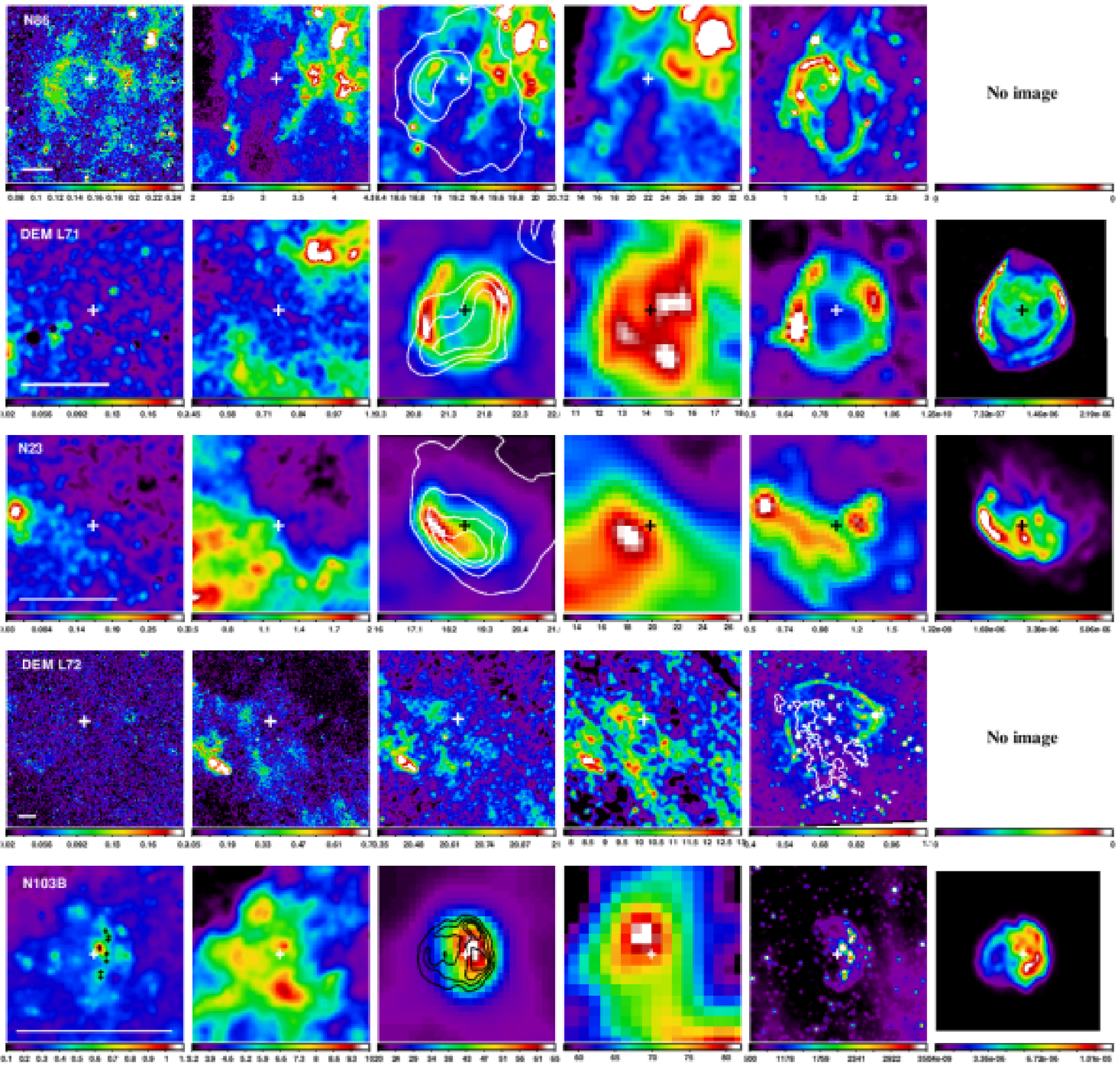}
\caption{From top to bottom: same as Figure \ref{irimg1} but for N86, DEM L71, N23, DEM L72, and N103B. Contours of N23 and N103B are from different 4.8 GHz ATCA data observed by \citet[HPBW: 10\arcsec]{dic98} and \citet[HPBW: 3\arcsec.0]{dic95}, respectively. As DEM L72 is outside of the ATCA survey, no 4.8 GHz contour is overlaid. Instead, contours in the H$\alpha$ image represent 0.2 MJy sr$^{-1}$ level of star-subtracted 8.0 \micron~image. Black crosses in the 4.5 \micron~image of N103B are marked on the positions of knots discernible in H$\alpha$. The H$\alpha$ image of N103B is from the Very Large Telescope archive.  \label{irimg2}}
\end{figure}

\begin{figure}
\epsscale{1}
\plotone{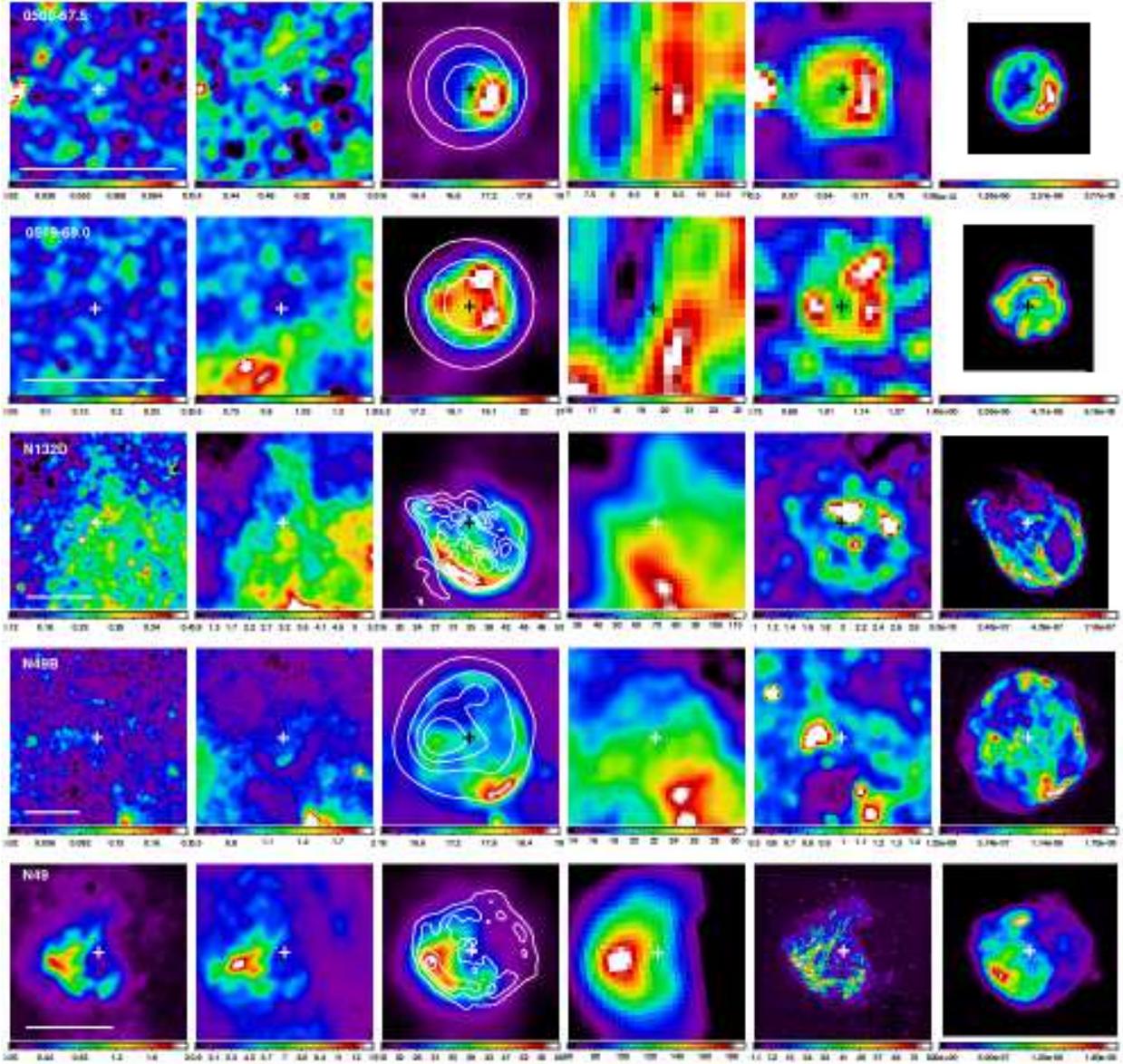}
\caption{From top to bottom: same as Figure \ref{irimg1} but for SNR 0509--67.5, SNR 0519--69.0, N132D, N49B, and N49. Contours of N132D and N49 are from different 4.8 GHz ATCA data observed by \citet[HPBW: 3\arcsec.0]{dic95} and \citet[HPBW: 10\arcsec]{dic98}, respectively. The H$\alpha$ image of N49 is from the {\it Hubble Space Telescope} archive \citep{bil07}. \label{irimg3}}
\end{figure}

\begin{figure}
\epsscale{1}
\plotone{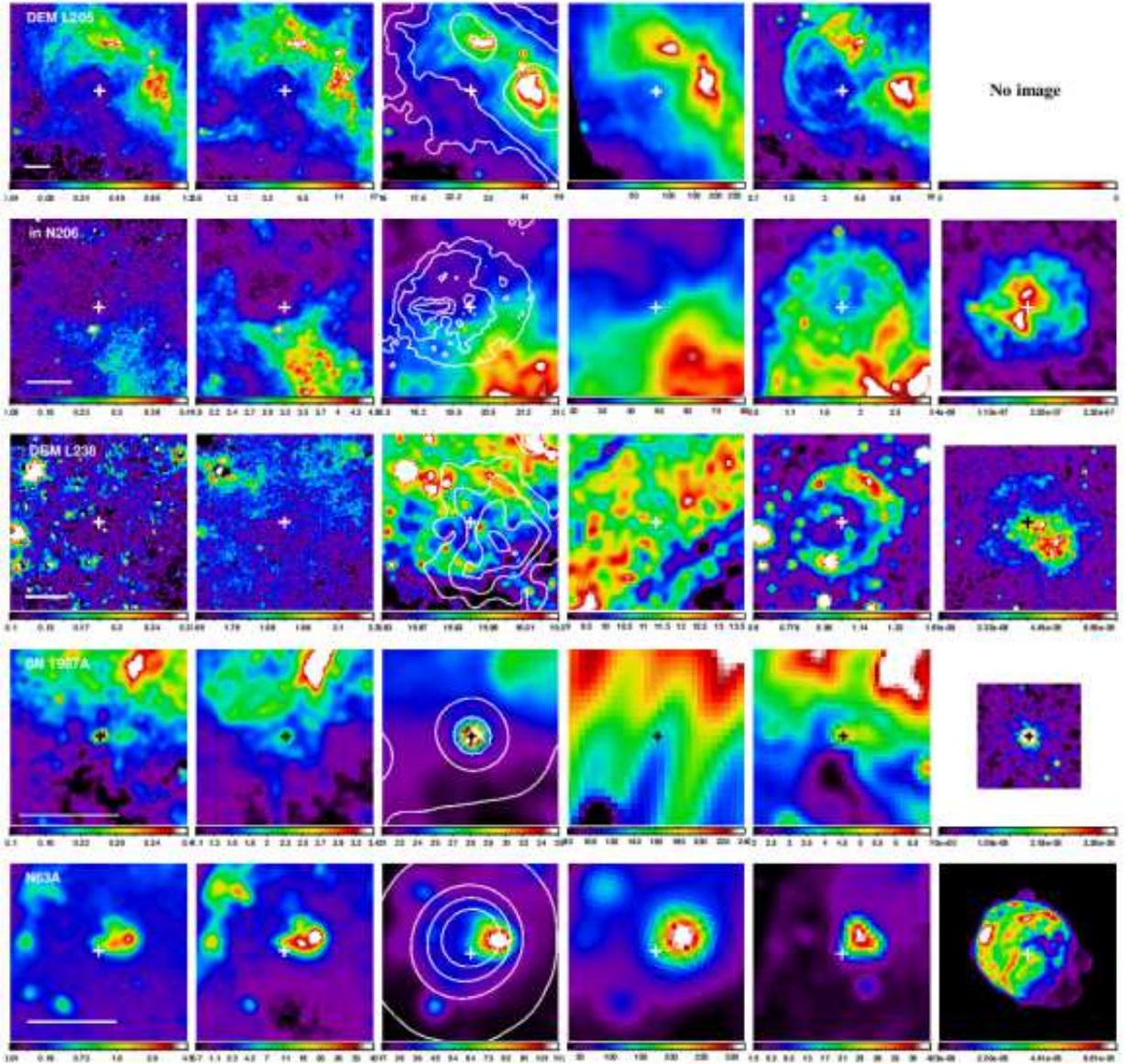}
\caption{From top to bottom: same as Figure \ref{irimg1} but for DEM L205, SNR in N206, DEM L238, SN 1987A, and N63A. Contours of SNR in N206 are from different 4.8 GHz ATCA data (HPBW: 1\arcsec.8) observed by \citet{klin02}.   \label{irimg4}}
\end{figure}

\begin{figure}
\epsscale{1}
\plotone{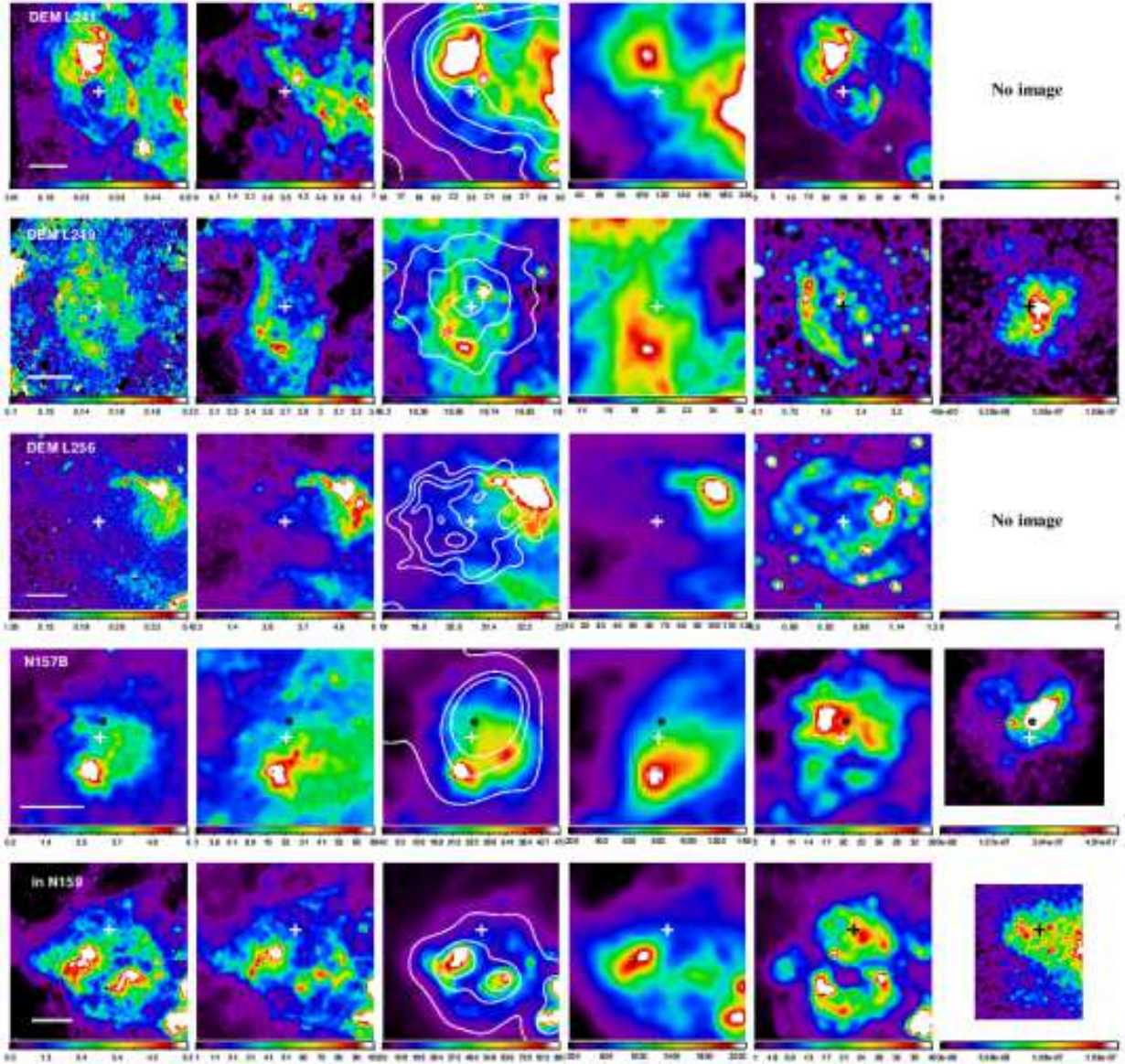}
\caption{From top to bottom: same as Figure \ref{irimg1} but for DEM L241, DEM L249, DEM L256, N157B, and SNR in N159. Black circle in the image of N157B indicates the position of pulsar. The MCELS images of DEM L241 and DEM L249 are flux-calibrated, so that they are in units of 10$^{-5}$ erg cm$^{-2}$ s$^{-1}$. \label{irimg5}}
\end{figure}

\begin{figure}
\epsscale{1}
\plotone{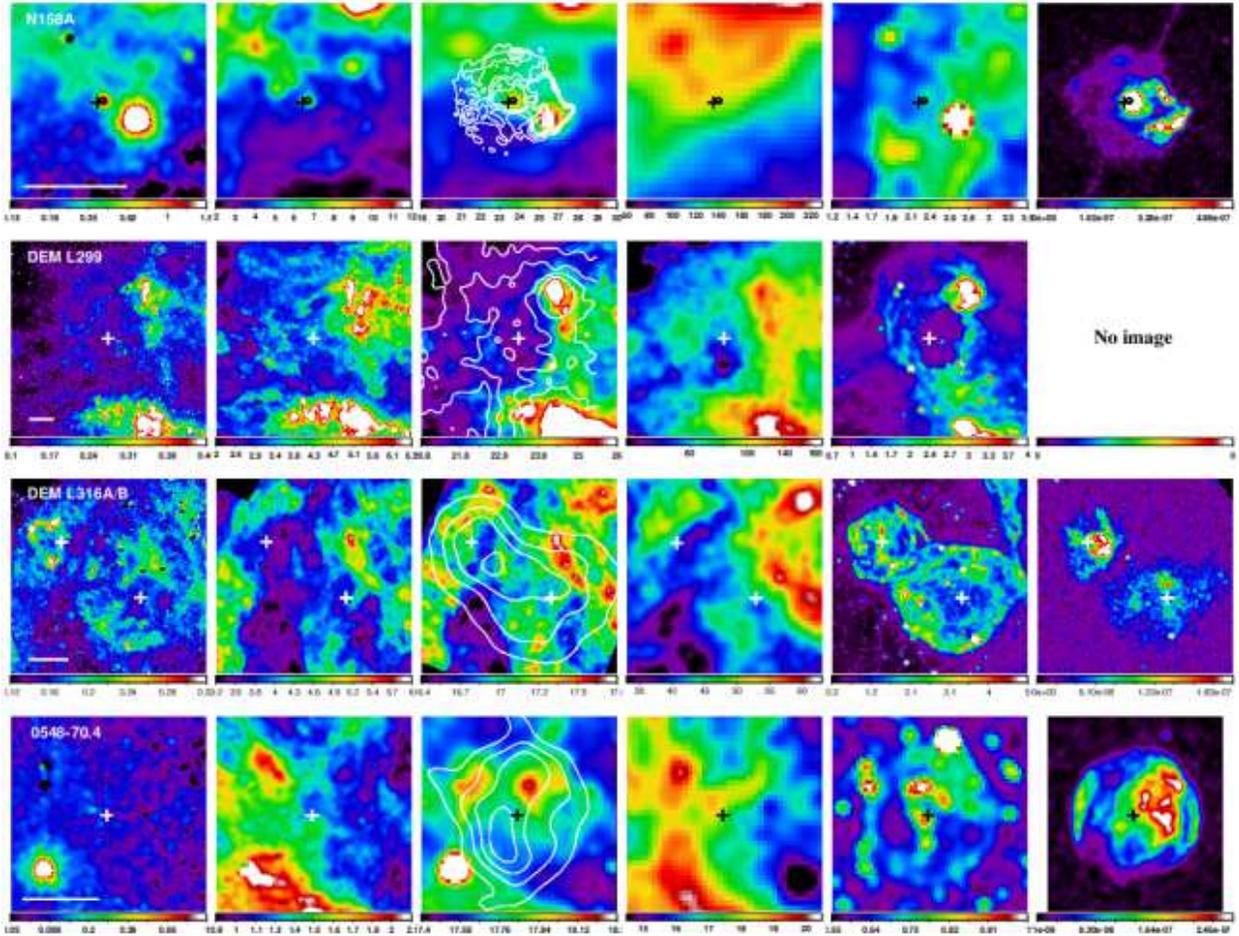}
\caption{From top to bottom: same as Figure \ref{irimg1} but for N158A, DEM L299, DEM L316A/B, and SNR 0548--70.4. Black circle in the image of N158A indicates the position of pulsar. Contours of N158A are from different 4.8 GHz ATCA data (HPBW: 2\arcsec.7) observed by \citet{man93}. The MCELS image of DEM L316A/B is flux-calibrated, so that it is in units of 10$^{-5}$ erg cm$^{-2}$ s$^{-1}$. \label{irimg6}}
\end{figure}
\begin{figure}
\plotone{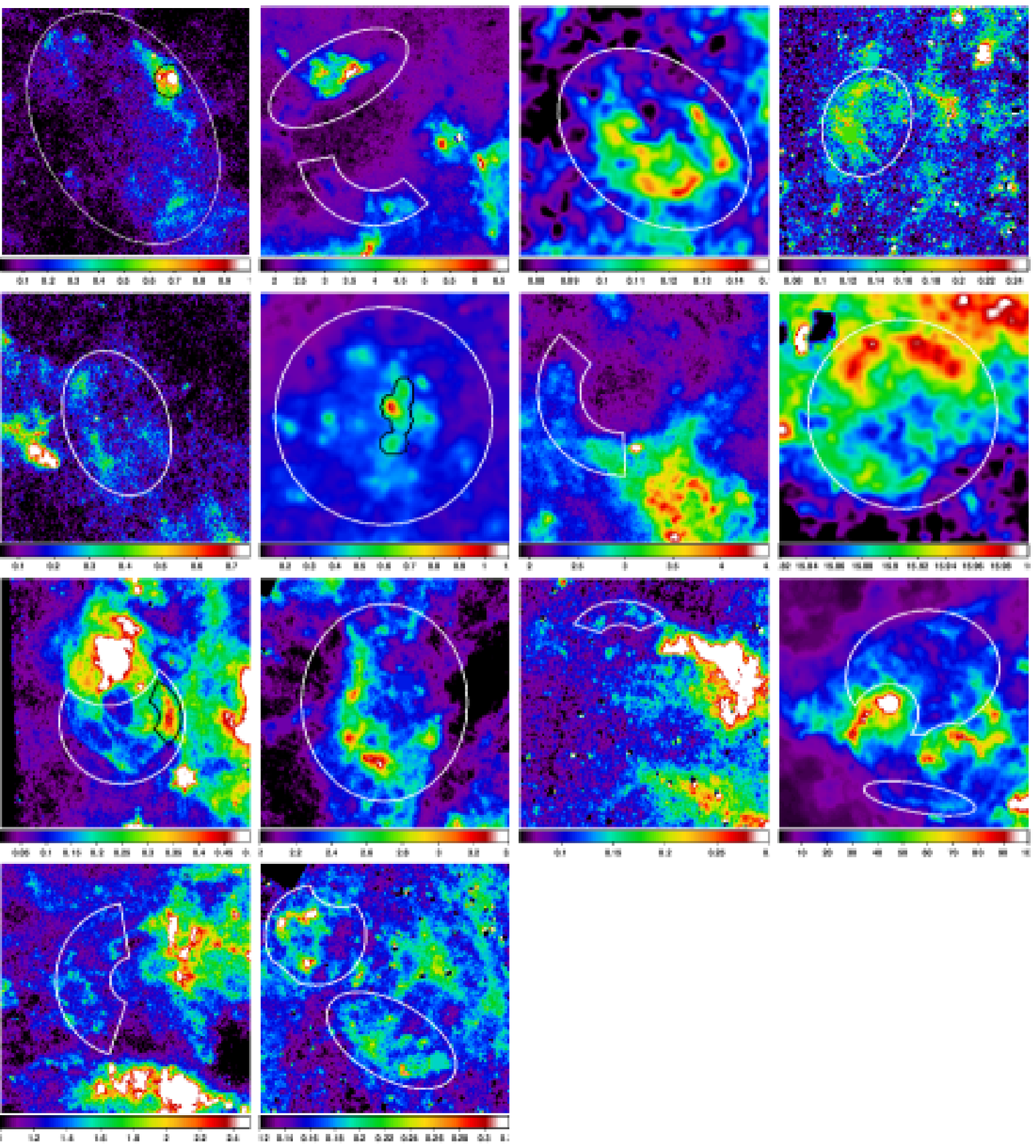}
\caption{Star-subtracted $Spitzer$ images of 14 SNRs of which IR fluxes are measured in this paper. Top row: SNR 0450--70.9 (8.0 \micron), SNR in N4 (8.0 \micron), N11L (4.5 \micron), and N86 (4.5 \micron). Second row: DEM L72 (8.0 \micron), N103B (4.5 \micron), SNR in N206 (8.0 \micron), and DEM L238 (24 \micron). Third row: DEM L241 (4.5 \micron), DEM L249 (8.0 \micron), DEM L256 (3.6 \micron), and SNR in N159 (8.0 \micron). Bottom row: DEM L299 (5.8 \micron) and DEM L316A/B (4.5 \micron). Regions where fluxes are extracted are marked in individual figures (white). In N103B and DEM L241, fluxes are also measured for some specific areas which are marked by solid black lines. In SNR 0450--70.9, the northwestern region excluded for the flux estimation is also marked (black dashed circle). The units on the colorbar are MJy sr$^{-1}$. For all images, north is up, and east is to the left. \label{scimg}}
\end{figure}


\begin{figure}
\epsscale{1}
\plotone{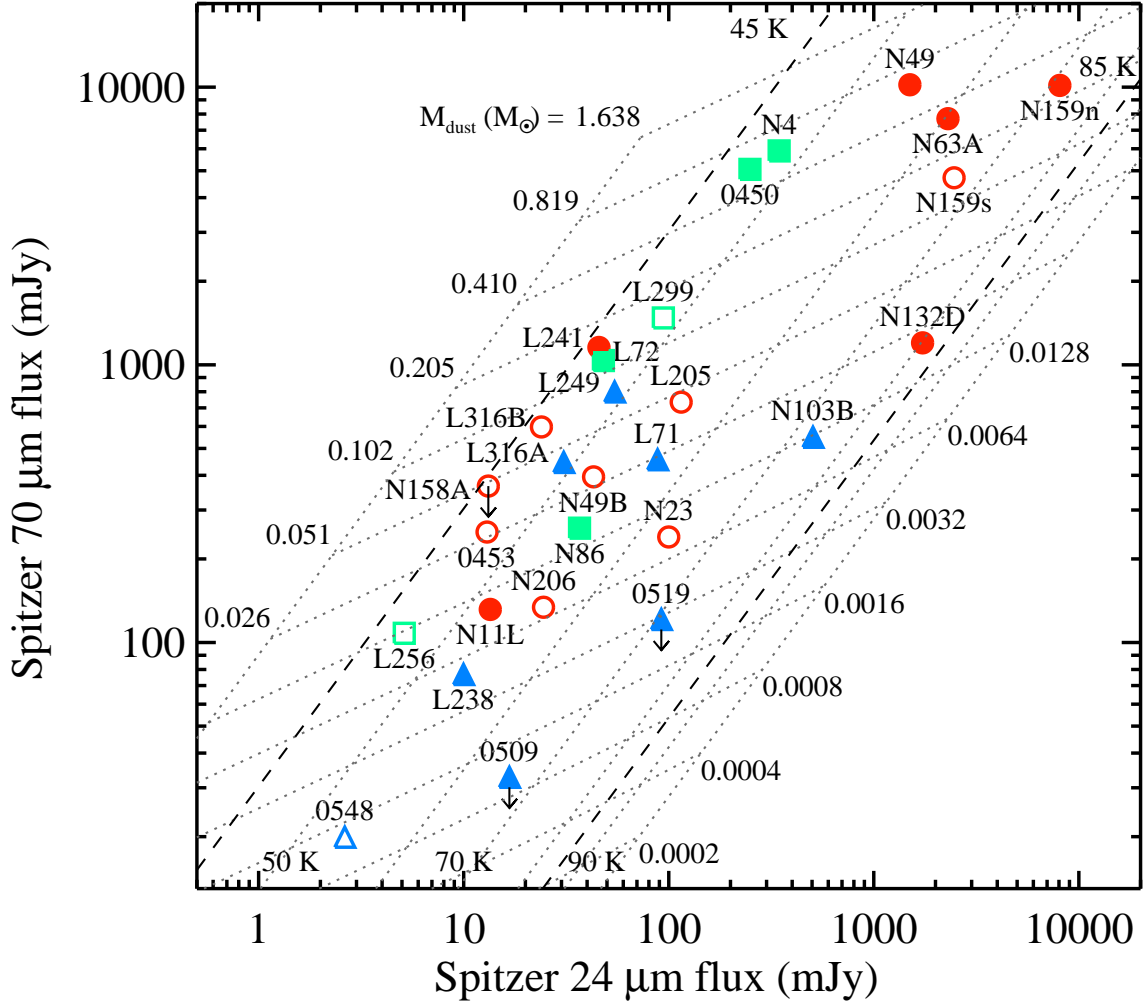}
\caption{70 \micron~flux vs. 24 \micron~flux for all IR SNRs in the LMC. Fluxes are measured from either whole SNRs (filled symbol) or limited areas of SNRs (open symbol). Symbols represent SN types (see Table \ref{tab:lmcsnr}): Type Ia (triangle), CCSNR (circle), and SNRs of unknown type (square). Hereafter, symbol designation is applied in the same way for Figure \ref{fig:ccd1}-\ref{fig:irra}. Dotted lines show a grid of constant dust mass and constant dust temperature for modified black body emission. The grid ranges of dust temperature and dust mass are from 40 to 100 K and from 0.0002 to 1.6 $M_\sun$, respectively. For comparison, the range of dust temperature from Galactic SNRs \citep[45 and 85 K,][]{pin11} is marked (dashed line). \label{fig:mmd}}
\end{figure}

\begin{figure}
\epsscale{1}
\plotone{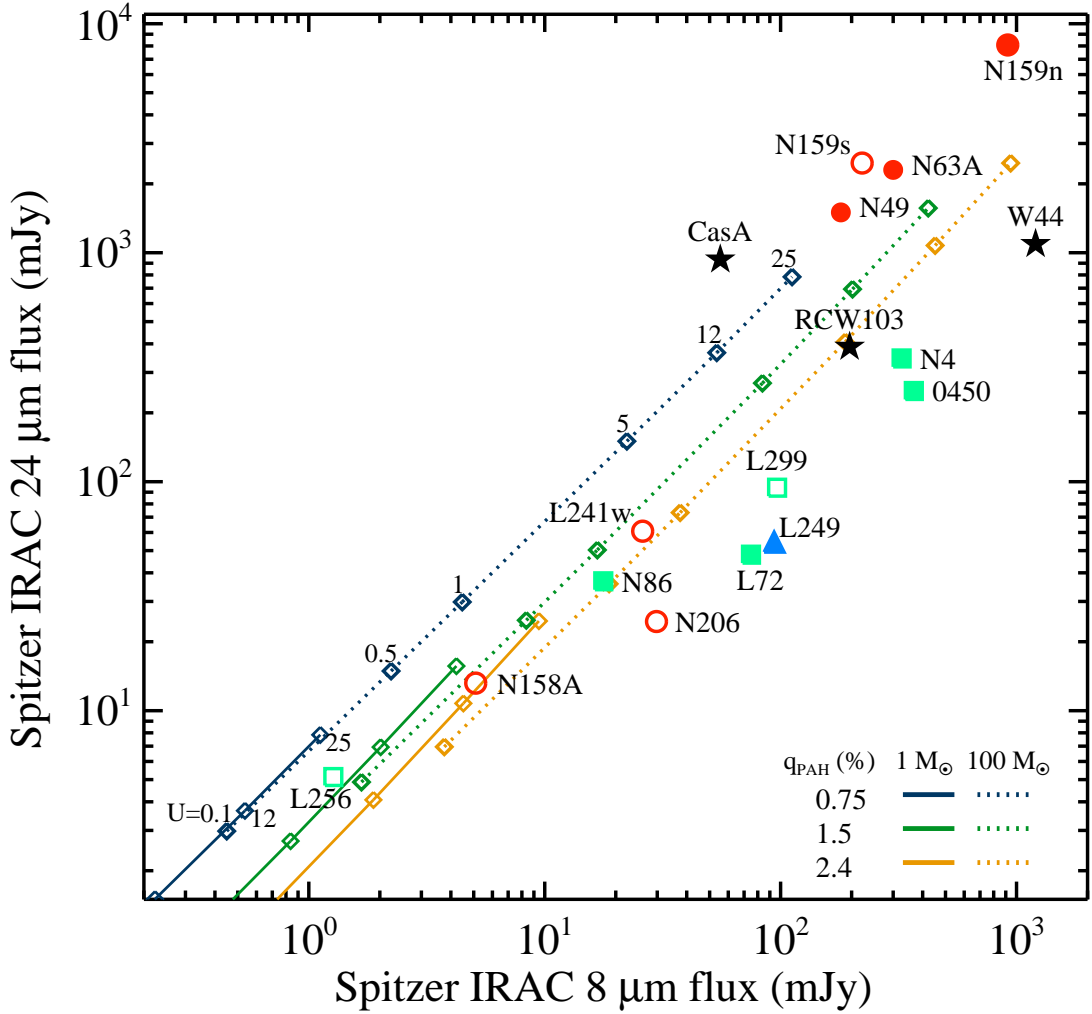}
\caption{Distribution of LMC SNRs in the 8 \micron~and 24 \micron~flux plane. Symbol designation is the same as Figure \ref{fig:mmd}. We exclude SNRs with upper limits of 8 \micron~fluxes in the diagram for simplicity. For comparison, fluxes of three Galactic SNRs are also shown when they are at the distance of the LMC (star). Three theoretical models of \cite{DL07} for the LMC are overlaid varying dust masses, $M_{d}=1~M_{\sun}$ (solid line) and 100 $M_{\sun}$ (dotted line). The three models have different PAH mass fractions, $q_{\rm{PAH}}=0.75\%$ (blue line), 1.49\% (green line), and 2.37\% (yellow line) with a range of starlight intensities (diamonds; $U=0.1, 0.5, 1, 5, 12,$ and 25). 
\label{fig:mmd3}}   

\end{figure}

\begin{figure}
\plotone{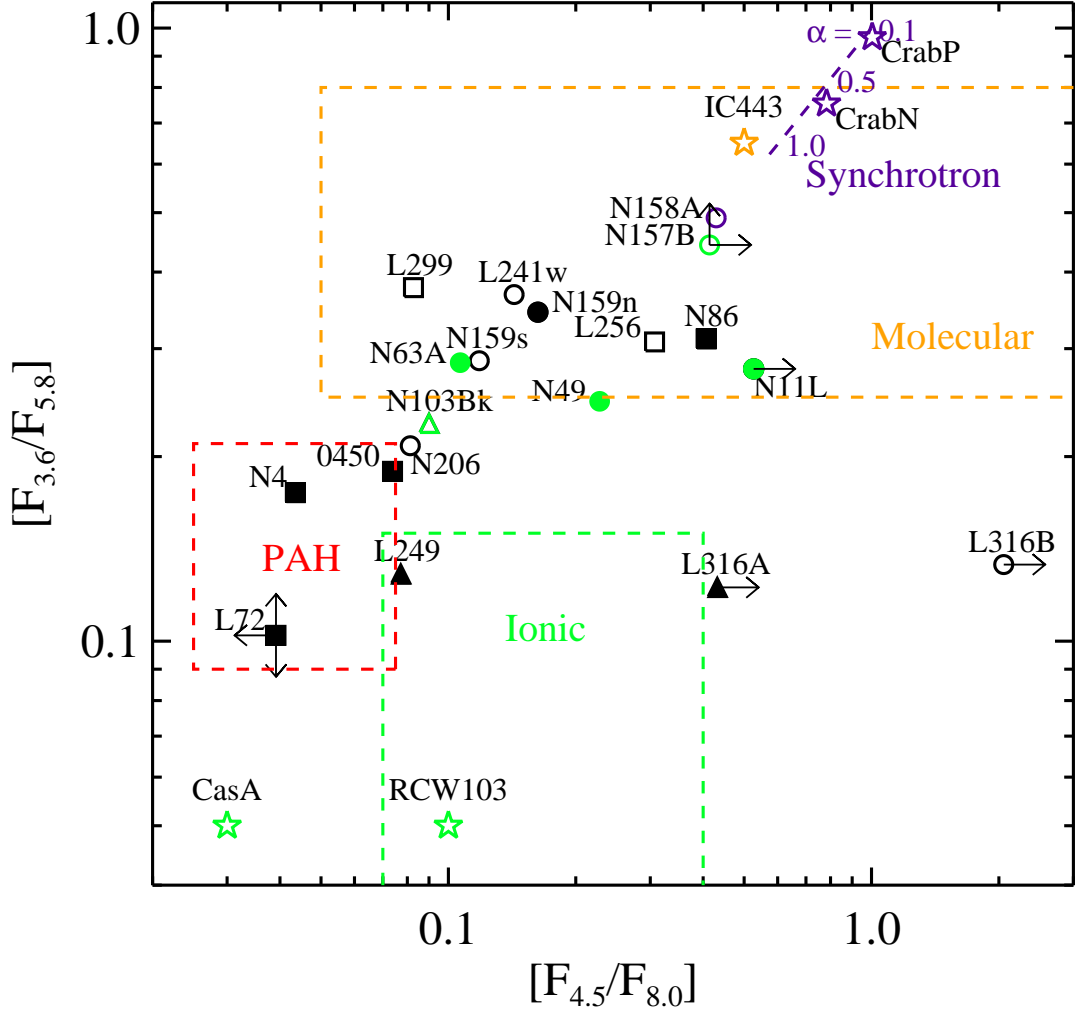}
\caption{IRAC color-color diagram ($[F_{3.6}/F_{5.8}]$ and $[F_{4.5}/F_{8}$]) of the LMC SNRs. Colors of the Galactic SNRs (Cas A, RCW 103, IC 443, and Crab Nebula/Pulsar) are shown for comparison (open star). Symbol designation is the same as Figure \ref{fig:mmd}. Expected IRAC color ranges for ionic (green) and molecular (yellow) shocks, PAH emission (red), and synchrotron emission (purple) are depicted based on \citet{reach06}. For SNRs of which dominant mechanisms of IR emission are previously known, their symbols are colored according to the above color classification.    \label{fig:ccd1}}
\end{figure}
\begin{figure}
\epsscale{1}
\plotone{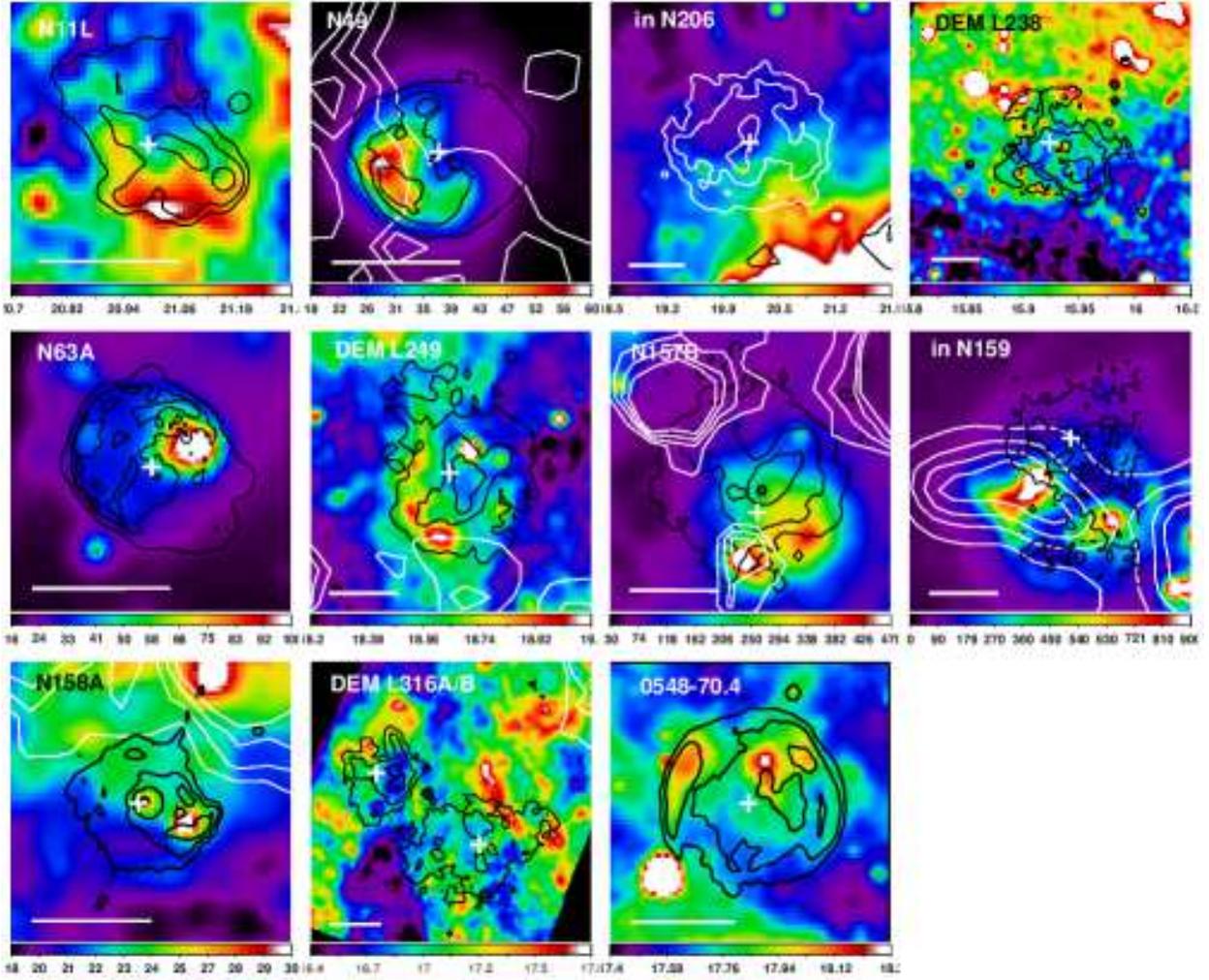}
\caption{SNRs showing spatial discrepancy between IR and X-ray emissions. $Spitzer$ 24 \micron~images are shown with contours from $Chandra$ X-ray images (black contours), and CO contours from MAGMA data are also overlaid (white contours). For SNR in N206, X-ray and CO contours are white and black, respectively. CO contour levels are 0.25 to 1 K km s$^{-1}$ at intervals of 0.25 K km s$^{-1}$, but CO levels in SNR in N159 are different as there is very strong CO emission around it (5 to 25 K km s$^{-1}$ at intervals of 5 K km s$^{-1}$). The cross marks the center position of each SNR listed in Table \ref{tab:lmcsnr}, and the scale bar corresponds 1\arcmin~(i.e., 15 pc at 50 kpc). Pulsars in N157B and N158A are denoted with circle. North is up, and east is to the left.\label{fig:corr_irxr}}
\end{figure}

\begin{figure}
\plotone{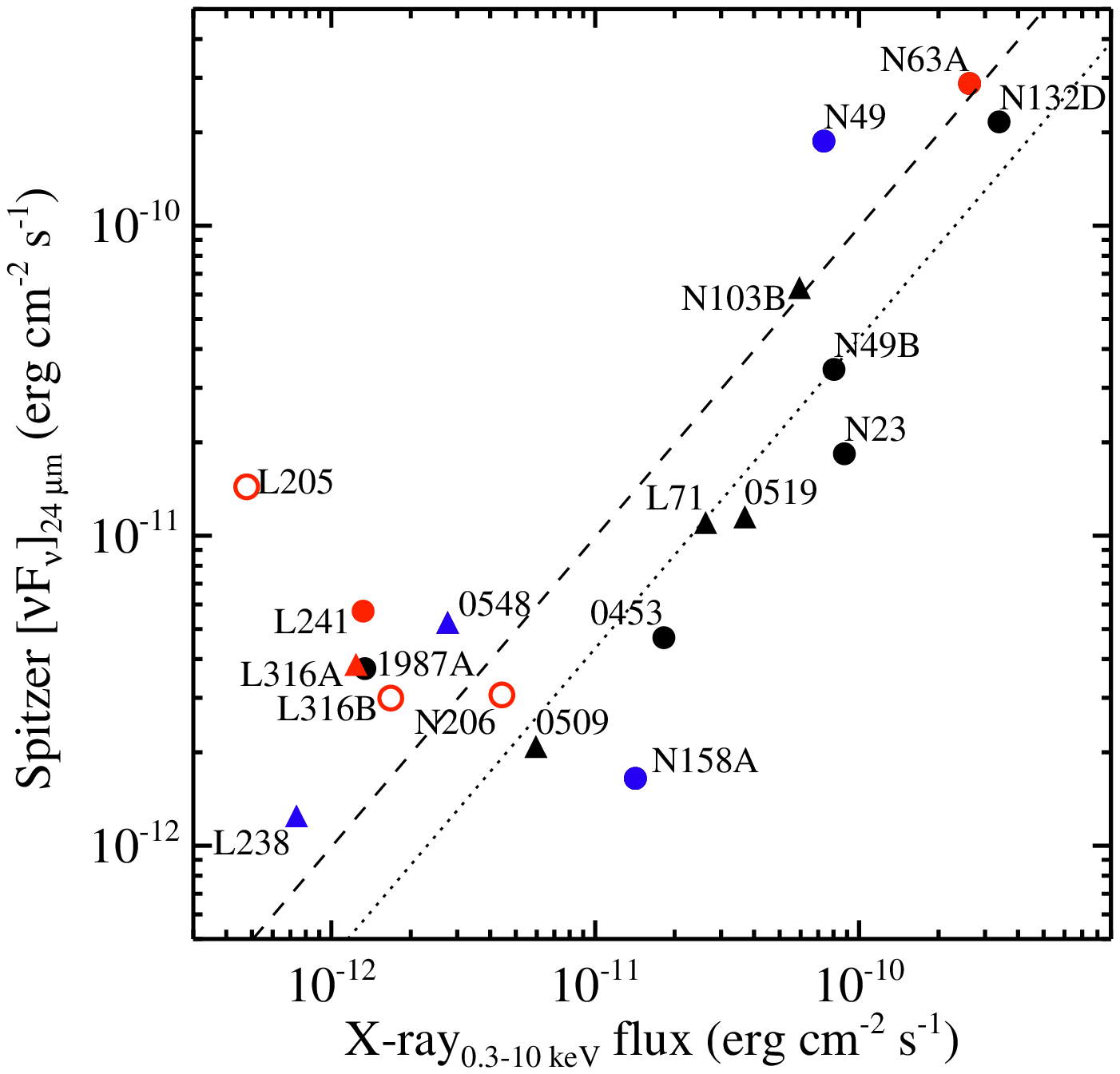}
\caption{Correlation between $Spitzer$ 24 \micron~fluxes and X-ray fluxes. All X-ray fluxes are $Chandra$ (0.3-10 keV) fluxes from the $Chandra$ SNR catalog, but we adopt {\it XMM-Newton} fluxes of DEM L205 (0.2-5 keV) and DEM L241 (0.5-10 keV) \citep{mag12,bam06}. Symbol designation except color is the same as Figure \ref{fig:mmd}, but the symbol of N158A is filled here because both the IR and X-ray fluxes are extracted from its PWN only. The color of symbols indicates whether there is a strong (black), moderate (showing both similarities and dissimilarities, blue), or poor (red) correlation between IR and X-ray morphologies. While all data with total fluxes (solid symbol) can be fitted by a linear fit with a coefficient of $0.99^{+0.40}_{-0.14}$ (dashed line), the fluxes from those with a good spatial correlation is fitted by a linear fit with a coefficient of $0.43^{+0.14}_{-0.11}$ (dotted line).\label{fig:irxr}}
\end{figure}

\begin{figure}
\epsscale{1}
\plotone{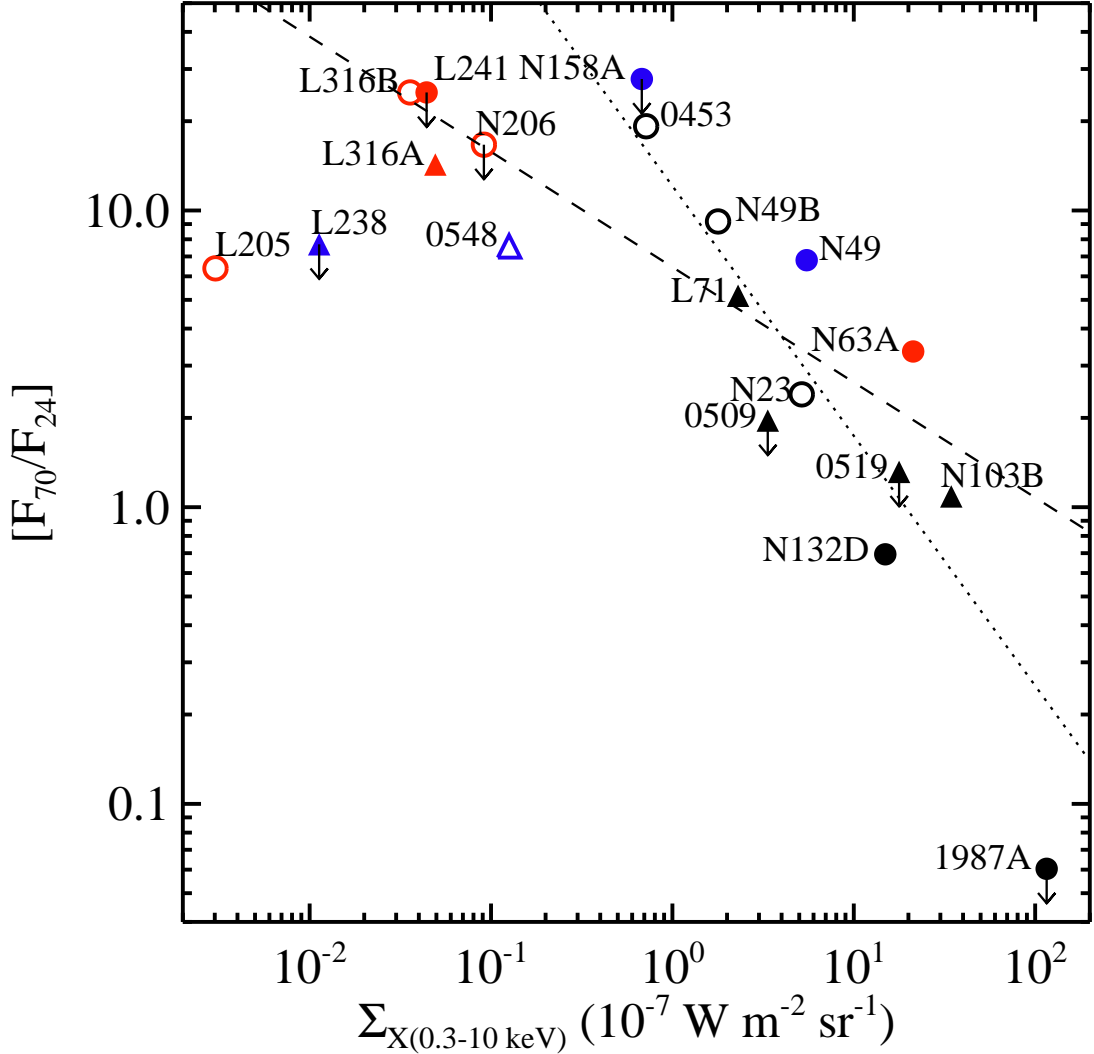}
\caption{$Spitzer$ $[F_{70}/F_{24}]$ ratios vs. X-ray surface brightness. Symbol designation is the same as Figure \ref{fig:irxr}, but some filled symbols in Figure \ref{fig:irxr} become open because their flux ratios are measured from the limited areas of the SNRs. The correlation between the ratios and the surface brightnesses can be fitted by a function of $[F_{70}/F_{24}]=7.2^{+1.7}_{-1.4}\times \Sigma_X^{-0.43\pm0.09}$ for all data with total fluxes (dashed line) and by a function of $[F_{70}/F_{24}]=12.1^{+3.9}_{-3.0}\times \Sigma_X^{-0.84\pm0.14}$ for those with a good spatial correlation (dotted line).
\label{fig:m2m1_xr}}
\end{figure}

\begin{figure}
\plottwo{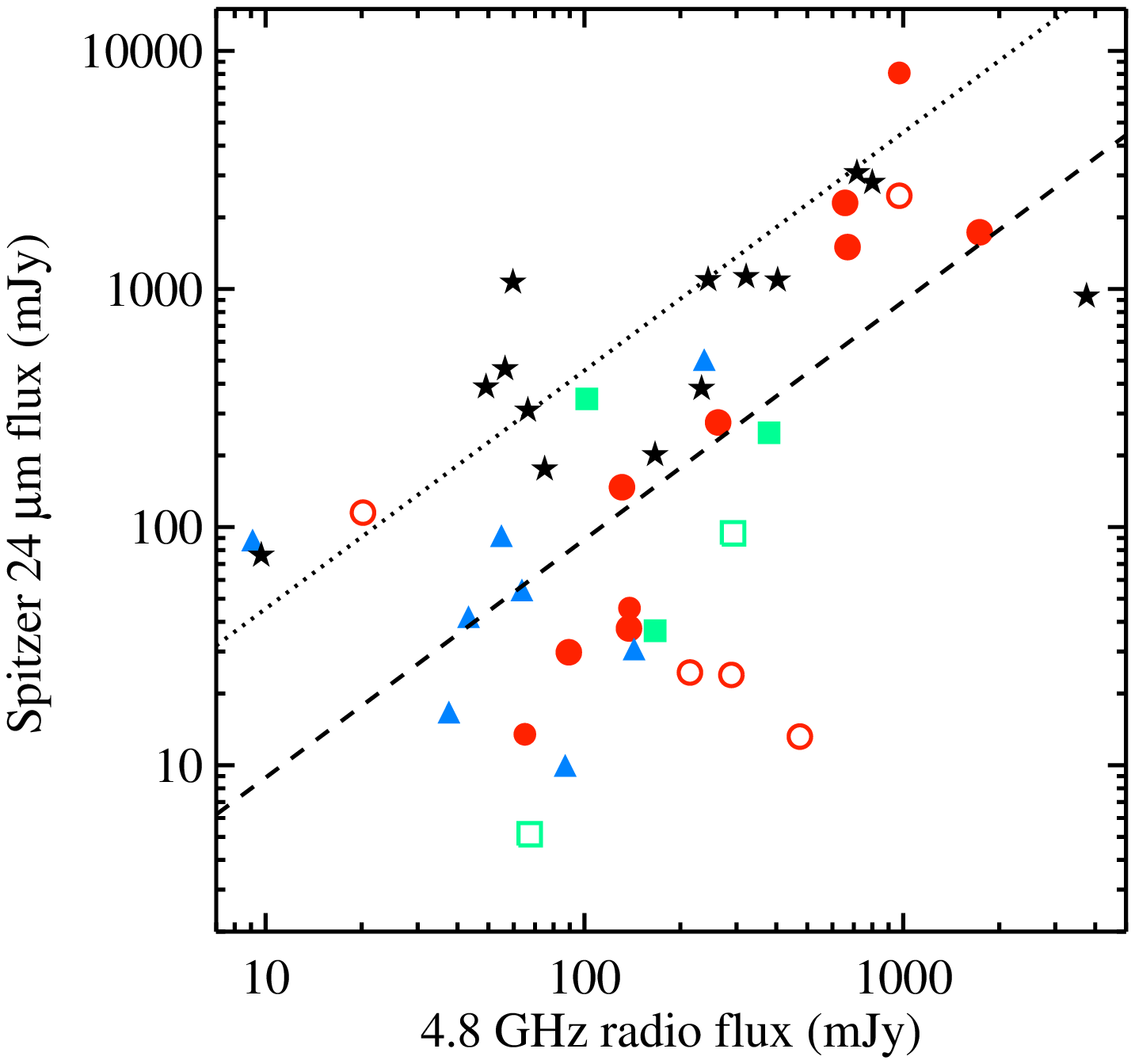}{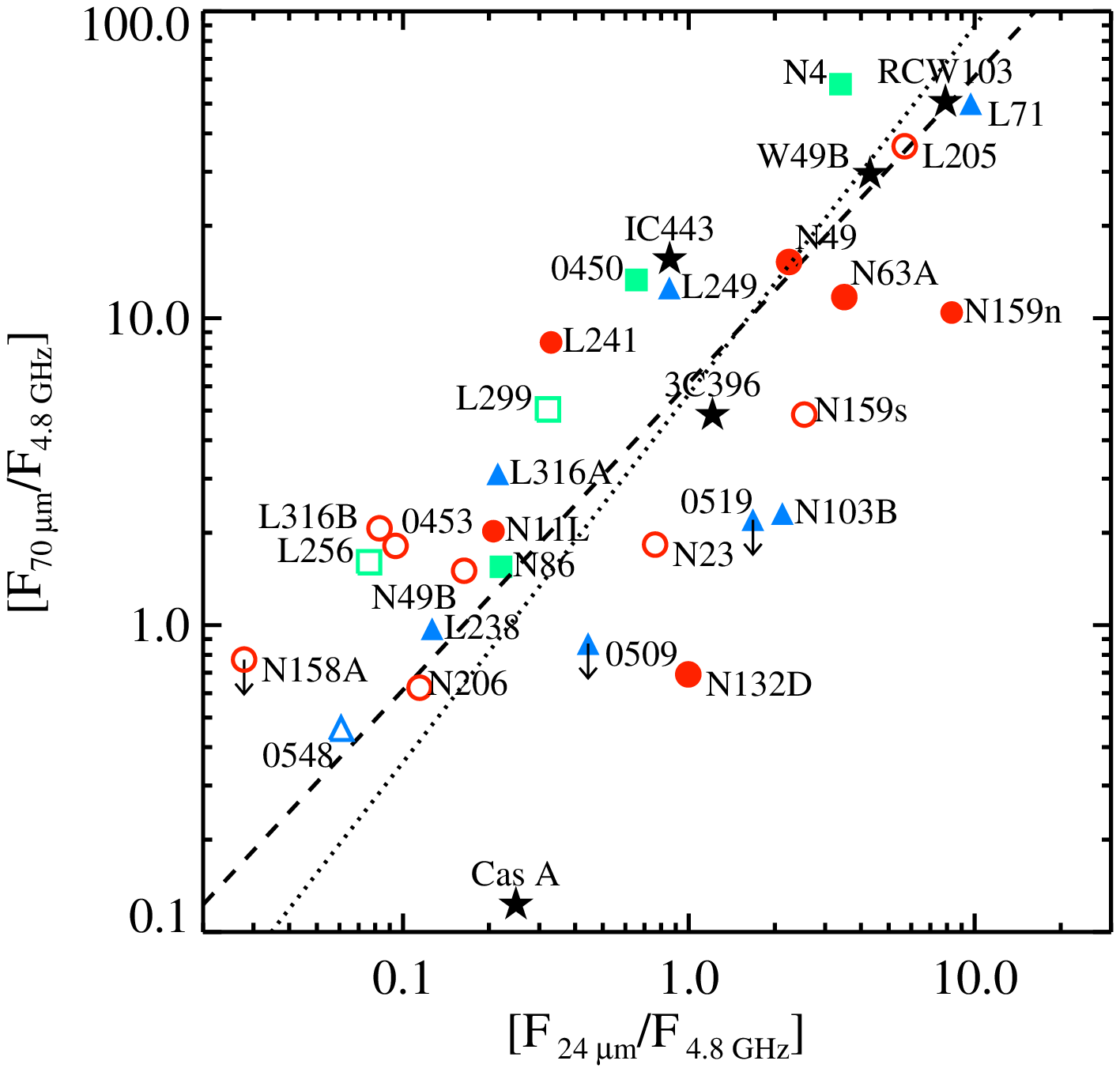}
\caption{Left: correlation between $Spitzer$ 24 \micron~fluxes and ATCA 4.8 GHz radio fluxes. The fluxes from whole SNRs (solid symbol) are described by a linear fit of $F_{24 \micron}=0.89^{+0.27}_{-0.21}\times F_{4.8\rm{GHz}}$ in a logarithmic scale (dashed line). For comparison, fluxes of the Galactic SNRs (star) are overlaid which are scaled to the distance of the LMC (50 kpc). A correlation of Galactic SNRs, $F_{24 \micron}=4.55^{+1.31}_{-1.02}\times F_{4.8\rm{GHz}}$, is derived from their average $q_{24}$ of $0.39\pm0.11$ for upper trend remnants \citep{pin11} assuming their spectral indices as 0.5 (dotted line). Right: correlation between 24 \micron~and 70 \micron~fluxes normalized to 4.8 GHz radio flux of each SNR. For SNRs having total IR fluxes, we perform a linear fitting of $[F_{70\micron}/F_{4.8\rm{GHz}}]=6.1^{+2.2}_{-1.6}\times [F_{24\micron}/F_{4.8\rm{GHz}}]$ (dashed line). As a comparison, ratios of five Galactic SNRs (RCW 103, W49B, IC443, 3C396, and Cas A) are shown (star). The correlation of Galactic SNRs in the MIPSGAL survey (dotted line) is also plotted assuming that the spectral index is 0.5 \citep[slope $=1.2\pm0.1$;][]{pin11}. 
\label{fig:irra}}
\end{figure}

\begin{figure}
\plotone{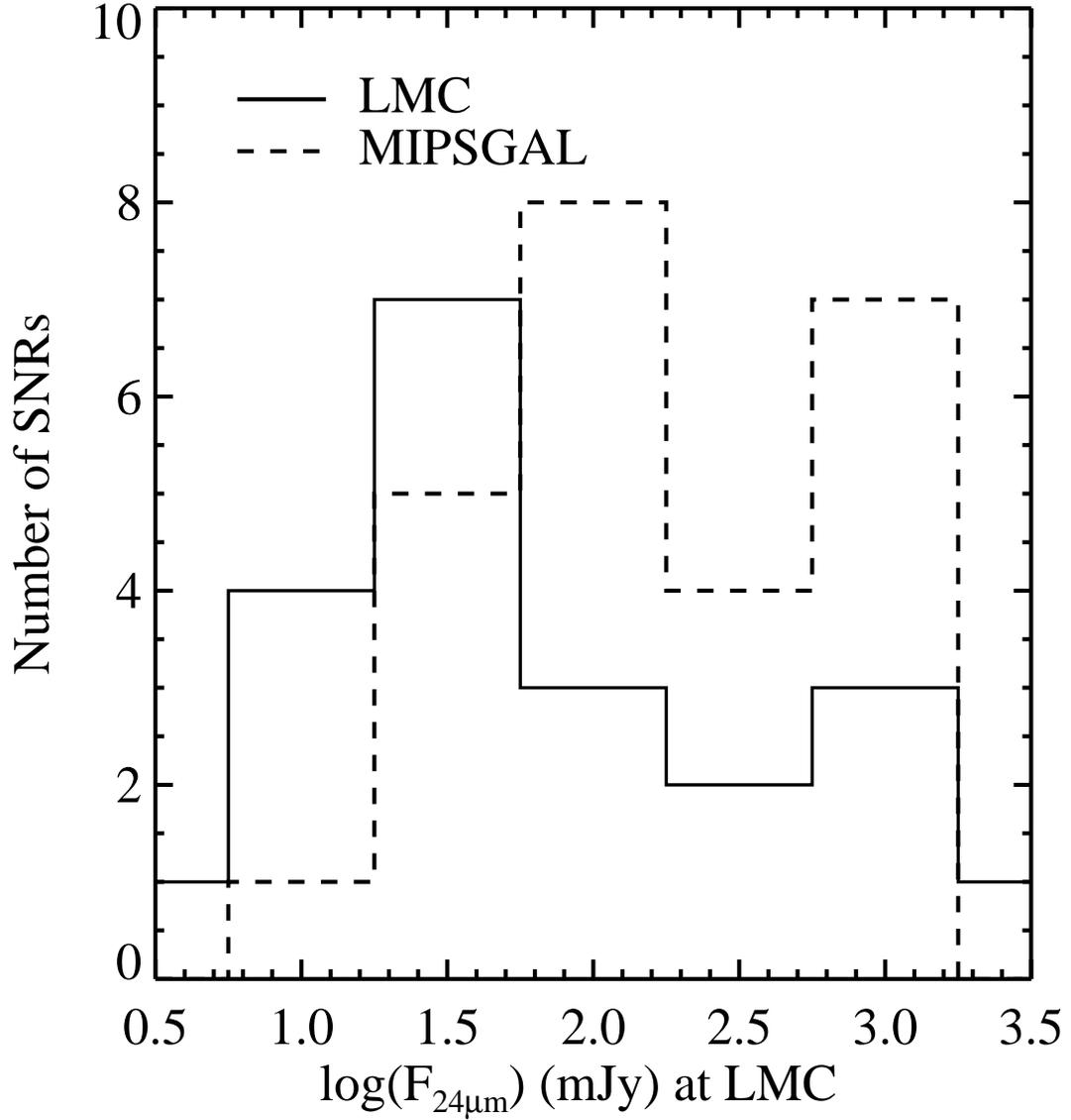}
\caption{Histogram of MIPS 24 \micron~fluxes estimated from LMC and Galactic SNRs. Only fluxes measured from whole SNRs are used (i.e., 21 LMC SNRs and 25 Galactic SNRs). For the Galactic SNRs, their fluxes are scaled to the distance of the LMC (50 kpc) either using their known distances or assuming their distances as 3 kpc.    
\label{fig:irhis}}
\end{figure}

\begin{figure}
\epsscale{.8}
\plotone{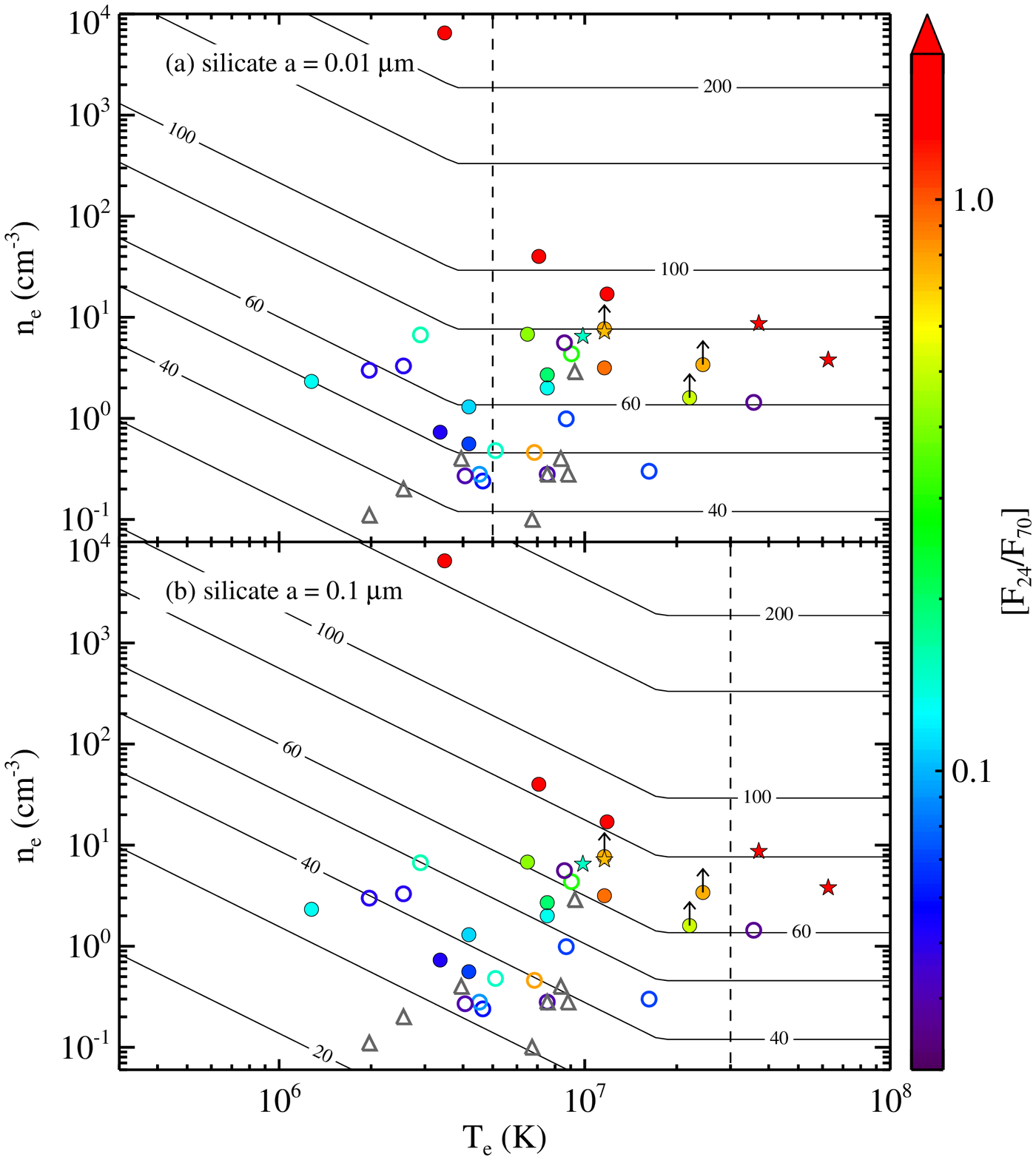}
\caption{Equilibrium temperature as a function of electron density ($n_e$) and temperature ($T_e$) for silicate grains with a size of (a) 0.01 \micron~and (b) 0.1 \micron. This figure mimics Figure 1 of \cite{dwek08} in a simple way. A critical temperature ($T_c$) where dust temperature only depends on electron density is marked at $5\times10^6$ K and $3\times10^7$ K for $a=0.01~\micron$ and $0.1~\micron$, respectively (dashed line). The electron densities and temperatures of LMC SNRs are plotted based on X-ray observations from literatures, which are listed in Table \ref{tab:nete}. Filled and open circles designate SNRs with good and bad morphological correlations between IR and X-ray emission, respectively. The color of a symbol represents a $[F_{24}/F_{70}]$ ratio as notified in a color bar. As a comparison, four Galactic SNRs (Cas A, Kepler, Tycho, and W49B) are shown (star). LMC SNRs without detected MIR emission are also superposed (triangle). 
\label{fig:nete}}
\end{figure}

\begin{figure}
\epsscale{1}
\plotone{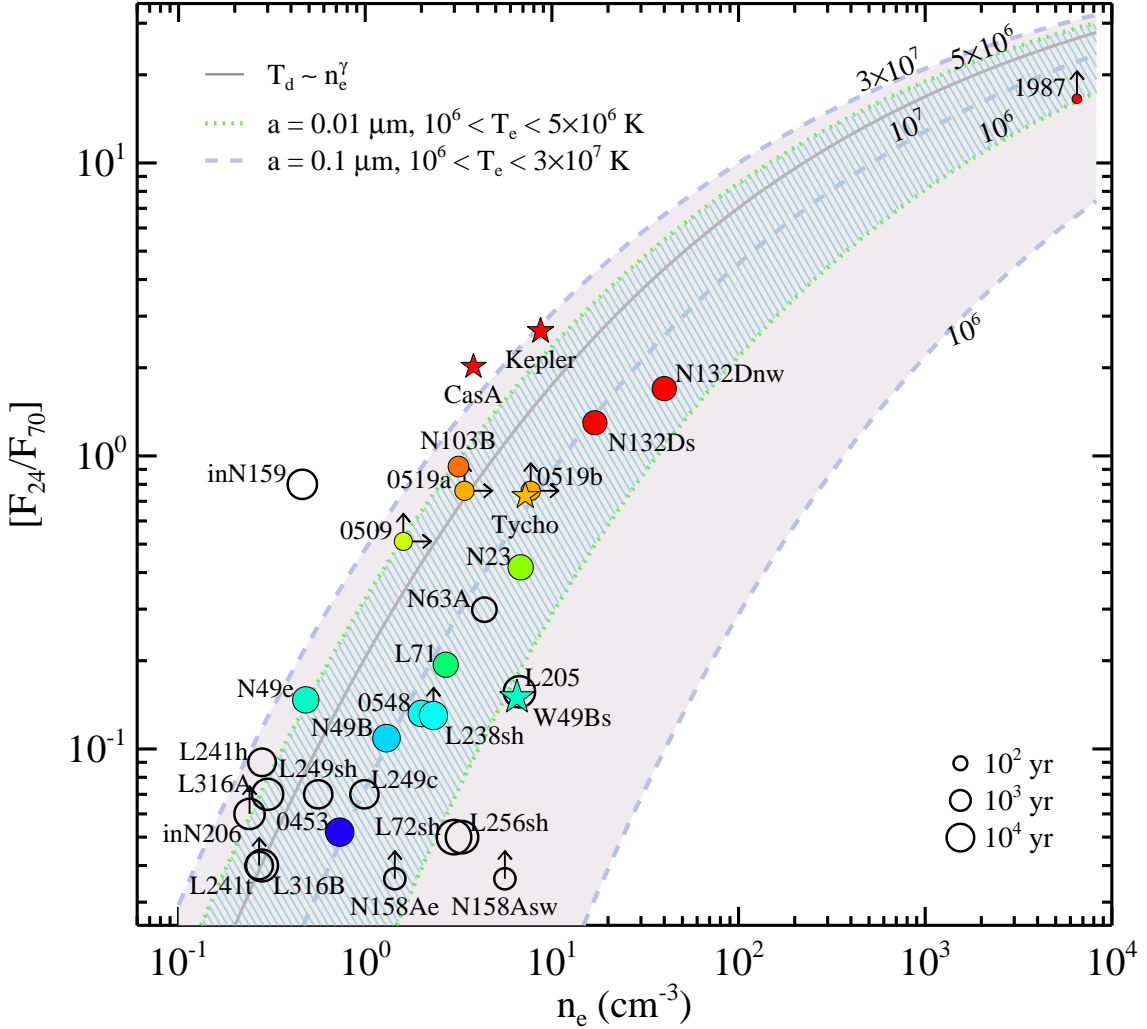}
\caption{Correlation between $Spitzer$ $[F_{24}/F_{70}]$ ratios and electron densities of LMC SNRs. Four Galactic SNRs are also plotted (star). Designation and colors of symbols are the same as Figure \ref{fig:nete}, but note that only filled symbols (i.e., good spatial correlation between IR and X-ray emission) are colored for emphasis. Size of a symbol represents the estimated age of an SNR, and note that DEM L241 is marked with a symbol for 10$^4$ yr assuming it is middle-aged (Table \ref{tab:lmcsnr}). Theoretical ratios are overlaid for three cases: (I) when electrons go through grains (i.e., $T_e\ge T_c$), $T_{\rm dust}\sim n_e^{\gamma}$ (solid line), (II) grain size $a=0.01~\micron$ at $10^6\lesssim T_e\lesssim5\times10^6$ K (line-filled region with a dotted line boundary), and (III) grain size $a=0.1~\micron$ at $10^6\lesssim T_e\lesssim3\times10^7$ K (solid-filled region with a dashed line boundary). $T_e$ corresponding each line for Case II and III are shown on the top-right side. See text for details.  
\label{fig:irne}}
\end{figure}
\clearpage

\begin{figure}
\plotone{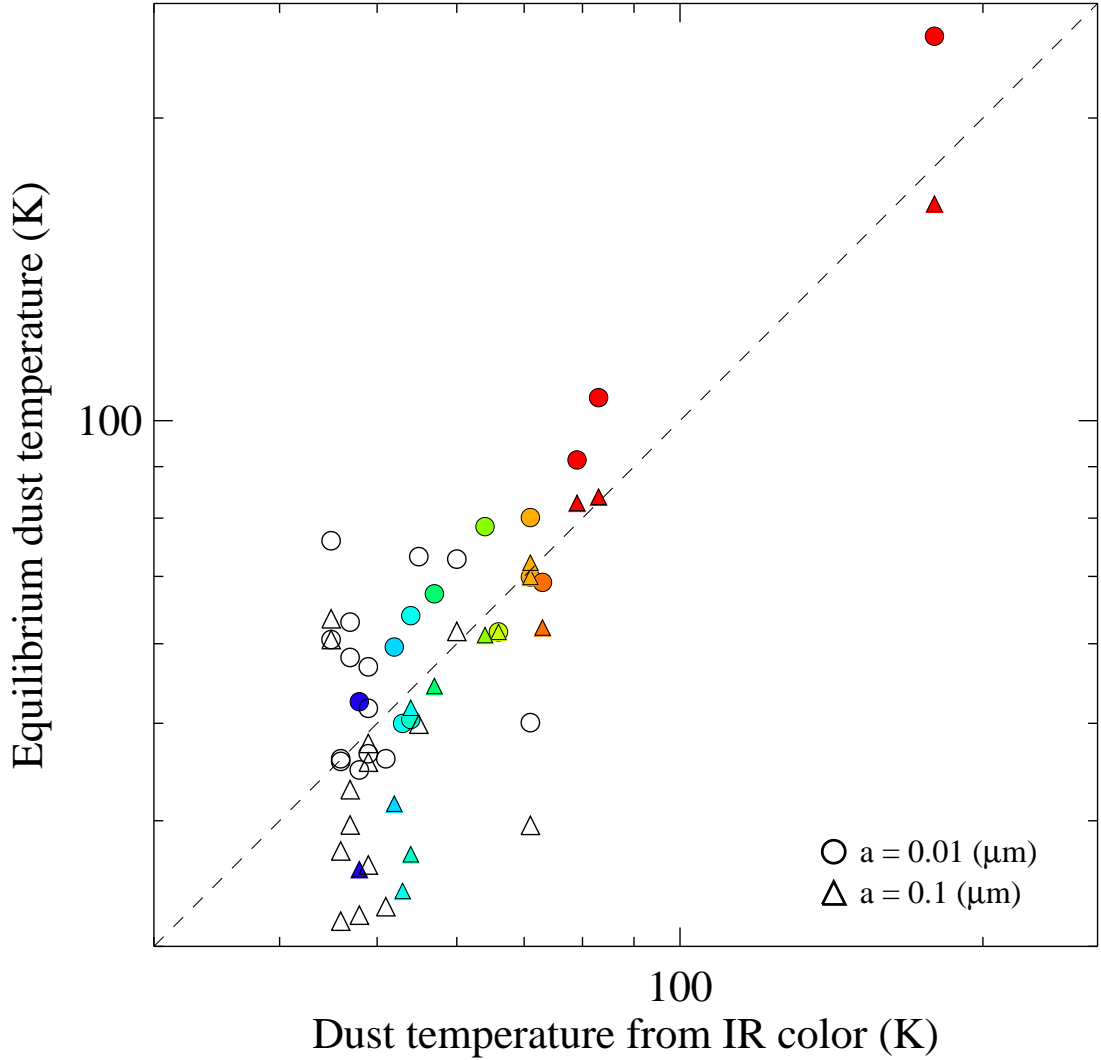}
\caption{Comparison between equilibrium dust temperature derived from plasma properties and dust temperature estimated from IR colors ($[F_{24}/F_{70}]$). For the equilibrium dust temperature, two cases with a grain size of 0.01 \micron~(circle) and 0.1 \micron~(triangle) are derived from Equation (\ref{eq:tequil}) (same as Figure \ref{fig:nete}). The dust temperatures from $[F_{24}/F_{70}]$ are estimated adopting $S_\nu\propto B_\nu (T)\nu^{\beta}$ where $\beta =2$. As Figure \ref{fig:irne}, colors of symbols represent the observed IR colors, and only those with a good spatial correlation between IR and X-ray are colored. For a reference, we overlaid a linear function of $y=x$ in the diagram (dashed line).
\label{fig:tdust}
}
\end{figure}
\clearpage

%% file: table1.tex
\clearpage
\begin{deluxetable}{llcccrcccccc}
\tablecolumns{12}
\tabletypesize{\scriptsize}
\rotate
\tablewidth{0pt}
\tablecaption{SNRs in the Large Magellanic Cloud \label{tab:lmcsnr}}
\tablehead{ \colhead{SNR} & \colhead{Other} &\colhead{R.A.} & \colhead{Dec.}& \colhead{Size} & \colhead{4.8 GHz} & \colhead{SN} & \colhead{Age} & \colhead{$AKARI$} & \colhead{$Spitzer$} & \colhead{Data} & \colhead{} \\ 
\colhead{B1950} & \colhead{Name} & \colhead{(J2000)} & \colhead{(J2000)} & \colhead{(arcmin)} & \colhead{(mJy)} & \colhead{Type} &  \colhead{(kyr)} &\colhead{Detection} & \colhead{Detection} & \colhead{(I1-I4/M1/M2)} & \colhead{Reference} \\
\colhead{(1)} & \colhead{(2)} & \colhead{(3)} & \colhead{(4)} & \colhead{(5)} & \colhead{(6)} & \colhead{(7)} & \colhead{(8)} & \colhead{(9)} & \colhead{(10)} & \colhead{(11)} & \colhead{(12)}}
\startdata
0448--67.1 & J0448--6659 & 04:48:25 & -67:00:12 & 4.5$\times$3.4 &$\dots$&$\dots$& $\dots$&N &U& S/S/S &\dots\\	
0449--69.4 & J0449--6921 & 04:49:22 & -69:20:25 & 2.0 &56&$\dots$&$\dots$&N&U& S/S/S&\dots\\	
0450--70.9 & 0450--709 & 04:50:30 & -70:50:05 & 7.7$\times$5.3 & 380 & $\dots$ &$\ge45$&N&I1-M2 & S/S/S& 1\\	
0453--66.9 & SNR in N4 & 04:53:14 & -66:55:42 & 4.3	&102 &$\dots$&$\dots$ & N & I1-M2 & S/B/S & \dots\\
0453--68.5 & 0453--68.5 & 04:53:38 & -68:29:27 & 2.0 &138& II(C)&$\sim13$& N & M1-M2 & S/B/S& 2\\
0454--67.2 & SNR in N9 & 04:54:33 & -67:13:00 & 2.8$\times$2.2 & 39 & Ia & $\sim30$ & N & U & S/B/S& 3\\	
0454--66.5 & N11L & 04:54:50 & -66:25:37 & 1.4$\times$1.0 & 65 & II &$7-15$ & N & I1-I3, M1-M2& CS/S/S & 4\\ 
0455--68.7 & N86  & 04:55:44 & -68:38:23 & 6.5$\times$3.5 & 167 &$\dots$&$20-86$ &N& I1-M2 & S/BS/S& 4\\	
0500--70.2 & N186D	 & 04:59:56 & -70:07:58 & 2.6$\times$2.3 & 62 & II &$\sim11$ &N&U& S/B/S & 5\\	
0505--67.9 & DEM L71  & 05:05:42 & -67:52:39 & 1.5$\times$1.2 & 9 & Ia &$\sim4.36$ &N&M1-M2& B/S/S & 2\\
0506--68.0 & N23 & 05:05:55 & -68:01:47 & 1.2$\times$0.8 & 131 & II & $\sim4.6$ &N& M1-M2& S/B/B &2\\
0506--65.8 & DEM L72 & 05:06:06 & -65:41:08 & 6.4$\times$4.7 &$\dots$ &$\dots$&$\sim100$&N&I4-M2 & S/S/S&6\\	
0509--68.7 & N103B & 05:08:59 & -68:43:35 & 0.50 &238 & Ia & $\sim1$ & N& I1-M2& B/S/S & 7\\ 
0509--67.5 & 0509--675 & 05:09:31 & -67:31:17 & 0.56 & 38 & Ia & $\sim0.41$& S11-L24 & M1-M2 & B/B/B &7 \\ 
0513--69.2 & 0513--692 & 05:13:14  & -69:12:20  & 4.5$\times$3.2 &178 & $\dots$&$\dots$ & U &U& S/S/S&\dots\\ 
0519--69.7 & SNR in N120 & 05:18:44 & -69:39:09 &1.6$\times$1.3 & 181 & II &$\dots$& N & U & S/S/S& 8\\ 
0519--69.0 & 0519--690  & 05:19:35  & -69:02:09 & 0.55& 55 &Ia & $\sim0.61$ & S11-L24 & M1-M2 & B/B/B & 7\\ 
0520--69.4 & 0520--694  & 05:19:44  & -69:26:08 &2.4$\times$2.1 & 99 &$\dots$ &$\dots$& U & U& S/B/S &\dots\\ 
0522--65.8 & J0521--6542 & 05:21:39 & -65:43:10 & 3.0$\times$2.4 & $\dots$ &$\dots$&$\dots$&N&U& S/S/S&\dots\\
0523--67.9 & SNR in N44 & 05:23:07 & -67:53:12 &3.5&126 & II & $\sim18$ & U & U & C/C/C & 5, 9\\ 
0524--66.4 & DEM L175A & 05:24:20 & -66:24:23 & 4.1$\times$2.8 & 98 & II  &$\dots$ & U & U& S/B/S & 8\\ 
0525--69.6 & N132D & 05:25:04 & -69:38:20 &2.0$\times$1.5& 1737 & II(O) & $\sim3.15$ & S11-L24 & M1-M2& Rh/B/S & 2\\ 
0525--66.0 & N49B  & 05:25:25 & -65:59:19 & 2.5$\times$2.3 & 263 & II & $\sim10$ & S11-L24 & M1-M2& B/B/S & 2\\ 
0525--66.1 & N49  & 05:26:00 & -66:04:57 & 1.5$\times$1.3& 669 & II  & $\sim6.6$ & N3-L24 & I1-M2& G/G/G & 9 \\ 
0528--69.2 & 0528--692 & 05:27:39 & -69:12:04 & 2.7$\times$2.0 & 87 & II & $\dots$& U & U & S/B/S& 8\\ 
0527--65.8 & DEM L204 & 05:27:54 & -65:49:38 & 4.5 & 88 & $\dots$ & $\dots$ & U & U & S/B/S&\dots \\ 
J0528--6727\tablenotemark{a} & DEM L205 & 05:28:05 & -67:27:20 & 5.4$\times$4.4 & $\ga22$ & II & $\sim35$ & U & I3-M2& CS/C/C & 10 \\
0531--70.2\tablenotemark{b} & J0530--7007 & 05:30:40 & -70:07:27 & 3.6$\times$3.0 & 96 & Ia &$\dots$& N & U& S/S/S & 11 \\
0532--71.0 & SNR in N206 & 05:31:56 & -71:00:19 & 3.0 & 214 & II(C) & $\sim25$ & N & I1-M1& GoS/RB/RS &2, 12 \\
0532--67.5 & 0532--675 & 05:32:30 & -67:31:33 & 4.5 & 207 & $\dots$ & $\dots$ & N & U& I/BI/I &\dots\\
0534--69.9 & 0534--699 & 05:34:02 & -69:55:03 & 1.7$\times$1.4 & 74 & Ia & $\sim10$ & U & U & B/B/S&2\\ 
0534--70.5 & DEM L238 & 05:34:18  & -70:33:26 & 2.9$\times$2.5 & 79 & Ia & $10-15$ & U & M1& B/B/S &13\\
0535--69.3 & SNR 1987A & 05:35:28 & -69:16:11 & $<0.1$& 89 & IIpec & 0.025 & N3-L24 & I1-M2 & D/DG/DG &\dots\\ 
0535--66.0 & N63A & 05:35:44 & -66:02:14 & 1.4$\times$1.2 & 657 & II & $2-5$ & N & I1-M2& C/C/C & 9\\ 
0536--69.3 & Honeycomb & 05:35:48 & -69:18:04 &1.4$\times$0.6 & 108 & $\dots$ & $\dots$ & U & U & B/B/S &\dots\\
0536--67.6 & DEM L241  & 05:36:03 & -67:35:04 & 2.4 & 139 & II(C) &$\dots$ & N & I1-M2 & I/B/B & 14\\
0536--70.6 & DEM L249  & 05:36:07 & -70:38:37 & 3.0$\times$2.0 & 64 & Ia & $10-15$ & N & I1-M2& B/BS/BS & 13\\ 
0538--66.5 & DEM L256  & 05:37:30 & -66:27:47 & 3.6$\times$2.8 & 67 & $\dots$ & $\le50$ & N & I1-M2 & S/S/S& 6\\
0538--69.1 & N157B	 & 05:37:48 & -69:10:35 & 1.7$\times$1.2 & 1923 & II(C) & $\sim5$ & N3, S11-L24 & I1-M1 & Br/B/S & 9\\ 
0540--69.7 & SNR in N159 & 05:39:59 & -69:44:02  & 1.8 & 972 & II &$\ge18$ & U & I1-M2 & S/B/S& 15\\ 
0540--69.3 & N158A	& 05:40:12 & -69:19:55 & 1.3$\times$1.1 & 474 & IIP(C) & $0.76-1.66$ & N3-L24 & I1-M2& B/B/S &16, 17 \\ 
J0541.8--6659\tablenotemark{a} & [HP99] 456 & 05:41:52 & Ð66:59:03 & 5.0$\times$4.6 & 44 & $\dots$ & $\sim23$ & N & U& S/S/S & 18 \\
0543--68.9 & DEM L299 & 05:43:10 & -68:58:49 & 5.8$\times$4.0 & 293 & $\dots$ & $\dots$ & N & I1-M2 & MS/S/S &\dots\\
0547--69.7 & DEM L316B & 05:46:59 & -69:42:50 & 3.4$\times$2.8 & 289 & II & $\ge42$ & U & I1-I3, M1-M2 & B/B/S &19, 20\\ 
0547--69.7 & DEM L316A & 05:47:22 & -69:41:26  & 2.0 & 143 & Ia & $27-39$ & U & I1-I3, M1-M2 & B/B/S& 19, 20\\ 
0548--70.4 & 0548--704 & 05:47:49 & -70:24:54  & 2.0$\times$1.8 & 43 & Ia & $\sim7.1$ & S11-L24 & M1-M2 & B/B/BS & 2\\ 
0551--68.4 & J0550--6823 & 05:50:30 & -68:23:22 & 5.2$\times$3.5 & 316 & II(O) & $\dots$ & N & U& S/S/S& 21 \\	
\enddata

\tablecomments{Columns 1--5: SNR names, alternative names, positions, and angular sizes from \citet{desai10}. Angular sizes are mostly measured in optical, but some are in X-ray/IR.  Column 6: 4.8 GHz radio fluxes of SNRs measured in this work using the ATCA data \citep{dic05}. See text for an explanation on the flux measurement. See also note ``a'' below. Column 7: SNR type from literatures. Except Crab-like SNRs marked as ``II(C)'', Type II SNRs are shell type SNRs of core-collapse SN origin. Oxygen-rich SNRs are marked as ``II(O)''. Column 8: SNR age from literatures. Columns 9--10: detection status by $AKARI$ and $Spitzer$. If IR emission is detected from the SNR, the detected filter name is given ($AKARI$: N3, S7, S11, L15, and L24, $Spitzer$: I1, I2, I3, I4, M1, and M2). Otherwise, U: undetected and N: no data. Column 11: $Spitzer$ data set we used for IRAC four bands, MIPS 24 \micron, and MIPS 70 \micron, respectively (I1-I4/M1/M2). In some SNRs, two data sets are combined to achieve higher signal to noise ratios. Program ID and PI of each data set are given as follow; $B$: ID 3680, PI: K. Borkowski, $Br$: ID 1032, PI: B. Brandl, $C$: ID 3565, PI: Y.-H. Chu, $D$: ID 30067 PI: E. Dwek, $G$: ID 124, PI: R. Gehrz, $Go$: ID 1061, PI: V. Gorjian, $I$: ID 249, PI: R. Indebetouw, $M$: ID 3578 PI: K. Misselt, $R$: ID 717 PI: G. Rieke, $Rh$: ID 3483 PI: J. Rho, $S$: SAGE survey, PI: M. Meixner. Column 12: references for SNR ages and types; (1) \citet{will04}; (2) \citet{lope10}; (3) \citet{sew06}; (4) \citet{will99}; (5) \citet{jas11}; (6) \citet{kli10}; (7) Ages estimated from light echoes by \citet{rest05}; (8) \citet{chu88}; (9) \citet{will06}; (10) \citet{mag12}; (11) \citet{horta12}; (12) \citet{will05a}; (13) \citet{bork06}; (14) \citet{bam06}; (15) \citet{sew10}; (16) \citet{park10}; (17) \citet{bwill08}; (18) \citet{gro12}; (19) \citet{will05}; (20) \citet{nishi01}; (21) \citet{bozz11}. 
}
\tablenotetext{a}{These SNRs are newly identified by recent observations, which are not included in \citet{desai10}. Their radio fluxes are measured in this work, but the flux of DEM L205 is only extracted from where nearby H\textsc{ii} regions do not interrupt. 
The area for the radio flux measurement include the most part of the eastern shell for the IR flux measurement. As the radio flux does not represent the total flux, it gives a lower limit. }
\tablenotetext{b}{The information is revised based on a multi-frequency study by \citet{horta12}.}
 \end{deluxetable}
 \clearpage

%% file: table2.tex
\clearpage
\begin{deluxetable}{lccccccccc}
\tablecolumns{10}
\linespread{1.2}
\tabletypesize{\scriptsize}
\rotate
\tablewidth{0pt}
\tablecaption{$Spitzer$ Fluxes of LMC SNRs \label{tab:sstflux}}
\tablehead{ \colhead{}&\colhead{I1} & \colhead{I2}& \colhead{I3} & \colhead{I4} & \colhead{M1} & \colhead{M2} & \colhead{} & \colhead{Area}& \colhead{}\\
\colhead{SNR}&\colhead{(mJy)} & \colhead{(mJy)}& \colhead{(mJy)} & \colhead{(mJy)} & \colhead{(mJy)} & \colhead{(mJy)} & \colhead{Region} & \colhead{(arcmin$^2$)} & \colhead{Reference}  }
\startdata
0450--70.9 & 28.2$\pm$0.31 & 27.1$\pm$0.35 &149$\pm$1.4 &367$\pm$1.4 &249$\pm$4.6 & 5057$\pm$63 & Whole &76.56 & this work \\
SNR in N4 NE & 18$\pm$0.1 & 12$\pm$0.1 & 102$\pm$0.4 & 276$\pm$0.5 & 307$\pm$1.6 &5316$\pm$42 & Northeastern region &  6.25 & this work \\
SNR in N4 S& 3.6$\pm$0.07 & 2.3$\pm$0.09 & 22$\pm$0.3 & 50$\pm$0.6 & 39$\pm$0.9 &578$\pm$32 & Southern shell & 3.22 & this work \\
0453--68.5\tablenotemark{a} &\dots& \dots& \dots& \dots& 37.5$\pm$4 & \dots & Whole & 4.71 & this work \\
	&\dots& \dots& \dots& \dots& 13$\pm$1.3 & 250$\pm$50 &Northern shell & \dots & 1\\
N11L & 1.14$\pm$0.04 & 1.78$\pm$0.04 & 4.10$\pm$0.16 & $\le3.4$ & 13$\pm$0.75 & 132$\pm$21 & Whole & 1.68 & this work \\ 
 & \dots & 1.5 & \dots  & $<8.5$& \dots &\dots & Whole &0.73 & 2\\ 
N86 & 7.5$\pm$0.18 & 7.2$\pm$0.09 & 24$\pm$0.31 & 17.7$\pm$0.47 & 37$\pm$1.35 & 258$\pm$25 & Northeastern region & 3.98 & this work \\
DEM L71&\dots &\dots & \dots & $<1.06$ & 88.2$\pm$8.8 & 455$\pm$94 & Whole shell & \dots & 3 \\
N23\tablenotemark{a} & \dots&\dots &\dots &\dots & 147$\pm$15 & \dots & Whole  & 1.69 & this work \\
	& \dots&\dots &\dots &\dots & 100$\pm$10 & 240$\pm$50 & Southeastern shell  & \dots & 1 \\
DEM L72 &$\le3.2$ &$\le2.9$ &$\le31$ &74.9$\pm$0.69&48.0$\pm$2.2&1036$\pm$30&Whole&19.48&this work\\
N103B &$\le7.5$ & $\le4.5$ & $\le41$ &$\le115$ &505$\pm$2 & 549$\pm$28 & Whole & 0.78& this work \\
N103B kn\tablenotemark{b}&0.8$\pm$0.02 &0.9$\pm$0.01 & 3.4$\pm$0.06 & 10$\pm$0.2 & \dots & \dots  & knots  & 0.04 & this work \\
0509--67.5\tablenotemark{c}&\dots &\dots &\dots & $<0.2$ & 16.7$\pm$1.7 & $<32.7$ & Whole & \dots &  3 \\
0519--69.0\tablenotemark{c}&\dots &\dots &\dots & $<0.9$ & 92.0$\pm$9.2 & $<121$ & Whole  & \dots & 3 \\
N132D NW\tablenotemark{c} &\dots & \dots& \dots& \dots& 730$\pm$73 & 430$\pm$96 & Northwestern shell  & \dots& 1 \\ 
N132D S\tablenotemark{c} &\dots &\dots &\dots & \dots& 1000$\pm$100 & 770$\pm$170 & Southern shell  &\dots& 1 \\
N49B\tablenotemark{a,c} &\dots &\dots &\dots &\dots & 275$\pm$27 & \dots & Whole & 7.43 & this work \\
	&\dots &\dots &\dots &\dots & 43$\pm$4.3 & 395$\pm$79 & Northern shell &\dots  &1 \\
N49\tablenotemark{c}& 32 & 41 & 130 &180 & 1500 & 10200 & Whole&1.11 & 2 \\ 
DEM L205 & \dots & \dots & \dots & \dots & $115\pm4$ & $734\pm129$ & eastern shell & 3.43 & this work \\
 & \dots & \dots & \dots & \dots & $660\pm198$ & $3400\pm1020$ & eastern shell & \dots & 4 \\
SNR in N206 & 3.35$\pm$0.04 & 2.42$\pm$0.05& 16.1$\pm$0.19 & 30$\pm$0.3 & 24.5$\pm$1.3 & $\le134$ & eastern shell & 1.64 & this work \\
DEM L238 & $\le1.04$ & $\le0.92$ & $\le0.49$ & $\le0.48$ & 10$\pm$0.4 & $\le77$ & Whole & 7.56 & this work \\ 
SNR 1987A\tablenotemark{c,d}& \dots& \dots& 2.44$\pm$0.19 & 7.48$\pm$0.19 & 29.8$\pm$1.7 & \dots & Whole& $\le0.06$ & 5 \\
N63A& 37 & 32 & 130 & 300 & 2300 & 7700 &Whole &1.31 & 2\\ 
DEM L241&0.48$\pm$0.15 & 7.57$\pm$0.11 &$\le2.2$ &$\le5.8$ &45.7$\pm$14&$\le1155$& Whole &3.47 & this work\\
DEM L241 W&3.5$\pm$0.06&3.7$\pm$0.04& 9.6$\pm$0.3 & 26$\pm$0.8 & \dots & \dots & western shell & 0.5 & this work\\
DEM L249& 6.1$\pm$0.07 & 7.2$\pm$0.06 & 47$\pm$0.2 & 94$\pm$0.39 & 54$\pm$1.1 & 798$\pm$40 &  Whole &6.63 & this work\\
DEM L256 &0.40$\pm$0.03&0.39$\pm$0.03&1.30$\pm$0.12&1.27$\pm$0.16&5.1$\pm$1.6&108$\pm$14& Northern shell &0.57&this work\\
N157B\tablenotemark{c}& 62 & 87 & $<140$ & $<210$ & \dots & \dots & w/o southern clouds  & 1.20 & 2\\ 
SNR in N159 N &128$\pm$1.3 & 149$\pm$1.2 & 371$\pm$7 & 918$\pm$19 & 8078$\pm$551 &10145$\pm$2858 &Northern lobe &3.25 &this work\\
SNR in N159 S & 22$\pm$0.76 & 26$\pm$0.55 &76.9$\pm$1.8 & 222$\pm$5.1 & 2464$\pm$133 & 4717$\pm$490 &Southern shell &0.66 &this work\\
N158A\tablenotemark{c}& 1.77$\pm$0.23 & 2.19$\pm$0.27 & 3.61$\pm$0.46 & 5.10$\pm$0.74 & 13.19$\pm$3.95 & $<366$ & PWN  &0.03 & 6 \\ 
DEM L299 &  6.67$\pm$0.11 & 7.99$\pm$0.12 & 17.7$\pm$0.4& 97$\pm$1 & 94$\pm$3 & 1475$\pm$69 & eastern shell & 7.52  & this work\\
DEM L316B &1.9$\pm$0.05& 5.0$\pm$0.04 & 14.6$\pm$0.2 &$\le2.4$ &24$\pm$1.4 &597$\pm$37 & Southern shell & 3.34 & this work \\
DEM L316A & 2.8$\pm$0.04 &6.1$\pm$0.06 &23$\pm$0.19 &$\le14$ &31$\pm$1 &445$\pm$26 &Whole & 2.88 & this work \\
0548--70.4\tablenotemark{a,c}&\dots &\dots &\dots & \dots & 42$\pm$4 & \dots & Whole & 3.43 & this work \\
	&\dots &\dots &\dots & $<3.82$ & 2.63$\pm$0.30 & 19.9$\pm$4.7 & Northeastern shell & \dots &3 \\
\enddata

\tablecomments{IR fluxes of detected SNRs in the $Spitzer$ bands. The uncertainty (1$\sigma$) is taking only statistical uncertainties of background fluctuation. Upperlimits are 3$\sigma$. When necessary information does not exist in references, we mark it as ellipsis. References for the fluxes adopted from previous works; [1] \citet{bwill06}: Fluxes only measured from the confined regions that are clearly bright at both 24 and 70 \micron. [2] \citet{will06}: Area on the sky (of SNR) for which flux densities are estimated and descriptions for some cases (N49, N63A, and N157B) are given in Table 2 of the reference. [3] \citet{borko06} [4] \citet{mag12} [5] \citet{bouch06}: [6] \citet{bwill08}}

\tablenotetext{a}{24 \micron~fluxes from the whole SNRs are newly measured by this work. Their 24 and 70 \micron~fluxes from the literatures are the fluxes from limited areas where both band fluxes can be clearly extracted. 
}
\tablenotetext{b}{Knots are no longer distinguishable in the MIPS bands.}
\tablenotetext{c}{IR fluxes of these SNRs are measured by using the $AKARI$ data, too \citep{seok08}.}
\tablenotetext{d}{The IRAC fluxes and the MIPS flux are observed on 6487 and 6184 days after the SN explosion. Radii used for aperture photometry were 4\arcsec$.8$, 6\arcsec$.0$, and 14$\arcsec$.7 at 5.8, 8.0, and 24 \micron, respectively \citep{bouch06}.}
 \end{deluxetable}
\clearpage

%% file: table3.tex
\begin{deluxetable}{lccl}
\tablecolumns{4}
\linespread{1.2}
\tabletypesize{\scriptsize}
\tablewidth{0pt}
\tablecaption{15 LMC SNRs with Morphological Differences between IR and X-Ray \label{tab:corr_irxr}}
\tablehead{ \colhead{SNR}&\colhead{NANTEN\tablenotemark{a}} &\colhead{MAGMA\tablenotemark{b}}&\colhead{Note\tablenotemark{c}}  }
\startdata
 
N11L &  N & N/A & CO emission adjacent to the southern boundary \\
DEM L72\tablenotemark{d} & N & N/A & No CO emission observed \\
N49 & Y & Y & CO emission come from northeast of the SNR \\
DEM L205\tablenotemark{d} & N/A & N/A & No CO emission observed \\
SNR in N206 & N &  N/A & CO emission adjacent to the southern boundary \\
DEM L238 & N & N/A & No CO emission observed \\
N63A & N & N/A & No CO emission observed \\
DEM L241\tablenotemark{d} & Y & Y & CO emission seen inside the SNR \\
DEM L249 & N & Y & Weak CO emission at the southern part seen by MAGMA \\
DEM L256\tablenotemark{d} & Y & N/A & CO emission at the northern part detected by NANTEN \\
N157B & Y & Y & CO emission at the southern part \\
SNR in N159 & Y & Y & Strong CO emission across the SNR  \\
N158A & N & Y & CO emission near the northern boundary seen by MAGMA \\
DEM L316A/B & N & N/A & CO emission adjacent to the northern boundary of B shell\\
SNR 0548--70.4 & N & N/A & No CO emission observed \\
\enddata
\tablecomments{Among 29 SNRs with IR emission, 16 SNRs do not show a strong spatial IR-X-ray correlation. Four out of 16 (N49, DEM L238, N158A, SNR 0548--70.4) show both similarities and differences. 
}
\tablenotetext{a}{Presence or absence of CO emission near an SNR based on the NANTEN survey from \cite[Table 1 in][]{desai10}}
\tablenotetext{b}{Presence or absence of CO emission near an SNR based on the MAGMA survey (Y: Exist, Y: No emission seen, N/A: No available data). Although N11L, SNR in N206, and DEM L316B are not directly covered, CO emission is detected around them by the MAGMA survey.   }
\tablenotetext{c}{Short description about CO emission around SNRs.}
\tablenotetext{d}{{\it XMM-Newton} X-ray images from literatures are used for comparison; DEM L72 and DEM L256 from \citet{kli10}, DEM L205 from \citet{mag12}, and DEM L241 from \citet{bam06}.}

\end{deluxetable}

%% file: table4.tex
\begin{deluxetable}{llc}
\tablecolumns{3}
\tabletypesize{\footnotesize}
\tablewidth{0pt}
 \tablecaption{Origin of the IRAC-band IR Emission from SNRs\label{tab:origin}}
  \tablehead{ \colhead{Origin} & \colhead{SNR} & \colhead{Total} }
 \startdata
Ionic line & N11L, N103B, N49, SN 1987A, N63A, DEM L241\tablenotemark{a}, N157B, DEM L316A/B & 9 \\
Molecular line & N86, DEM L256, SNR in N159\tablenotemark{a}, DEM L299 & 4\\
PAH & 0450--70.9, SNR in N4, DEM L72, SNR in N206, DEM L249 & 5  \\
Synchrotron & N158A & 1 \\ 
 \enddata
 \tablenotetext{a}{The IRAC emissions of specific regions in these SNRs, the western region of DEM L241 and southern shell SNR in N159 are more likely to be dominated by molecular or ionic line emissions, respectively.}
\end{deluxetable}

%% file: table5.tex
\begin{deluxetable}{lcccccc}
\tabletypesize{\footnotesize}

\tablewidth{435pt}

\tablecaption{Properties of X-Ray Emitting Plasma in LMC SNRs \label{tab:nete}}


\tablehead{\colhead{SNR} & \colhead{$n_e$\tablenotemark{a}} & \colhead{$T_e$} & \colhead{$f$\tablenotemark{b}} & \colhead{IR\tablenotemark{c}} & \colhead{Note} & \colhead{Reference} \\ 
\colhead{} & \colhead{(cm$^{-3}$)} & \colhead{(keV)} & \colhead{Factor} & \colhead{Correlation} & \colhead{} & \colhead{}} 

\startdata
0453--68.5 & 0.73 & 0.29 &\dots& 0 &\dots&1\\
SNR in N9 & 0.40 & 0.34 &\dots& 2 & Average $n_e$ of several regions &2\\
N186D & 0.1 & 0.58 &\dots& 2 &\dots&3\\
DEM L71 & 2.7 & 0.65 &\dots& 0 &\dots&4\\
N23 & 6.8 & 0.56 &\dots& 0 &\dots& 1\\
DEM L72sh & 2.99 & 0.17 &\dots& 1 & Shell & 5 \\
DEM L72h & 0.1\dots1 & 0.17 &\dots& 2 & Hot center & 5\\
N103B & 3.16 & 1.0 &\dots& 0 & Overall value of $kT$ and $n_e$& 6 \\
0509--67.5 & 1.6 & 1.9 &\dots& 0 &\dots&4\\
0519--69.0a & 3.4 & 2.1 &\dots& 0 & Fainter portion of SNR &4\\
0519--69.0b & 7.7 & 1.0&\dots& 0 & Three bright knots & 4 \\
N132D NW & 40 & 0.61 &\dots& 0 &\dots&1 \\
N132D S & 17 & 1.02 &\dots& 0 &\dots& 1\\
N49B & 1.3 & 0.36 &\dots& 0 &\dots& 1 \\
N49 E& 0.48 & 0.44 &\dots& 0 & Average of eastern region  &7\\
DEM L205 & 6.7 & 0.25 &\dots& 1 &\dots& 8 \\
SNR in N206 & 0.24 & 0.4 & 1 & 1 & Hot gas region & 9\\
0534--69.9 & 0.28 & 0.76 &\dots& 2 & Average of central region & 10 \\
DEM L238sh & 2.32 & 0.11 &\dots& 0 & Shell & 11 \\
DEM L238c & 2.89 & 0.8 &\dots& 2 & Central region & 11\\
SNR 1987A & 6500 & 0.3 &\dots& 0 &\dots& 12\\
N63A & 4.35 & 0.78 &\dots& 1 &\dots& 13\\
DEM L241h & 0.28 & 0.39 &\dots& 1 &Head region of diffuse emission & 14\\
DEM L241t & 0.27 & 0.35 &\dots& 1 &Tail region of diffuse emission & 14\\
DEM L249sh & 0.56 & 0.36 &\dots& 1 & Shell & 11\\
DEM L249c & 0.99 & 0.75 &\dots& 1 & Central region & 11\\
DEM L256sh & 3.30 & 0.22 &\dots& 1 & Shell &5 \\
DEM L256c & 0.2 & 0.22 &\dots& 2  &Hot center & 5\\
N157B & 0.40 & 0.72 &1 & 1 &\dots& 15 \\
SNR in N159 & 0.46 & 0.59 & 0.3 & 1 &\dots& 16\\
N158A E & 1.44 & 3.08 & 1 & 1 & Eastern region &17\\
N158A SW & 5.6 & 0.74 & 1 & 1 & Bright southwestern region &17\\
J0541.8--6659 & 0.02 & 0.49 &\dots& 2 & Soft thermal emitting region & 18\\
DEM L316B & 0.28 & 0.65 &\dots& 1 &\dots& 19\\
DEM L316A & 0.30 & 1.4 &\dots& 1 &\dots& 19\\
0548--70.4 & 2.0 & 0.65 &\dots& 0 &\dots& 4\\

\enddata)

\tablenotetext{a}{In some literatures, an ionization timescale ($\tau$) is given instead of an explicit electron density. In those cases, $n_e$ is derived from $\tau=n_e\times t$ in this work. The value of $t$ used for the estimation is explained in each reference. }

\tablenotetext{b}{Some literatures give an expression of $n_e$ proportional to a volume filling factor ($f$). If a preferential $f$ is given, we adopt it or, otherwise, assume $f=1$.}

\tablenotetext{c}{Morphological correlation between IR emission and X-ray emission (0: consistent, 1: inconsistent, 2: no IR emission detected). Four SNRs showing similarities and dissimilarities listed in Table \ref{tab:corr_irxr} are categorized as ``consistent'' here. }

\tablerefs{(1) \citet[and references therein]{bwill06}; (2) \citet{sew06}; (3) \citet{jas11}: using an ionization age of 11,000 yr; (4) \citet[and references therein]{borko06}; (5) \citet{kli10}: {\it XMM-Newton} data; (6) \citet{lewis}: using $t\sim1000$ yr; (7) \citet{park03}: using a dynamical age of 6600 yr ; (8) \citet{mag12}; (9) \citet{will05a}; (10) \citet{hend03}: with maximum mixing; (11) \citet{bork06}: using the SNR ages ($t_{\mathrm{SNR}}$, DEM L238: 13,500 yr, DEM L249: 12,500 yr); (12) \citet{dwek08}; (13) \citet{warr03}: using the SNR age of 3500 yr; (14) \citet{bam06}: {\it XMM-Newton} data; (15) \citet{chen06}: using an ionization age of 2000 yr; (16) \citet{sew10}; (17) \citet{park10}; (18) \citet{gro12}: using the SNR age of 23,000 yr; (19) \citet{will05}. }

\end{deluxetable}

\clearpage